# Unified University Inventory System (UUIS)

# Software Design Document

## Authors

Yassine Amaiche
Virginia Cook
Ahmed Daoudi
Mariano Diaz
Gay Hazan
David Zerkler
William Nzoukou
Isabelle Toutant
René Toutant

## Revision History

| Revision | Date | Comments |
|---|---|---|
| 1.0 | 2010-03-20 | First draft |
| 1.1 | 2010-03-23 | Sections 5.3 added |
| 1.2 | 2010-03-24 | Updated ER model and data dictionary |
| 2.0 | 2010-03-26 | Added state diagrams to sections 5.2.1, 5.2.2 and 5.2.5 |
| 2.1 | 2010-04-21 | Added missing parts |
| 2.2 | 2010-04-26 | Proofreading, deployment management |
| 2.3 | 2010-04-27 | Updated references. Added configuration management section. |
| 3.0 | 2010-04-30 | Re-Formatted Detailed Design by Modules. Extended Class Diagrams and Sequence Diagrams. |

## Table of Content

# List of Figures

# List of Tables

# 1. Introduction

## 1.1 Purpose and Scope

The Software Design Documentation (SDD) is the formal document that explains the UUIS system architecture. It is detailed enough to give the reader a clear idea of what the system is designed to implement. Diagrams and tables which describe the three subcomponents of the system, Presentation, Business logic layer and Database layer are provided to clarify the UUIS design.

## 1.2 Acronyms

| | |
|---|---|
| E/R | Entity Relationship |
| IUfA | Imaginary University of Arctica |
| MVC | Model View Controller |
| SDD | Software Design Document |
| SRS | The UUIS Software Requirements Specification |
| SVN | Subversion |
| TBD | To Be Determined |
| UUIS | Unified University Inventory System |

## 1.4 Overview

In this document the system architecture is formally described in UML (Unified Modeling Language). UML diagrams such as ER Diagrams, State Diagrams, Sequence Diagrams, and Class Diagrams are used to model the system's expected behaviour and to ease the coding and development task later on in the project life cycle.

# 2. Design Consideration

## 2.1 Assumptions and Dependencies

Since we are using the PHP technology to build our dynamic websites, we are assuming all the users have access to an internet connection and to a modern computer or mobile phone device with a browser installed on it (Firefox, Explorer, Safari, Opera). At the server level, we are assuming that the server on which the website will be deployed is installed with a PHP interpreter, an web server package such as Apache, and a MySQL DBMS.

## 2.2 General Constraints

The system is designed to be user friendly due to the fact that it will be used by an audience with various backgrounds ranging from those with limited computing experience to computer programmers and database experts. UUIS is designed to be reliable, crash-free and secure. In addition to the previous constraints, the system's code base shall be comprehensively commented, conventions explained, and ambiguities noted to ease the task of maintenance for future developers.

## 2.3 System Environment

Our solutions and technologies are compatible with virtually any web server platform running on Microsoft Windows, MacOS, Linux, and UNIX. Development has been done on Windows, linux and Mac machines using the following software packages:

1. Xampp 1.7.3 bundle of PHP, Apache server, and MySQL.
2. Eclipse PDT 3.5
3. PHP version used is 5.3.1
4. Apache 2.2.14
5. MySQL 5.1.41
6. PHPMyAdmin 3.2.4

## 2.4 Deployment Management

The UUIS System can be deployed on any web server platform which supports PHP (version greater than 5.2) and MSQL (version greater than 5.1). To deploy the application enable SSL encryption in the configuration of the webserver.

### 2.4.1 Application Deployment

A current version of the code base can be downloaded from the following link: http://comp5541t3.svn.sourceforge.net/viewvc/comp5541t3/finalversion.tar.gz?view=tar

Simply un archive this directory under the root of the web server's html directory and rename it to "uuis."

### 2.4.2 SQL Deployment

Set up a user and password in the MySQL database service on the web server. Set the username and password in DB/config.inc.php to the chosen username and password.

Create a database (by default the database is named uuis, the default name can also be changed in DB/config.inc.php) assign read write privileges to the MySQL user. Insert the sample database dump provided in sqldata/uuis.sql into the created database.

Navigate to http://localhost/uuis/HomePage.php or https://localhost/uuis/HomePage.php (using SSL). Enter the sample username a_khan, password "wemooki" to explore the functionalities of the UUIS system.

## 2.5 Development Method

The UUIS development team uses a mixed-control structure. Team members have various backgrounds; every team member has a different set of strengths and weaknesses which makes the adopted structure ideal for exploiting the strengths and expertise that some team members have in specific fields. The Team is subdivided into small sub teams that are led by a Guru who guides the other sub team members and assists them in troubleshooting.

## 2.6 Configuration Management

To ease the coordination of software development and to control the change and evolution of the UUIS product, the UUIS development team used Subversion as a configuration management tool. Subversion not only stored the current version of the source code and documentation, but maintained historical version of each file.

In order to access Subversion repository via command line, Collabnet Subversion Client v1.6.9 had to be installed by all development team member.

The UUIS team 3 project SourceForge.net Subversion repository could be checked out through SVN with the following instruction set:

```
svn co https://comp5541t3.svn.sourceforge.net/svnroot/comp5541t3 comp5541t3
```

# 3. System Architecture

## 3.1 Rationale

The team uses the MVC model due to its popularity and its separation between what the user sees as interface GUI, the business logic of the application, and the relational database level. As an addition to the MVC model, an extra layer has been added – MAPPER -, a layer which maps database tables into respective domain objects. The rationale behind this is to enable the domain logic to be Database Service independent, allowing the system to exchange database technology without having to re-write the whole system. In such a case, only the Mapper layer would need to be re-written.

## 3.2 Logical View

The UUIS is a web-based client application that can be accessed via any web browsers over a network. It is a 3-tier solution in which the user interface, the business logic and the data management are developed and maintained as independent modules:

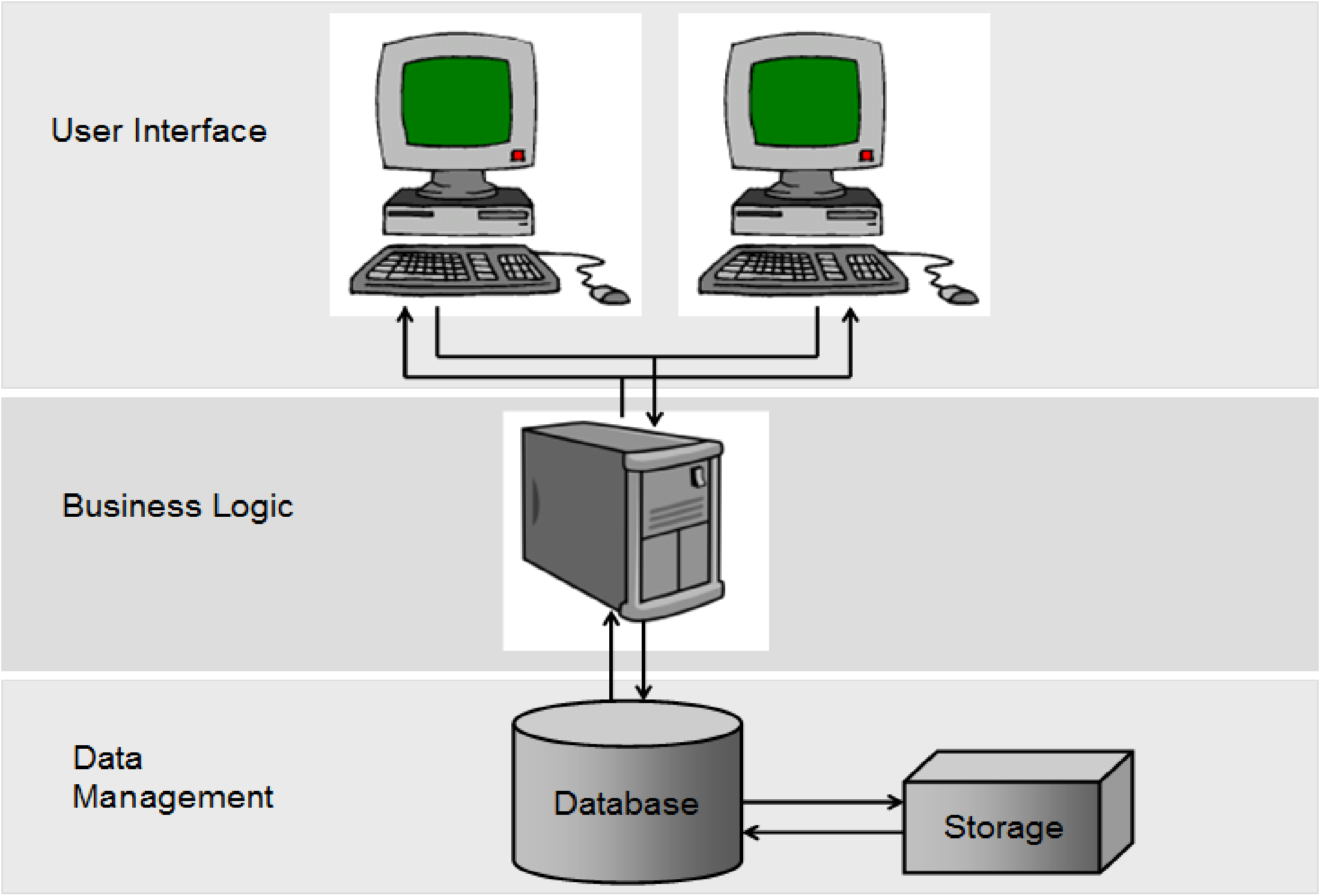

**Figure 1 -** UUIS Tier Architecture

*1. User Interface*

Topmost layer of the application, the interface main function is to display information received from other tiers in a user-friendly format.

*2. Business Logic*

This layer coordinates the application by handling and processing information exchange between the database and the user interface, and by making logical decisions and performing calculations. The logic layer consists of four primary modules (Assets Inventory, Locations Inventory, Software Inventory and Requests Service) as well as two supporting modules (Authentication and Format).

*3. Data Management*

As the bottommost layer of the application, the data management consists of a database server that stores all system's data. Consisting of module specific Mappers and the DB and Query classes the data management layer keeps the data independent from application servers or business logic.

# 4. Interface Description

## 4.1 User Friendly

1. Individual user's history and application history are tracked so as to provide data to discover repeated tasks and to make such tasks easier in later versions of the software. The expiry age and sorting features of the histories are also customizable to allow users to customize their experience.

2. The interface is designed to provide optimal functionality to users of BlackBerry, iPhone or similar mobile devices, and users of Screen Readers (for the visually impaired, or for use of the device without the screen to save battery). This is achieved in two ways:

   a. The Side Menu is located on the right of the page so that the main content area is displayed first in a mobile device browser, and read first by a Screen Reader.

   b. The Footer contains frequent links (generated by the user's history) to reduce the number of page loads. This serves to reduce the data download by the mobile device (mobile device users often pay for the amount of data transferred, and fewer page loads increases battery life) and reducing the wait time for the user to access the pages which they personally access frequently.

3. In addition to the concerns specific to mobile device users and screen readers, other usability constraints were considered. The software uses color coding and highly imageable representations of the data, in addition to providing the traditional tabular display it also uses clickable floor plans for the Locations Module and generated statistic charts for the Assets Module. Given the magnitude of the items inventoried, this allows the user to better visualize the inventory.

Refer to Appendix A.

## 4.2 Security

1. All user input (including URLs) is validated to be sure that executable code or other malicious information is not saved/run/etc (specifically aimed at preventing cross-site scripting and denial of service attacks).
2. Concatenation of SQL queries with un-processed user input should not be allowed to prevent SQL injection attacks.

## 4.3 CSS

1. In order to keep the appearance of the various pages consistent, all the modules included in this project use Cascading Style Sheets (css). This presentational technology minimized the amount of code necessary, resulting in shorter and more streamlined pages. It maintained a consistent style throughout the website content, and also enabled our pages to download faster.

# 5. Detailed Modular Design

## 5.1 Physical Assets Module

The Physical Assets module is responsible for all possible actions performed in the assets inventory. There are four types of Physical Assets: Equipment, Computers, Furniture, and Storage Units.

### 5.1.1 Controller Package

The controller package controls the data flow between the user interface and the model/data/mapper packages of the Physical Assets module. It includes the following files:

*GroupUpdateDispatcher*

Controls the control flow of the EditGroup. It called by View Pages where the form data is directed through validation through database connection and back to the appropriate presentation layer page.

*PhysicalAssetDispatcher*

Controls the control flow of PhysicalAsset INSERTS and UPDATES. It called by View Pages where the form data is directed through validation through database connection and back to the appropriate presentation layer page.

*SelectAssetTypeDispatcher*

Adds extra data to the Assets according to asset Type - e.g Furniture, computer, Equipment, StorageUnit-. It called by SelectPhysicalAssetDispatcher, then passes control to the respective Mapper, according with asset type.

*SelectGroupDispatcher*

Controls the control flow of the Group SELECTS. It called by View Pages where the form data is directed through validation through database connection, then passes control to the respective Mapper.

*SelectPhysicalAssetDispatcher*

Controls the control flow of the PhyssicalAssets SELECTS. It called by View Pages where the form data is directed through validation through database connection, then passes control to the respective Mapper when called from SearchAsset or to SelectAssetTypeDispatcher when called from SearchResults.

ValidateEntries

Validates forms entries. If no error, follows to respective dispatcher, according to which page called it.

### 5.1.2 Model Package

Represents the Business logic of the Physical Assets Module. It includes the following classes:

*PhysicalAssets.Class*

Contains Private object variables of and member functions for PhysicalAssets objects.

| | | | |
|---|---|---|---|
| **Class Name** | PhysicalAssets.Class | | |
| **Inherits From** | None | | |
| **Attributes** | **Visibility** | **Name** | **Description** |
| | Private | $assetId | id of the asset |
| | Private | $locationId | location where is stored |
| | Private | $groupId | groupid the asset is assigned to |
| | Private | $barCode | asset barcode |
| | Private | $legacyCode | asset legacy code |
| | Private | $datePurchased | date the asset was purchased |
| | Private | $warrantyExpiration | date in which the warraty expires |
| | Private | $manufacturer | asset manufacturer |
| | Private | $model | asset model |
| | Private | $category | category of te asset (Furniture, Equipment, etc.) |
| | Private | $status | status of the asset (in-stock, broken, stolen, etc.) |
| | Private | $poNumber | PO number in which the asset was purchased |
| | Private | $pRequest | Request Number. |
| | Private | $departmentId | department id that owns the asset |
| **Methods** | **Visibility** | **Name** | **Description** |
| | Public | getAssetId() | returns assetID |
| | Public | getLocationId() | returns locationID |
| | Public | getCategory() | returns category |
| | Public | getGroupId() | returns groupid |
| | Public | getBarCode() | returns barcode |
| | Public | getDepartmentId() | returns department id |
| | Public | getDatePurchased() | returns date purchased |
| | Public | getWarrantyExpiration() | returns warranty expiration date |
| | Public | getLegacyCode() | returns legacy code |
| | Public | getPRequest() | returns PRequest |
| | Public | getPoNumber() | returns PONumber |
| | Public | getManufacturer() | returns manufacturer |
| | Public | getModel() | returns model |
| | Public | getStatus() | returns status |
| | Public | setAssetId($assetId) | sets assetID |
| | Public | setLocationId($locationId) | sets location id |
| | Public | setbarCode($barCode) | sets barcode |
| | Public | setGroupId($groupId) | sets groupid |
| | Public | setLegacyCode($legacyCode) | sets legacycode |
| | Public | setDatePurchased($datePurchased) | sets date purchased |
| | Public | setManufacturer($manufacturer) | sets manufacturer |
| | Public | setWarrantyExpiration($warrantyExpiration) | sets warranty expiration date |
| | Public | setModel($model) | sets model |
| | Public | setCategory($category) | sets category |
| | Public | setStatus($status) | sets status |
| | Public | setPoNumber($poNumber) | sets PONumber |
| | Public | setPRequest($pRequest) | sets PRequest |
| | Public | setDepartmentId($departmentId) | sets department id |
| | Public | displayAsset() | display fix number of assets details to the screen |
| | Public | displayAssetDetails() | Needs to be overrriden by children classes. |

*Equipment.Class*

Contains Private object variables of and member functions for Equipment objects.
Extends PhysicalAssets Class.

| | | | |
|---|---|---|---|
| **Class Name** | Equipment.Class | | |
| **Inherits From** | PhysicalAsset.Class | | |
| **Attributes** | **Visibility** | **Name** | **Description** |
| | Private | $userId | user id the equiment is assigned to |
| | Private | $serialNo | serial number of the equipment |
| | Private | $type | type of equipment (e.g. projector, laptop, etc) |
| **Methods** | **Visibility** | **Name** | **Description** |
| | Public | setUserId($userId) | sets userid |
| | Public | setSerialNo($serialNo) | sets equipment serial number |
| | Public | setType($type) | sets equipment type |
| | Public | getType() | returns equipment type |
| | Public | getSerialNo() | returns equipment serial number |
| | Public | getUserId() | returns userid assigned to |
| | Public | displayAssetDetails() | Displays equipment details to the screen. |

*Computer.Class*

Contains Private object variables of and member functions for Computer objects.
Extends Equipment Class.

| | | | |
|---|---|---|---|
| **Class Name** | Computer.Class | | |
| **Inherits From** | Equipment.Class | | |
| **Attributes** | **Visibility** | **Name** | **Description** |
| | Private | $processor | Type of processor |
| | Private | $macAddress | MacAddress of the computer |
| | Private | $hardDriveCap | Hard Drive Capacity |
| | Private | $rom | Type of Rom |
| | Private | $ram | Ram capacity |
| **Methods** | **Visibility** | **Name** | **Description** |
| | Public | setProcessor($processor) | sets Processor variable |
| | Public | setMacAddress($macAddress) | sets MacAddress variable |
| | Public | setHardDriveCap($hardDriveCap) | sets HardDriveCap variable |
| | Public | setRom($rom) | sets Rom variable |
| | Public | setRam($ram) | sets Ram variable |
| | Public | getProcessor() | Returns processor Variable |
| | Public | getMacAddress() | Returns MacAddress Variable |
| | Public | getHardDriveCap() | Returns HardDriveCap Variable |
| | Public | getRom() | Returns Rom Variable |
| | Public | getRam() | Returns Ram Variable |
| | Public | displayAssetDetails() | Displays asset details to the screen. |

*Furniture.Class*

Contains Private object variables of and member functions for Furniture objects.
Extends PhysicalAssets Class.

| | | | |
|---|---|---|---|
| **Class Name** | Furniture.Class | | |
| **Inherits From** | PhysicalAssets.Class | | |
| **Attributes** | **Visibility** | **Name** | **Description** |
| | Private | $height | heigth of the furniture |
| | Private | $depth | depth of the furniture |
| | Private | $width | width of the furniture |
| | Private | $color | color of the furniture |
| | Private | $type | type of the furniture (e.g. chair, desk, etc) |
| | Private | $finish | finish type of the furniture |
| **Methods** | **Visibility** | **Name** | **Description** |
| | Public | setDepth($depth) | sets depth variable |
| | Public | setHeight($height) | sets height variable |
| | Public | setWidth($width) | sets width variable |
| | Public | setColor($color) | sets color variable |
| | Public | setType($type) | sets type variable |
| | Public | setFinish($finish) | sets finish variable |
| | Public | getType() | Returns type Variable |
| | Public | getFinish() | Returns finish Variable |
| | Public | getColor() | Returns processor Variable |
| | Public | getWidth() | Returns width Variable |
| | Public | getDepth() | Returns depth Variable |
| | Public | getHeight() | Returns height Variable |
| | Public | displayAssetDetails() | Displays furniture details to the screen. |

*StorageUnit.Class*

Contains Private object variables of and member functions for StorageUnit objects. Extends Furniture Class.

| | | | |
|---|---|---|---|
| **Class Name** | StorageUnit.Class | | |
| **Inherits From** | Furniture.Class | | |
| **Attributes** | **Visibility** | **Name** | **Description** |
| | Private | $numberOfCompartments | Amount of compartment the storage unit has. |
| **Methods** | **Visibility** | **Name** | **Description** |
| | Public | setNumberOfCompartments ($numberOfCompartments) | sets numberOfCompartments variable |
| | Public | getNumberOfCompartments() | returns numberOfCompartments variable |
| | Public | displayAssetDetails() | Displays asset details to the screen. |

*Group.Class*

Contains Private object variables of and member functions for Group objects.

| | | | |
|---|---|---|---|
| **Class Name** | Group.Class | | |
| **Inherits From** | None | | |
| **Attributes** | **Visibility** | **Name** | **Description** |
| | Private | $groupId | id of the group |
| | Private | $groupName | name of the group |
| | Private | $userId | user id the group is assigned to |
| | Private | $userName | user name the group is assigned to |
| | Private | $locationId | location id the group is assigned to |
| | Private | $locationName | location name the group is assigned to |
| | Private | $status | status of the group (active, inactive) |
| | Private | $assets | assets that conform the group. Ussually an array |
| **Methods** | **Visibility** | **Name** | **Description** |
| | Public | getGroupID() | returns groupID |
| | Public | getGroupName() | returns groupname |
| | Public | getUserID() | returns UserID assigned to |
| | Public | getUserName() | returns userName asigned to |
| | Public | getLocationID() | returns locationID assigned to |
| | Public | getLocationName() | returns locationName assigned to |
| | Public | getStatus() | returns status of the group |
| | Public | getAssets() | returns array of assets the group is made of |
| | Public | setGroupID($groupId) | sets group id |
| | Public | setGroupName($groupName) | sets group name |
| | Public | setUserID($userId) | sets user id |
| | Public | setUserName($userName) | sets user name |
| | Public | setLocationID($locationId) | sets location id |
| | Public | setLocationName($locationName) | sets location name |
| | Public | setStatus($status) | sets group status |
| | Public | setAssets($assets) | sets Array of assets |
| | Public | displayGroup() | displays group details to the screen |

### 5.1.3 Mapper Package

The Mapper Package provides interaction between the database and the model of the Physical Assets Module. It includes the following files:

*ComputerMapper*

Provides a MySQL mapper between the MySQL database and Computer objects.

*EquipmentMapper*

Provides a MySQL mapper between the MySQL database and Equipment objects.

*FurnitureMapper*

Provides a MySQL mapper between the MySQL database and Furniture objects.

*GroupMapper*

Provides a MySQL mapper between the MySQL database and Group objects.

*PhysicalAssetMapper*

Provides a MySQL mapper between the MySQL database and PhysicalAsset objects.

*StorageUnitMapper*

Provides a MySQL mapper between the MySQL database and StorageUnit objects.

### 5.1.4 View Package

The View package provides all the user interfaces for the Physical Assets Module. It includes the following files:

*AddAsset*

Presents Forms to add a physical asset. Then goes to AdditionalAssetInformation to complete asset Type Form.

*AdditionalAssetInformation*

Presents additional form to add assets subTypes - e.g. Furniture, Computer, Equipment, Storage Unit.

*AdditionalGroupInformation*

Presents additional Form for Grouping objects - e.g. how many assets it contains -.

*AssetDetailsPage*

Displays complete asset details and allows user to update the asset.

*AssetMenu*

Presents the available functions for the Asset module.

*ConfirmMessage*

Presents Confirmation message when Transaction was successfully executed.

*CreateGroup*

Present General Form to create a group. Then goes to AdditionalGroupInformation to add the group asset details.

*EditGroup*

Presents available search options to View/Updategroups.

*GroupCriteriaError*

Displays error in EditGroup criteria selection.

*GroupResults*

Displays group search results, and allow user to select a group and update certain data on it.

*Reports*

Displays the reports, according with the selected criteria from ReportsMenu.

*ReportsMenu*

Presents the available functions for the Reports module.

*SearchAsset*

Presents general criteria to search for assets.

*SearchAssetResults*

Displays the general search results for the Search Asset page. If user wants details, clicks on view details to go to AssetDetailsPage.

### 5.1.5 Physical Assets Class Diagram

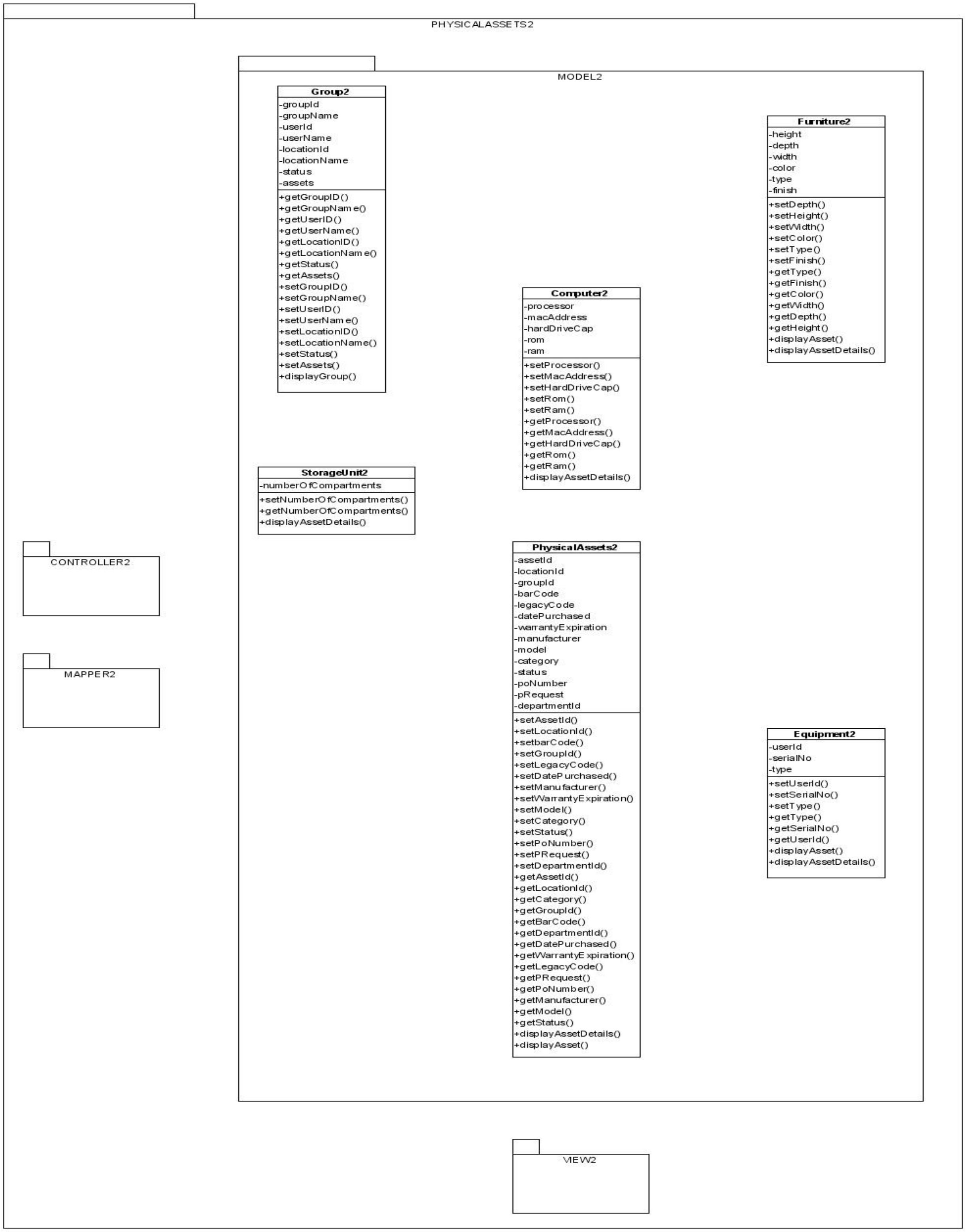

Figure 2: Physical Assets Class Diagram

## 5.2 Locations Module

Module to perform actions (search, edit, view) in the Locations Inventory.

### 5.2.1 Locations.config

A config file used to customize the Locations Object parameters which are shown in the search, allowing the system admin to add a new item to the locations search without having to modify the php. The locations.config is also used to change the the display language from English to French.

### 5.2.2 Controller Package

The controller package controls the data flow between the user interface and the model/data/mapper packages of the Locations module. It includes the following files:

*LocationDispatcher*

Controls the control flow of the general Location actions. It called by View Pages where the form data is directed through validation through database connection and back to the appropriate presentation layer page. This dispatcher was replaced by the SelectLcationDispatcher.

*SelectLocationDispatcher*

Controls the control flow of the SearchLocation AddLocation and EditLab actions. It called by View Pages where the form data is directed through validation through database connection and back to the appropriate presentation layer page (SearchLocationResults, AddLocation and AddLabMember respectively).

*ValidateEntries*

Validates all forms which are sent which might change the database. Confirms that the data has been entered correctly and doesn't contain any SQL injections. It then directs the control flow back to the Dispatcher.

### 5.2.3 Model Package

Represents the Business logic of the Locations Module. It includes the following classes:

*Locations.Class*

Contains Private object variables of and member functions for Locations objects.

| Class Name | Locations.Class | | |
|---|---|---|---|
| Inherits From | None | | |
| **Attributes** | **Visibility** | **Name** | **Description** |
| | Private | $LocationName | name of the location |
| | Private | $locationId | id of the location |
| | Private | $Type | type (room, locker, etc) |
| | Private | $SquareMeters | Surface of the location |
| | Private | $ResponsibleID | user id of responsible for the location |
| | Private | $ResponsibleName | name of user responsible for the location |
| | Private | $FloorID | id of floor the location is located |
| | Private | $FloorName | name of floor the locatin is located |
| | Private | $BuildingID | id of building the location is located |
| | Private | $BuildingName | name of building the locatin is located |
| | Private | $status | status of the location |
| | Private | $DepartmentName | name of department that owns the location |
| | Private | $departmentId | department id that owns the location |
| **Methods** | **Visibility** | **Name** | **Description** |
| | Public | getBuildingName() | returns building name |
| | Public | getLocationId() | returns locationID |
| | Public | getFacultyID() | returns faculty id |
| | Public | getFacultyName() | returns faculty name |
| | Public | getDepartmentName() | returns department name |
| | Public | getDepartmentId() | returns department id |
| | Public | getLocationName() | returns location name |
| | Public | getType() | returns location type |
| | Public | getSquareMeters() | returns location square meters |
| | Public | getResponsibleID() | returns id of responsible |
| | Public | getResponsibleName() | returns responsible name |
| | Public | getFloorID() | returns floor id |
| | Public | getFloorName() | returns floor name |
| | Public | getStatus() | returns status |
| | Public | getBuildingID() | returns building id |
| | Public | setLocationName($LocationName) | sets location name |
| | Public | setLocationId($locationId) | sets location id |
| | Public | setDepartmentName($DepartmentName) | sets department name |
| | Public | setDepartmentId($departmentId) | sets department id |
| | Public | setType($Type) | sets location type |
| | Public | setSquareMeters($SquareMeters) | sets location square meters |
| | Public | setResponsibleID($ResponsibleID) | sets responsible id |
| | Public | setResponsibleName($ResponsibleName) | sets responsible name |
| | Public | setFloorID($FloorID) | sets floor id |
| | Public | setBuildingID($BuildingID) | sets building id |
| | Public | setStatus($status) | sets status |
| | Public | setBuildingName($BuildingName) | sets building name |
| | Public | setFacultyID($FacultyID) | sts faculty id |
| | Public | setFacultyName($FacultyName) | sets faculty name |
| | Public | displayTableHeadingRow() | displays table headers to the screen |
| | Public | displayLocationInRow() | displays location general details to the screen |
| | Public | displayLocationDetails() | left to be implemented by children classes |

*Building.Class*

Contains Private object variables of and member functions for Building objects. Extends Locations.Class

| **Class Name** | Building.Class | | |
|---|---|---|---|
| **Inherits From** | Locations.Class | | |
| **Attributes** | **Visibility** | **Name** | **Description** |
| | Private | $buildingName | name of the building |
| | Private | $address | address of the building |
| | Private | $city | city of the building |
| | Private | $province | province of the building |
| | Private | $country | country of the building |
| | Private | $zipCode | building's zipcode |
| **Methods** | **Visibility** | **Name** | **Description** |
| | Public | getBuildingName() | returns building name |
| | Public | getAddress() | returns building address |
| | Public | getCity($city) | returns building's city |
| | Public | getProvince() | returns building's province |
| | Public | getCountry() | returns building's country |
| | Public | getZipCode() | returns building's zipcode |
| | Public | setBuildingName($buildingName) | sets building name |
| | Public | setAddress($address) | sets building address |
| | Public | setCity($city) | sets building's city |
| | Public | setProvince($province) | sets building's province |
| | Public | setCountry($country) | sets building's country |
| | Public | setZipCode($zipCode) | sets building's zipcode |

*Floor.Class (not implemented)*

Contains Private object variables of and member functions for Floor objects.  Extends Locations.Class

*Lab.Class*

Contains Private object variables of and member functions for Lab objects.  Extends Locations.Class

| **Class Name** | Lab.Class |
|---|---|
| **Inherits From** | Locations.Class |

| Attributes | Visibility | Name | Description |
|---|---|---|---|
| | Private | $ResponsibleLastName | first name of lab responsible |
| | Private | $ResponsibleFirstName | last name of lab responsible |
| | Private | $LabType | type of lab |
| | Private | $Capacity | max number of occupants |
| | Private | $LabMembers | users assigned to the lab (array) |
| | Private | $LabName | name of the lab |
| Methods | Visibility | Name | Description |
| | Public | getResponsibleFirstName() | returns responsible first name |
| | Public | getResponsibleLastName() | returns responsible last name |
| | Public | getLabType() | returns lab type |
| | Public | getCapacity() | returns lab capacity |
| | Public | getLabName() | returns lab name |
| | Public | getLabMembers() | returns array of lab members |
| | Public | setResponsibleFirstName($ResponsibleFirstName) | sets responsible first name |
| | Public | setResponsibleLastName($ResponsibleLastName) | sets responsible last name |
| | Public | setLabType($LabType) | sets lab type |
| | Public | setCapacity($Capacity) | sets lab capacity |
| | Public | setLabMembers($LabMembers) | sets lab name |
| | Public | setLabName($LabName) | sets array of lab members |

*Office.Class*

Contains Private object variables of and member functions for Office objects. Extends Locations.Class

| Class Name | Office.Class | | |
|---|---|---|---|
| Inherits From | Locations.Class | | |
| Attributes | Visibility | Name | Description |
| | Private | $officeNo | office number |
| Methods | Visibility | Name | Description |
| | Public | getOfficeNo() | returns office number |
| | Public | setOfficeNo($officeNo) | sets office number |

*Room.Class*

Contains Private object variables of and member functions for Room objects. Extends Locations.Class

| Class Name | Room.Class | | |
|---|---|---|---|
| Inherits From | Locations.Class | | |
| Attributes | Visibility | Name | Description |
| | Private | $roomNo | room number |
| Methods | Visibility | Name | Description |
| | Public | getRoomNo() | returns room number |
| | Public | setRoomNo($roomNo) | sets room number |

### 5.2.4 Mapper Package

The Mapper Package provides interaction between the database and the model of the Locations Module. It includes the following files:

*LabMapper*

Provides a MySQL mapper between the MySQL database and Lab objects.

*LocationMapper*

Provides a MySQL mapper between the MySQL database and Location objects.

### 5.2.5 View Package

The View package provides all the user interfaces for the Locations Module. It includes the following files:

*AddLabMember*

Displays the available labs in the inventory and allows the user to assign a new member to the lab (for example, a graduate student). It also allows the user to edit the Lab Head (person responsible) and/or the capacity of the Lab.

*AddLocation*

Allows the user to add a new location, if the location is a Lab the system prompts the user to enter the lab information as well.

*LocationDetailsPage*

Works in conjunction with the Location Search page, it displays the details of the location (beyond the parameters which are displayed in the search results table).

*LocationMenu*

Presents the available functions for the Location module.

*SearchLocation*

Allows the user to enter substrings which match values in the database. The user can search over more than one parameter, the intersection is returned in the SearchResults page.

*SearchLocationResults*

Displays the search results for the Search Location page.

### 5.2.6 Locations class Diagram

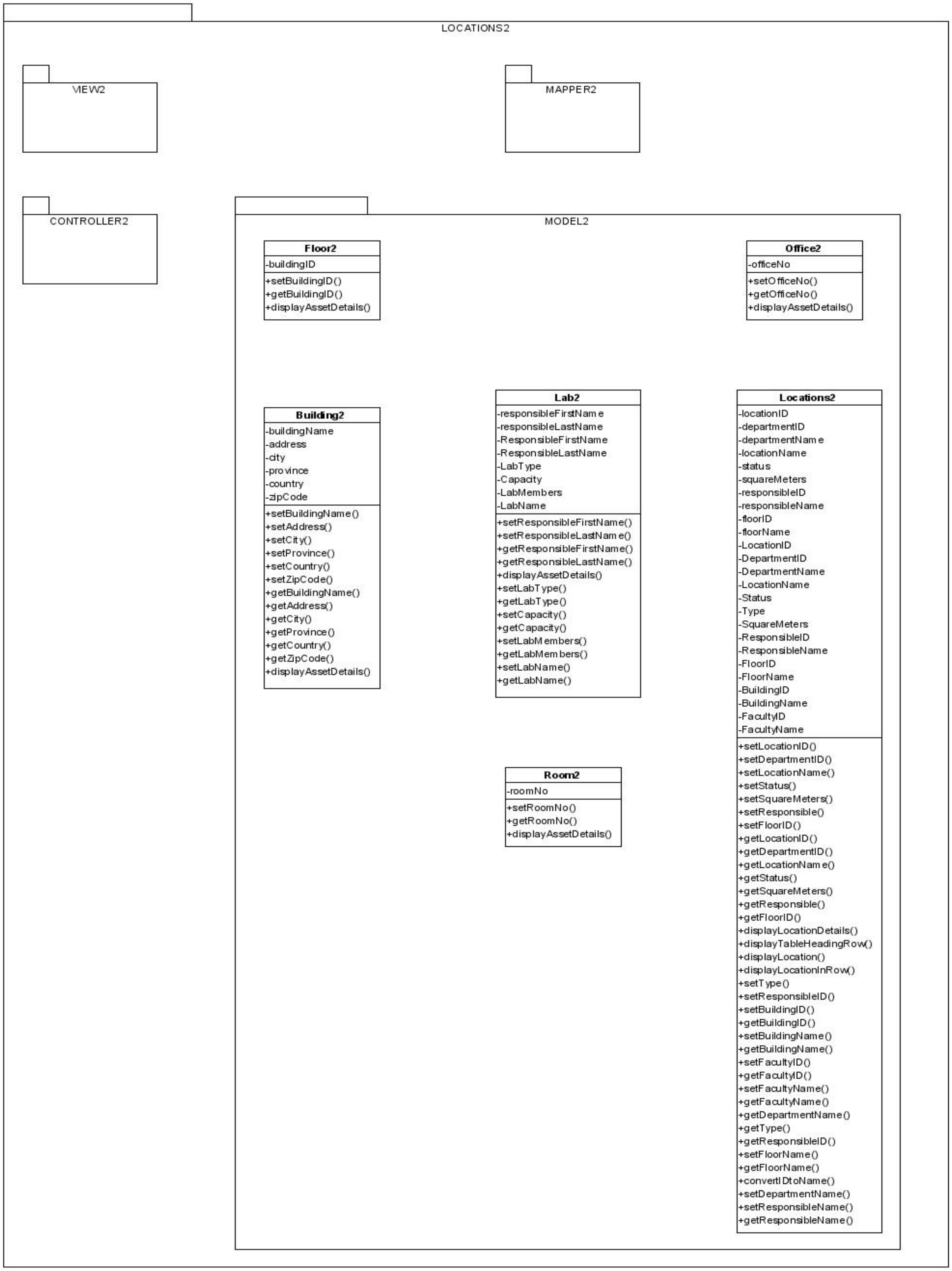

Figure 3: **Locations Class Diagram**

## 5.3 Software Module

Module to manage (search, edit, view, assign) software and licenses Inventory.

### 5.3.1 Controller Package

The controller package controls the data flow between the user interface and the model/data/mapper packages of the Software module. It includes the following files:

*SelectSoftwareDispatcher*

Perform the Sql and send Session to the mapper for display

*ValidateEntries*

Validate all the form inputs

### 5.3.2 Model Package

Represents the Business logic of the Software Module. It includes the following classes:

*License.Class*

Contains Private object variables of and member functions for License objects.

| **Class Name** | License.Class | | |
|---|---|---|---|
| **Inherits From** | None | | |
| **Attributes** | **Visibility** | **Name** | **Description** |
| | Private | $LicenseKey | license key |
| | Private | $PONumber | PONumber of the license |
| | Private | $Type | Type of license (site, research, etc.) |
| | Private | $NumberLicense | License number |
| | Private | $ExpirationDate | License's Expiration date |
| | Private | $datePurchased | date In which license was purchased |
| | Private | $SoftwareID | id of the software the license belongs to |
| | Private | $NumberLicenseRemaining | number of remaining licenses |
| | Private | $DepartmentName | Department name that owns the license |
| | Private | $FacultyName | faculty name of the department |
| **Methods** | **Visibility** | **Name** | **Description** |
| | Public | getLicenseKey() | returns license key |
| | Public | getDatePurchased() | returns date purchased |
| | Public | getPONumber() | returns PONumber |
| | Public | getType() | returns license Type |
| | Public | getNumberLicense() | returns license number |
| | Public | getExpirationDate() | returns expiration date |
| | Public | getSoftwareID() | returns software id |
| | Public | getDepartmentName() | returns department name |
| | Public | getFacultyName() | returns faculty name |
| | Public | getNumberLicenseRemaining() | returns number of remaining licenses |
| | Public | setLicenseKey($LicenseKey) | sets license key |
| | Public | setDatePurchased($DatePurchased) | sets date purchased |
| | Public | setPONumber($PONumber) | sets PONumber |
| | Public | setType($Type) | sets license Type |
| | Public | setNumberLicense($NumberLicense) | sets license number |
| | Public | setDatePurchased($datePurchased) | sets date purchased |
| | Public | setExpirationDate($ExpirationDate) | sets expiration date |
| | Public | setSoftwareID($SoftwareID) | sets software id |
| | Public | setDepartmentName($DepartmentName) | sets department name |
| | Public | setFacultyName($FacultyName) | sets faculty name |
| | Public | setNumberLicenseRemaining($NumberLicenseRemaining) | sets number of remaining licenses |
| | Public | displayLicense() | displays license data to the screen |

*Software.Class*

Contains Private object variables of and member functions for Software objects.

| **Class Name** | Software.Class | | |
|---|---|---|---|
| **Inherits From** | None | | |
| **Attributes** | **Visibility** | **Name** | **Description** |
| | Private | $name | name of the software |
| | Private | $vendorID | id of the vendor |
| | Private | $category | software category (OS, Antivirus, etc) |
| | Private | $version | software version (SP1, 5.2, etc) |
| | Private | $media | Media type (usb key, cd, etc.) |
| | Private | $SoftwareID | id of the software |
| | Private | $vendorName | name of the vendor |
| **Methods** | **Visibility** | **Name** | **Description** |
| | Public | getName() | returns SW name |
| | Public | getSoftwareID() | returns software id |
| | Public | getVendorID() | returns vendor id |
| | Public | getCategory() | returns category |
| | Public | getVersion() | returns SW version |
| | Public | getMedia() | returns SW media type |
| | Public | getVendorName() | returns vendor name |
| | Public | setVendorName($vendorName) | sets vendor name |
| | Public | setMedia($media) | sets SW media type |
| | Public | setVersion($version) | sets SW verdion |
| | Public | setCategory($category) | sets SW category |
| | Public | setVendorID($vendorID) | sets vendor id |
| | Public | setSoftwareID($SoftwareID) | sets SW id |
| | Public | setName($name) | sets SW name |
| | Public | displaySoftware() | displays software data to the screen |

### 5.3.3 Mapper Package

The Mapper Package provides interaction between the database and the model of the Software Module. It includes the following files:

*SoftwareMapper*

Provides a MySQL mapper between the MySQL database and Software objects.

### 5.3.4 View Package

The View package provides all the user interfaces for the Software Module. It includes the following files:

*AddLicense*

Displays a form where the user can enter details about a new license.

*AddLicenseResults*

Tell the users if the license was correctly entered or not.

*AddSoftware*

Display a form where user can enter details about a new software.

*AssignLicense*

Assign a license to a user according to your permissions.

*AssignLicenseResults*

A page telling the user if the license was correctly assigned or not.

*DisplaySoftwareDetails*

Display a software and all the license attached to it.

*EditSoftware*

Display a form in which the user can update details about a software

*EditSoftwareResults*

A page telling the user if the software was correctly edited or not

*ErrorPage*

Most of the errors a redirect on this page

*SearchSoftware*

A form a user uses to search for a software

*SearchSoftwareResults*

Displays the search results for the Search Software page.

*SoftwareMenu*

Presents the available functions for the Software module.

### 5.3.5 Software Class Diagram

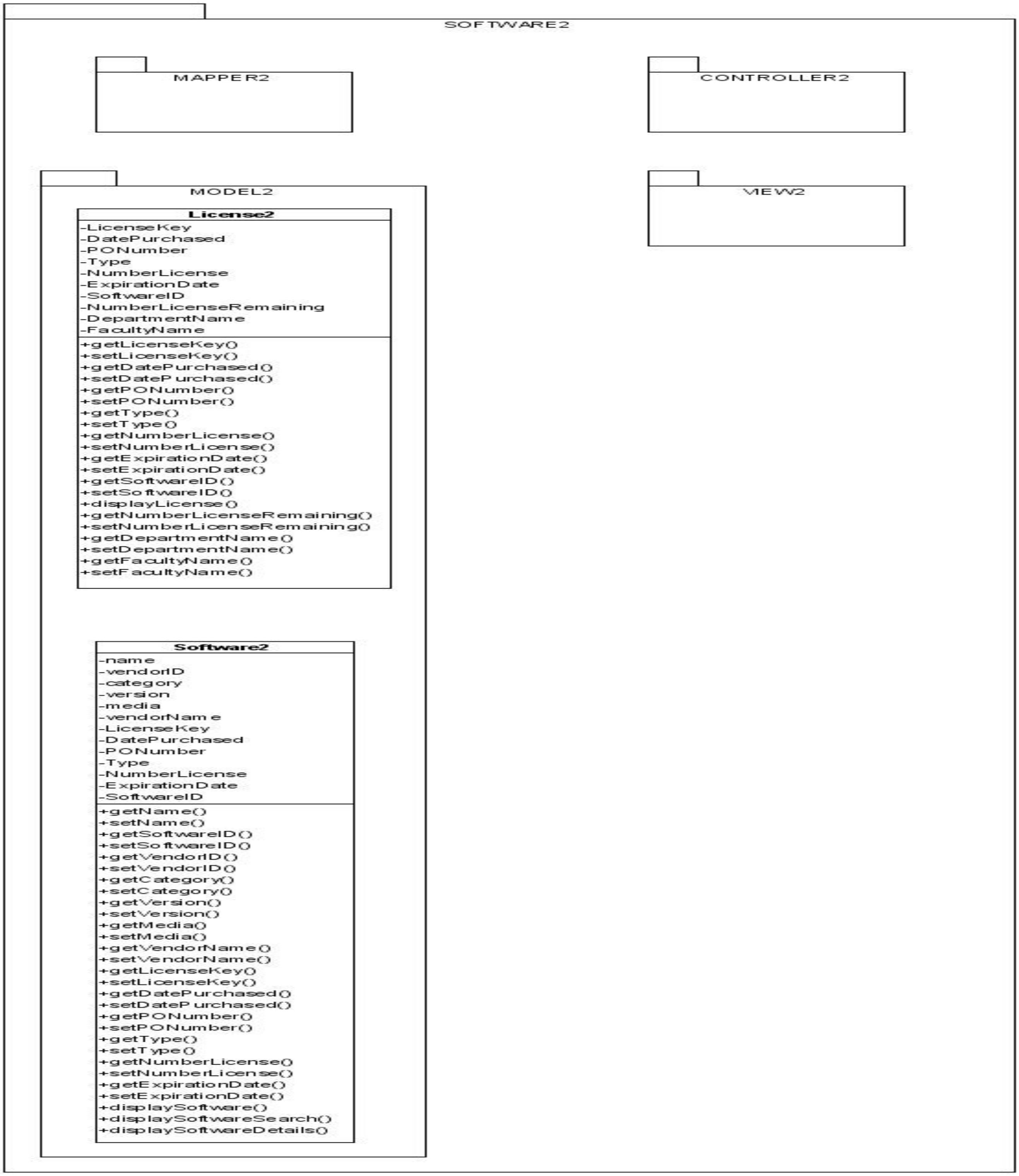

**Figure 4:** Software Class Diagram

## 5.4 Request Module

The Request module is responsible for all possible actions performed in the assets inventory. There are two types of Requests: General and specific.

### 5.4.1 Controller Package

The controller package controls the data flow between the user interface and the model/data/mapper packages of the Request module. It includes the following files:

*RequestDispatcher*

This dispatcher matches the request entry to the appropriate query in the sqlQuery class.

*SelectRequestDispatcher*

This dispatcher matches the request entry to the appropriate query in the selQuery class.

### 5.4.2 Model Package

Represents the Business logic of the Request Module. It includes the following classes:

*Request.Class*

Contains Private object variables of and member functions for Request objects.

| **Class Name** | Request.Class (Abstract) | | |
|---|---|---|---|
| **Inherits From** | None | | |
| **Attributes** | **Visibility** | **Name** | **Description** |
| | Private | $requestId | id of the request |
| | Private | $status | status of the request (approved, pending, closed, etc) |
| | Private | $closureNote | closure note |
| | Private | $category | category (Technical, Administrative) |
| | Private | $requester | id of user who made the request |
| | Private | $approver | id of the user who approved the request |
| **Methods** | **Visibility** | **Name** | **Description** |
| | Public | __construct($requestid,$category,$status, $closurenote, $requester,$approver) | constructor |
| | Public | closeRequest() | change status to closed |
| | Public | setClosureNote($closureNote) | sets closure note |
| | Public | getRequestId() | returns request id |
| | Public | getStatus() | returns request status |
| | Public | getCategory() | returns category |
| | Public | getClosureNote() | returns closure note |
| | Public | getRequester() | returns id of the requester |
| | Public | getApprover() | returns id of the approver |
| | Abstract | displayRequestDetails() | empty function |
| | Abstract | displayRequest() | empty function |

*GeneralRequest.Class*

Contains Private object variables of and member functions for GeneralRequest objects. Implements Request.Class.

| **Class Name** | GeneralRequest.Class | | |
|---|---|---|---|
| **Inherits From** | Request.Class | | |
| **Attributes** | **Visibility** | **Name** | **Description** |
| | Private | $description | text description of the general request |
| **Methods** | **Visibility** | **Name** | **Description** |
| | Public | __construct($requestid,$category,$status, $closurenote, $requester,$approver) | constructor |
| | Public | getDescription() | returns description |
| | Public | displayRequest() | implements parent fucntion, and displays general data about the request to the screen |
| | Public | displayRequestDetails() | implements parent fucntion, and displays specific data about the request to the screen |

*SpecificRequest.Class (not implemented)*

Contains Private object variables of and member functions for SpecificRequest objects. Implements Request.Class.

### 5.4.3 Mapper Package

The Mapper Package provides interaction between the database and the model of the Request Module. It includes the following files:

*RequestMapper*

Provides a MySQL mapper between the MySQL database and Request objects.

### 5.4.4 View Package

The View package provides all the user interfaces for the Request Module. It includes the following files:

*GeneralRequest*

This is the php page to enter new requests depending on your role ID.

*RequestDetailsPage*

Allows a user to view the details of a particular request.

*RequestMenu*

Presents the available functions for the Request module.

*RequestResult*

This page displays the results of a search for general request.

*SearchRequest*

This is the search page for request.

*SearchRequestResults*

Displays the search results for the Search Request page.

### 5.4.5 Request Class diagram

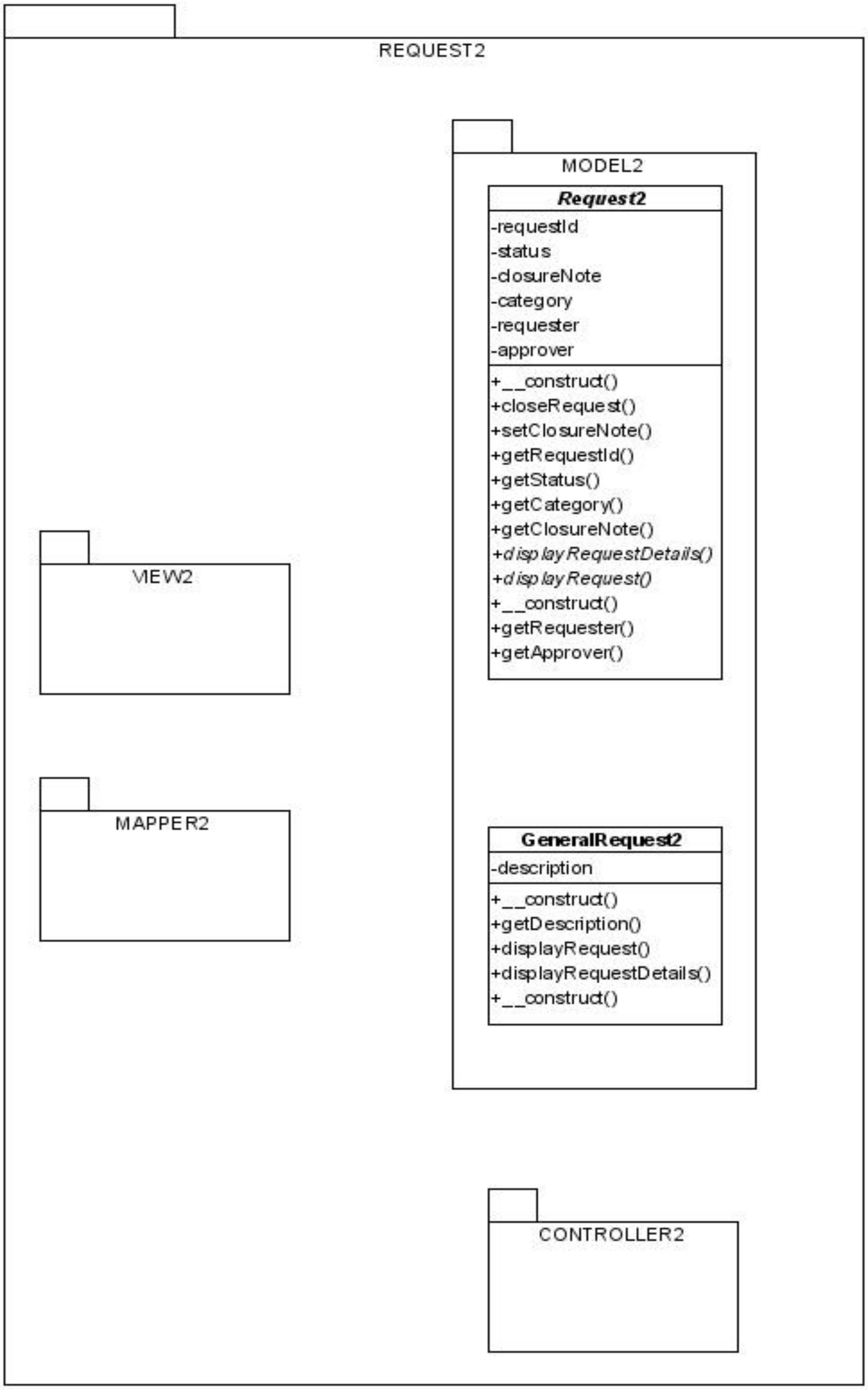

**Figure 5:** Request Class Diagram

## 5.5 Authentication Module

The Authentication module is responsible for all authentication when opening a session in the System. It also manages users.

### 5.5.1 Controller Package

The controller package controls the data flow between the user interface and the model/data/mapper packages of the Authentication module. It includes the following files:

*AccountDispatcher*

Controls the control flow of My Profile actions such as Update Account or Change Password). It is called by View Pages where the form data is directed through validation through database connection, then passes control to the respective Mapper.

*AuthenticationDispatcher*

Controls the control flow of Authentication actions such as Sign In, Reset Password or Sign Out. It is called by View Pages where the form data is directed through validation through database connection, then passes control to the respective Mapper.

*ChooseDepartmentDispatcher*

Controls the control flow of Department actions . It is called once by the AuthenticationDispatcher during Sign In operation.

### 5.5.2 Model Package

Represents the Business logic of the Authentication Module. It includes the following classes:

*Department.Class*

Contains Private object variables of and member functions for Department objects.

| **Class Name** | Department.Class | | |
|---|---|---|---|
| **Inherits From** | None | | |
| **Attributes** | **Visibility** | **Name** | **Description** |
| | Private | $departmentId | id of the department |
| | Private | $facultyId | id of the faculty the department belongs to |
| | Private | $departmentName | name of the department |

| Methods | Visibility | Name | Description |
|---|---|---|---|
| | Public | Department($departmentId, $facultyId, $departmentName) | constructor |
| | Public | setDepartmentId($departmentId) | sets department id |
| | Public | setFacultyId($facultyId) | sets faculty id |
| | Public | setDepartmentName($departmentName) | sets department name |
| | Public | getFacultyId() | returns faculty id |
| | Public | getDepartmentName() | returns department name |
| | Public | getDepartmentId() | returns department id |

*Faculty.Class*

Contains Private object variables of and member functions for Faculty objects.

| Class Name | Faculty.Class | | |
|---|---|---|---|
| Inherits From | None | | |
| Attributes | Visibility | Name | Description |
| | Private | $facultyName | name of the faculty |
| | Private | $facultyId | id of the faculty |
| | Private | $deanLastName | dean last name |
| | Private | $deanFirstName | dean first name |
| Methods | Visibility | Name | Description |
| | Public | Faculty($facultyId, $facultyName, $deanFirstName, $deanLastName) | constructor |
| | Public | setFacultyName($facultyName) | sets faculty name |
| | Public | setDeanFirstName($deanFirstName) | sets dean first name |
| | Public | setDeanLastName($deanLastName) | sets dean last name |
| | Public | setFacultyId($facultyId) | sets faculty id |
| | Public | getDeanFirstName() | returns dean first name |
| | Public | getDeanLastName() | returns dean last name |
| | Public | getFacultyId() | returns faculty id |
| | Public | getFacultyName() | returns faculty name |

*Menu.Class*

Contains Private object variables of and member functions for Menu objects.

| Class Name | Menu.Class | | |
|---|---|---|---|
| Inherits From | None | | |
| Attributes | Visibility | Name | Description |
| | Private | $menuId | id of the menu |
| | Private | $menuName | name of the menu |
| | Private | $menuAddress | address of the menu |
| Methods | Visibility | Name | Description |
| | Public | Menu($menuId, $menuName, $menuAddress) | constructor |
| | Public | setMenuId($menuId) | sets menu id |
| | Public | setMenuAddress($menuAddress) | sets menu address |
| | Public | setMenuName($menuName) | sets menu name |
| | Public | getMenuId() | returns menu id |
| | Public | getMenuName() | returns menu name |
| | Public | getMenuAddress() | returns menu address |

*User.Class*

Contains Private object variables of and member functions for User objects.

| Class Name | User.Class | | |
|---|---|---|---|
| Inherits From | None | | |
| Attributes | Visibility | Name | Description |
| | Private | $userId | id of the user |
| | Private | $roleId | id of the role the user has |
| | Private | $userName | username of the user |
| | Private | $password | user password |
| | Private | $firstName | user first name |
| | Private | $lastName | usesr last name |
| | Private | $email | user's email |
| Methods | Visibility | Name | Description |
| | Public | User($userId, $roleId, $userName, $password, $firstName, $lastName, $email) | constructor |
| | Public | setUserId($userId) | sets user id |
| | Public | setRoleId($roleId) | sets role id |
| | Public | setUserName($userName) | sets username |
| | Public | setPassword($password) | sets password |
| | Public | setFirstName($firstName) | sets user first name |
| | Public | setLastName($lastName) | sets user last name |
| | Public | setEmail($email) | sets user's email |
| | Public | getUserId() | returns user id |
| | Public | getRoleId() | returns role id |
| | Public | getUserName() | returns username |
| | Public | getPassword() | returns password |
| | Public | getFirstName() | returns user first name |
| | Public | getLastName() | returns user last name |
| | Public | displayUser() | displays user to the screen |
| | Public | getEmail() | returns user's email |

### 5.5.3 Mapper Package

The Mapper Package provides interaction between the database and the model of the Authentication Module. It includes the following files:

*DepartmentMapper*

Provides a MySQL mapper between the MySQL database and Department objects.

*FacultyMapper*

Provides a MySQL mapper between the MySQL database and Faculty objects.

*MenuMapper*

Provides a MySQL mapper between the MySQL database and Menu objects.

*UserMapper*

Provides a MySQL mapper between the MySQL database and User objects.

#### 5.5.4 View Package

The View package provides all the user interfaces for the Authentication Module. It includes the following files:

*ChangePassword*

Displays a form where the user has to enter old password and new password (2).

*ChooseDepartment*

Displays a form where the user has to choose his/her department before sign in operation is completed.

*ConfirmMessage*

Displays a confirmation message that a password has been successfully changed.

*Main*

Displays the UUIS main page content.

*MyProfile*

Presents the available sub-menu of My Profile

*ResetPassword*

Display a form in which user is asked to enter username and captcha. System will identify the user and e-mail a link to reset his/her password.

*UpdateAccount*

Display a form in which the user can update his personal account information.

*ViewAccount*

Displays the user information such has username, name, e-mail, etc.

### 5.5.5 Authentication Class Diagram

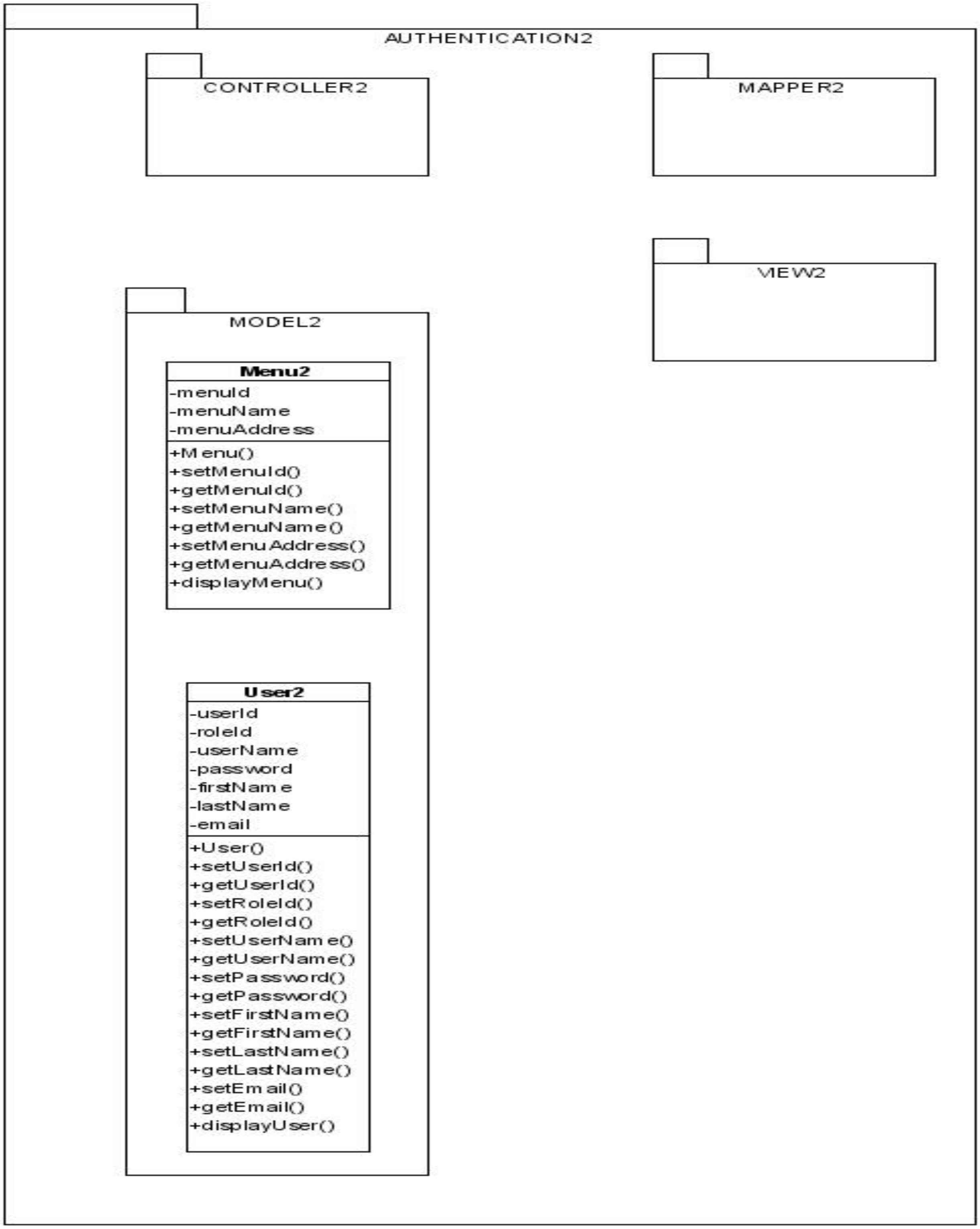

Figure 5: **Authentication Class Diagram**

## 5.6 Permissions Module

The Permissions module is responsible for managing user permissions .

### 5.6.1 Controller Package

The controller package controls the data flow between the user interface and the model/data/mapper packages of the Permissions module. It includes the following files:

*VerifyAuthorization*

Verifies that the user has enough permissions to perform a given action.

### 5.6.2 Model Package

Represents the Business logic of the Permissions Module. It includes the following classes:

*Permission.Class*

Contains Private object variables of and member functions for Permission objects.

| | | | |
|---|---|---|---|
| **Class Name** | Permission.Class | | |
| **Inherits From** | None | | |
| **Attributes** | **Visibility** | **Name** | **Description** |
| | Private | $permissionId | id of the permission |
| | Private | $permissionName | name of the permission |
| | Private | $authorized | indicates authorization |
| **Methods** | **Visibility** | **Name** | **Description** |
| | Public | Permission($permissionId, $permissionName, $authorized) | constructor |
| | Public | setPermissionId($permissionId) | sets permission id |
| | Public | setPermissionName($permissionName) | sets permission name |
| | Public | setAuthorization($authorized) | sets authorization |
| | Public | getAuthorization() | returns authorization |
| | Public | getPermissionName() | returns permission name |
| | Public | getPermissionId() | returns permission id |

### 5.6.3 Mapper Package

The Mapper Package provides interaction between the database and the model of the Permissions Module. It includes the following files:

*PermissionMapper*

Provides a MySQL mapper between the MySQL database and Permission objects.

### 5.6.4 View Package

The View package provides all the user interfaces for the Permissions Module. It includes the following files:

*SystemAdminMenu*

Presents the available functions for the SystemAdmin module.

## 5.7 DB & Query Modules

The DB & Query modules work together to provide a common interface for SQl commands. The SQL service implemented is MySql.

### 5.7.1 DB Package

The DB package controls the creation of Db objects. It includes the following files:

*Config.inc*

Provides the necessary configuration details to access the database, like User, Database name, Database Password, and Server. It also provides tables aliases to ease the access and not hardcode them in the code.

*UUIDB.Class*

Presents the available functions for the UUIDB objects.

### 5.7.2 Query Package

The Query package holds the select query class as well as sqlQuery class which includes update,insert and delete sql queries.

*SelQuery*

The selQuery class holds all the select queries that the system needs: select all fields with no conditions, select all fields with conditions, select one field with conditions, etc..

*SqlQuery*

The SQLQuery class has all the update, insert and delete queries the system needs.

### 5.7.3 DB Class Diagram

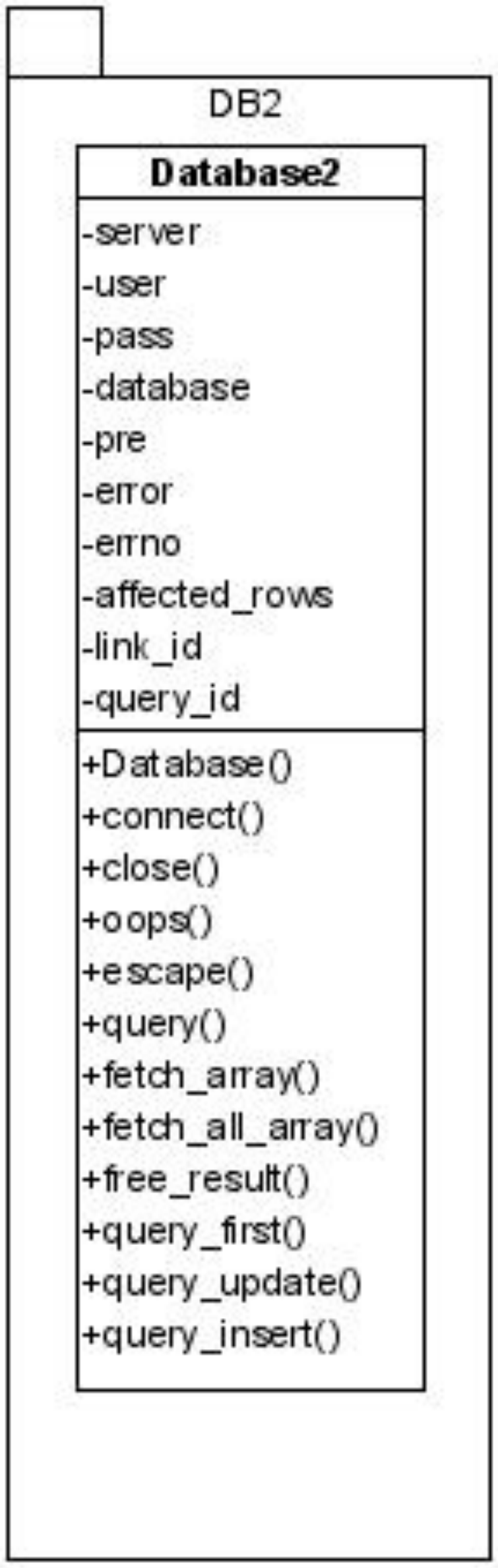

Figure 6: **DB Class Diagram**

### 5.7.4 Query Class Diagram

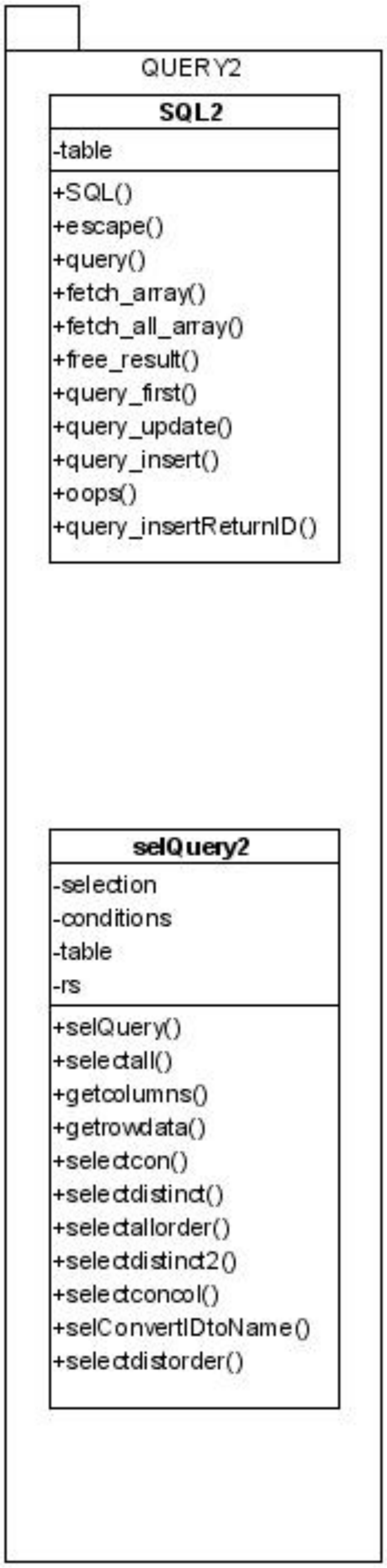

**Figure 7:** Query Class Diagram

## 5.8 Format & CSS Modules

The Format & CSS modules provide a common look to the whole system.

### 5.8.1 Format Package

The Format package provides common header & footer for all the system pages. Includes the following files:

*Footer*

Footer to be includes in all pages.

*Header*

Header to be included in all pages.

### 5.8.2 CSS Package

The CSS package is used to implement a common template for the system. It includes only one file.

*UUIS.css*

The UUIS template is responsible for the font formatting, table formatting, div placement and error messages. All the presentation format is controlled by this file.

## 5.9 Sequence Diagrams

### 5.9.1 Assign Responsible to Lab

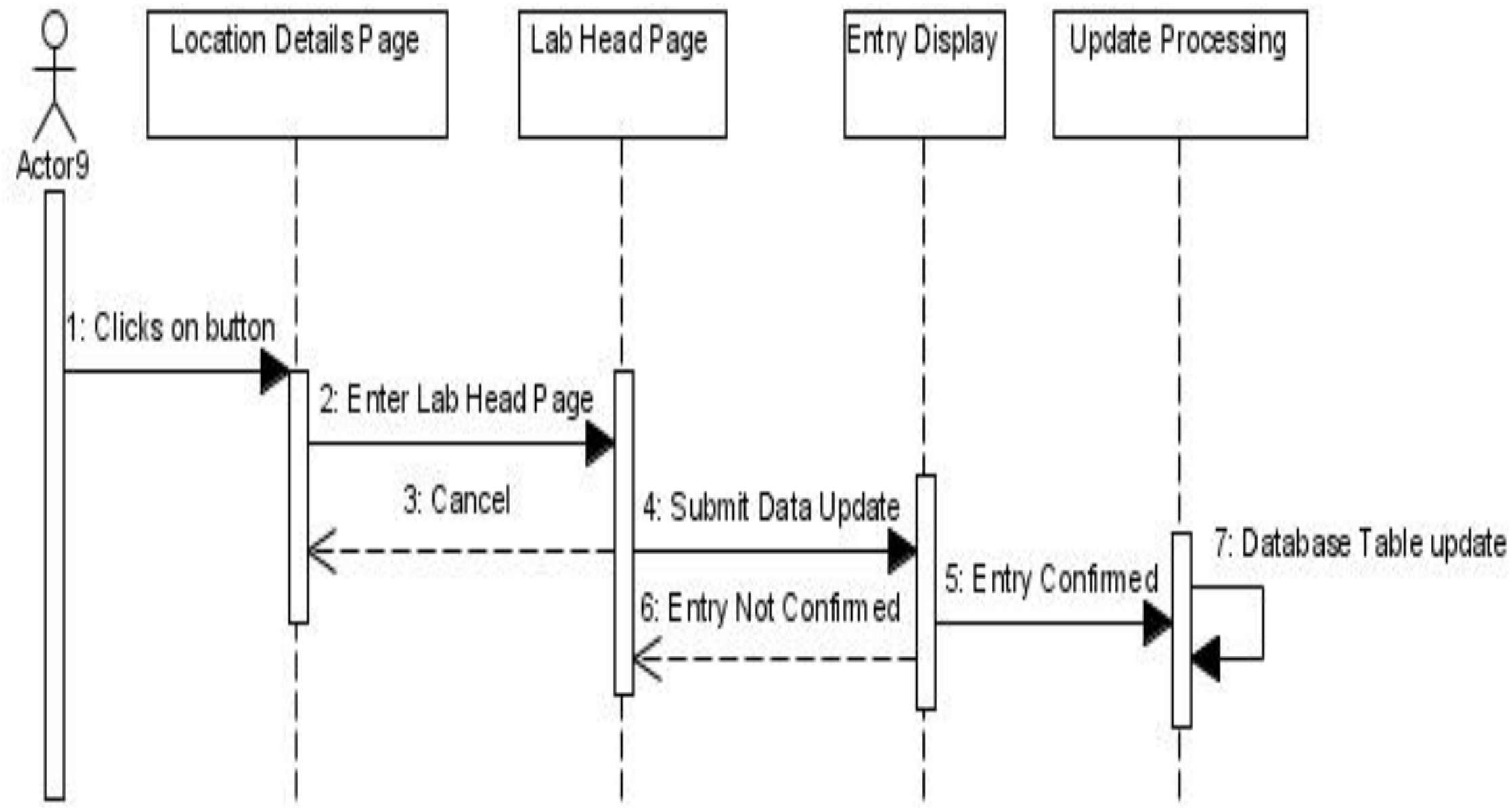

**Figure 8:** Assign Responsible to Lab – Sequence diagram

### 5.9.2 Change Password

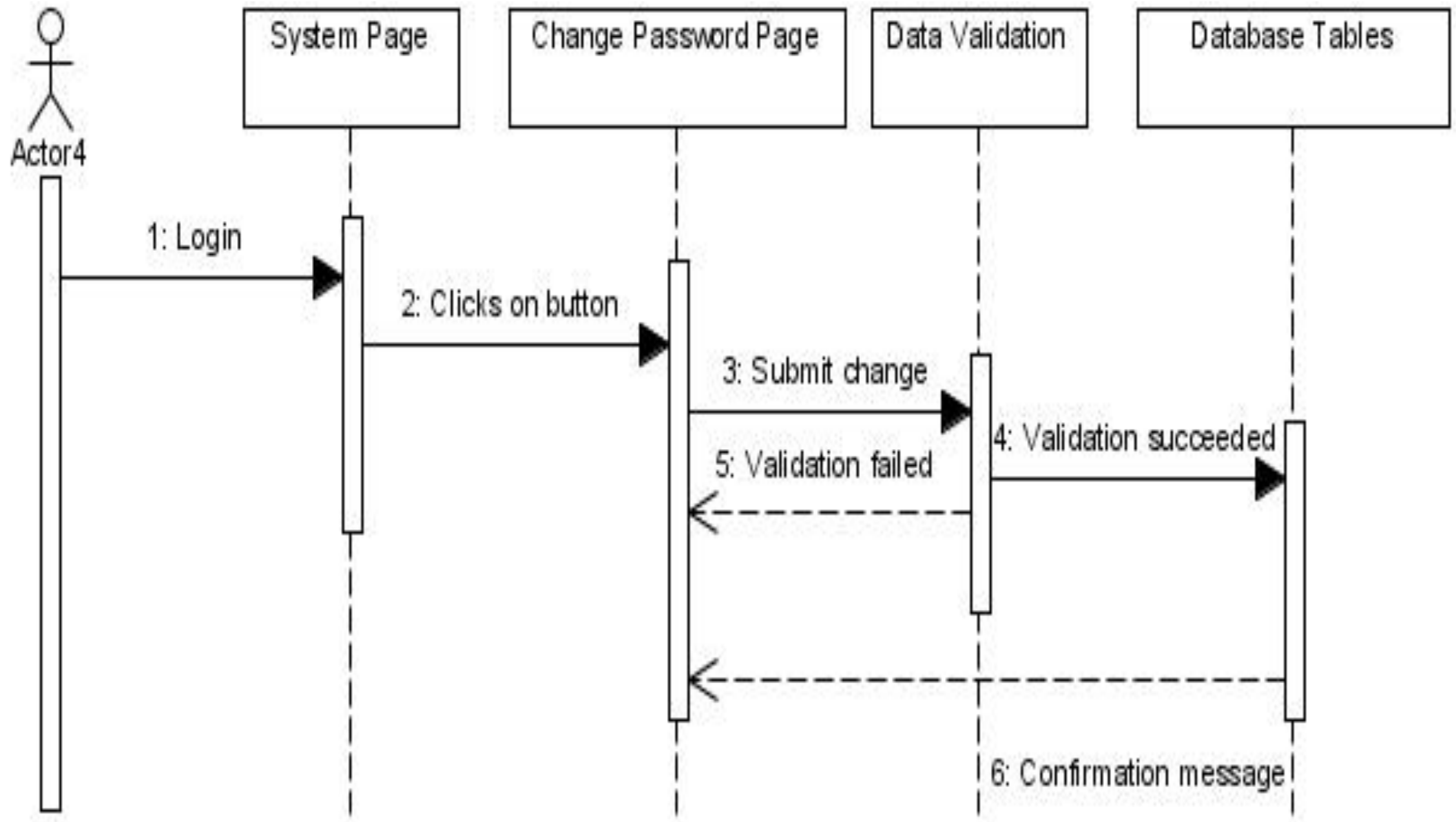

**Figure 9:** Change Password – Sequence diagram

### 5.9.3 Log in

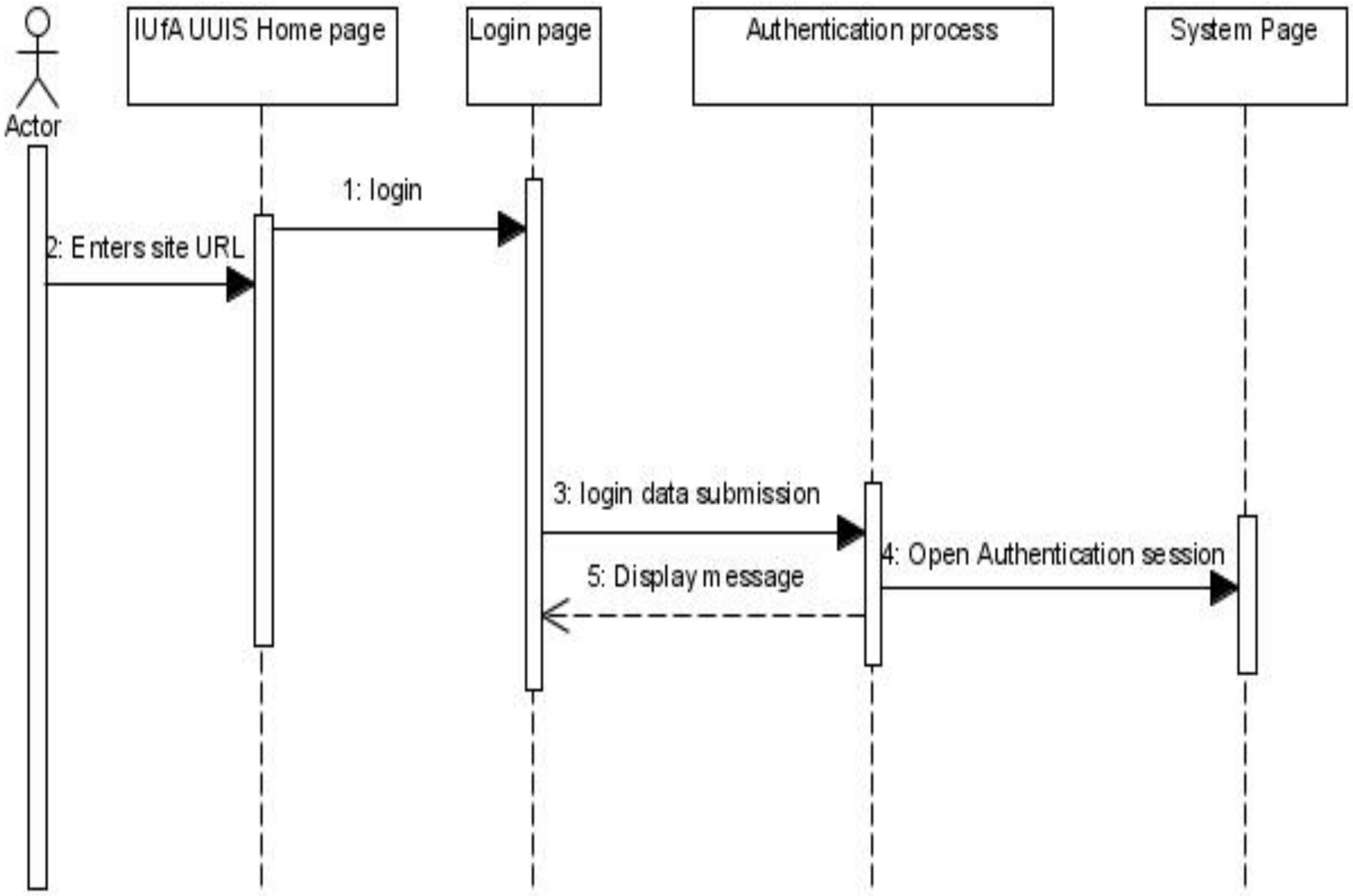

**Figure 10:** Login – Sequence diagram

### 5.9.4 Log Out

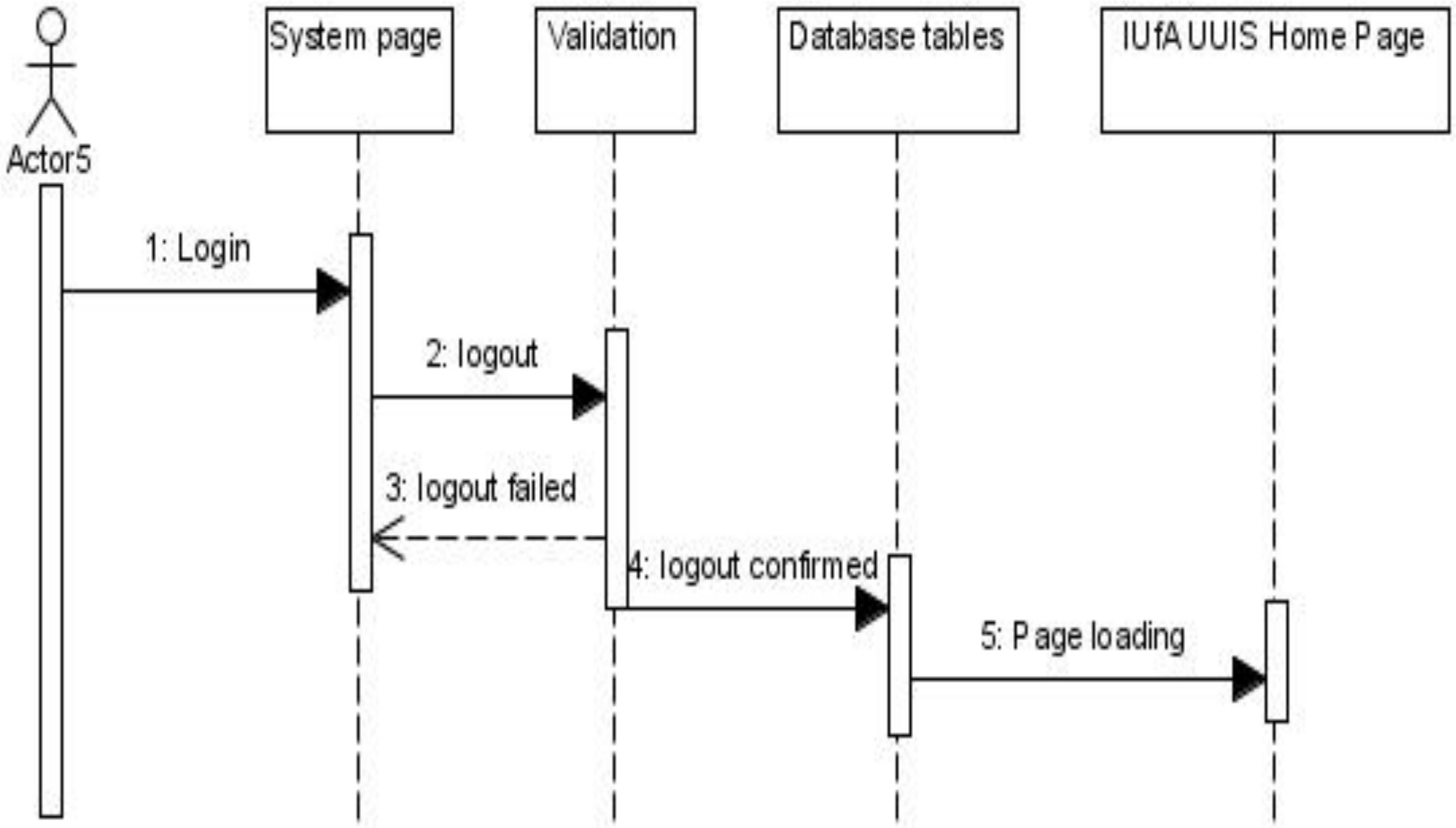

**Figure 11:** Logout – Sequence diagram

### 5.9.5 Search Physical Assets

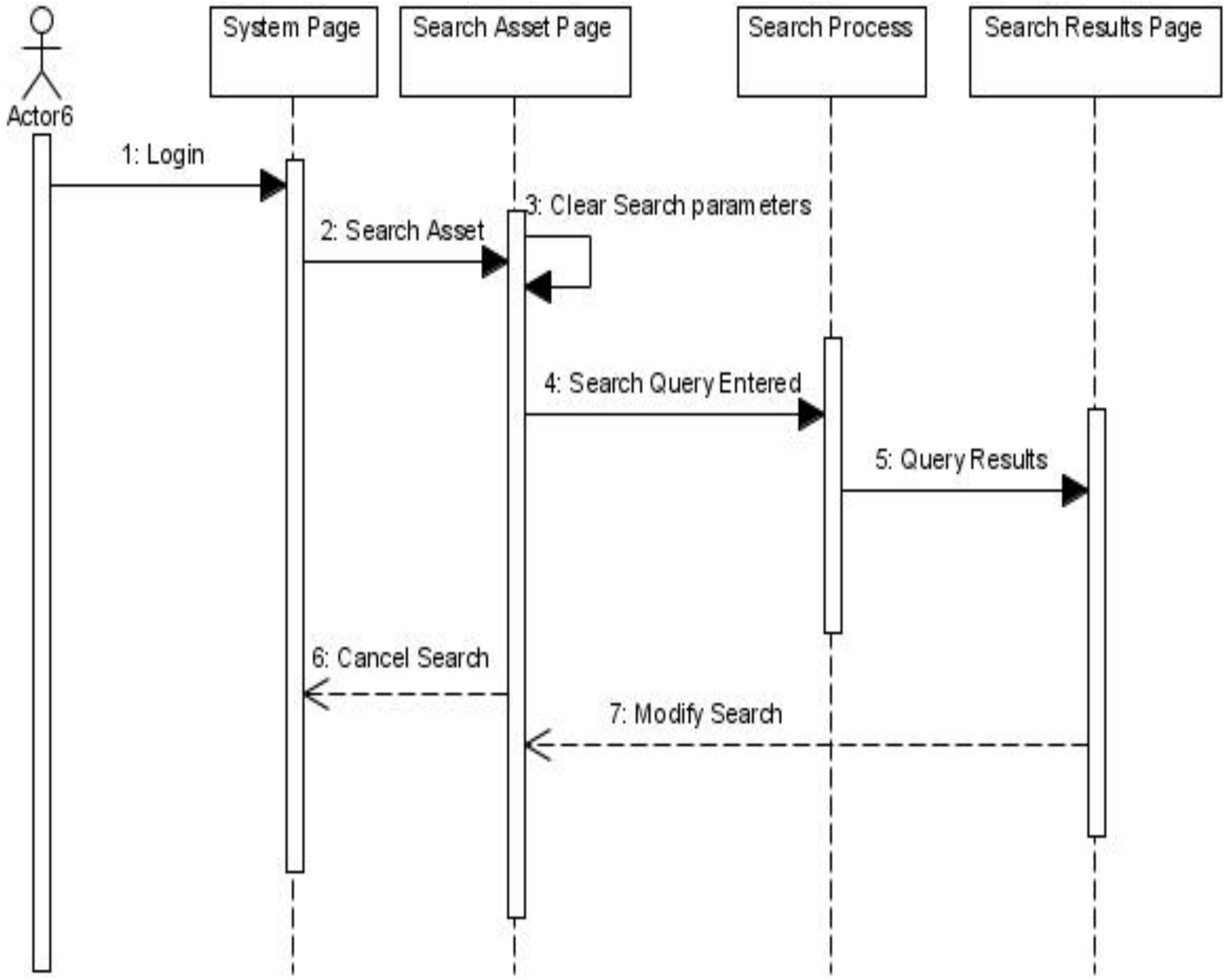

**Figure 12:** Search Physical Asset – Sequence diagram

### 5.9.6 Submit General Request

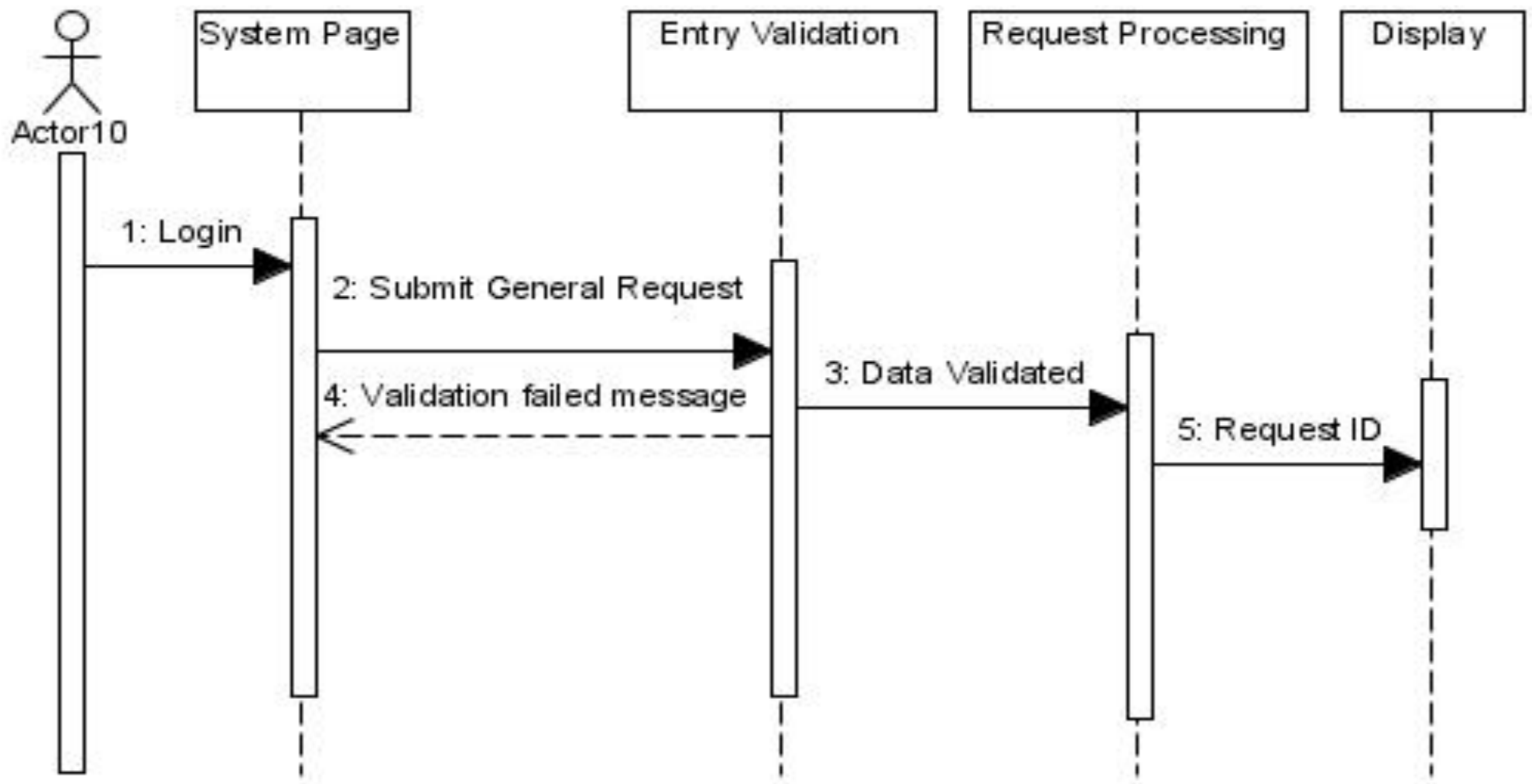

**Figure 13:** Submit General Request – Sequence diagram

### 5.9.7 Account Update information

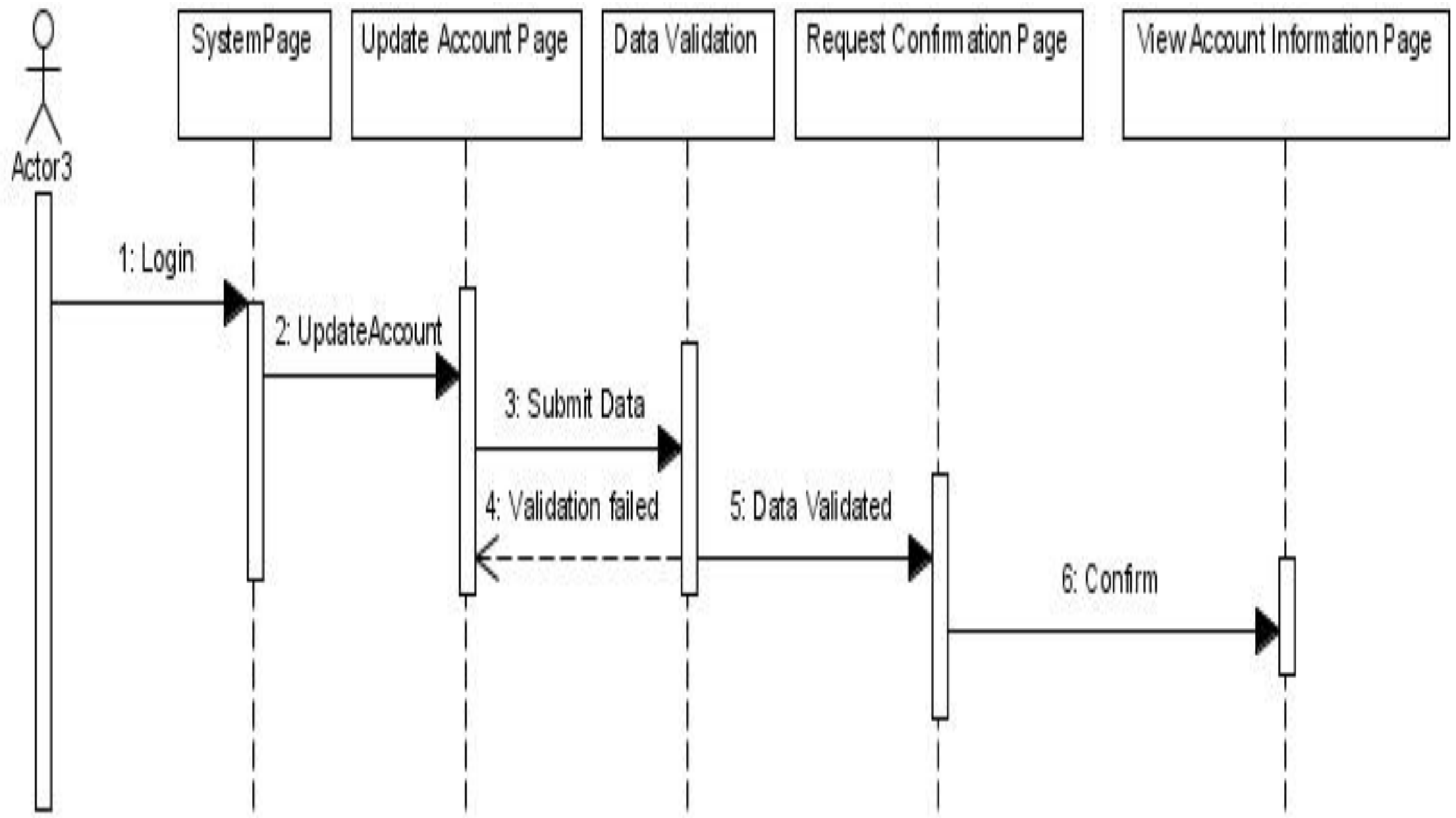

**Figure 14:** Account update information – Sequence diagram

### 5.9.8 Update Physical Asset

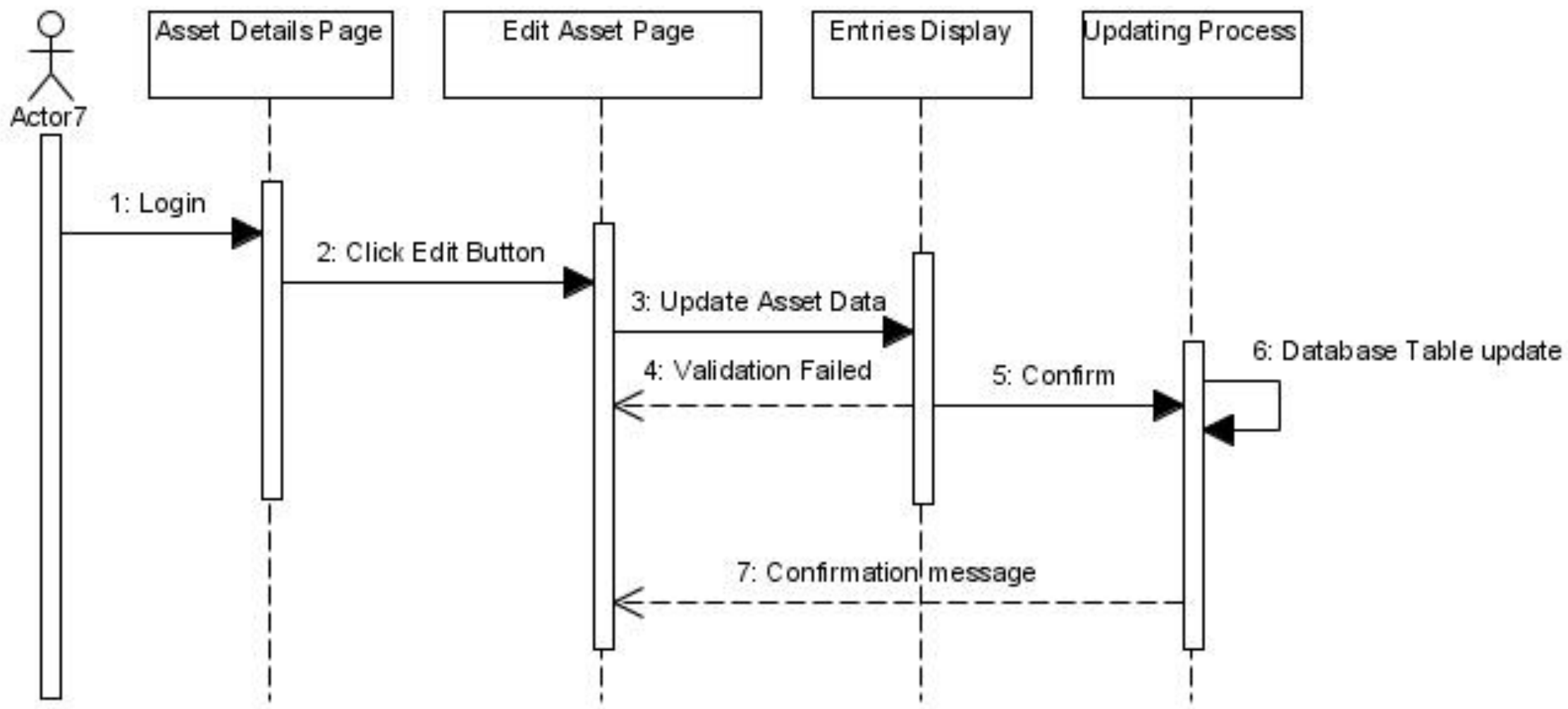

**Figure 15:** Update Physical Asset – Sequence diagram

### 5.9.9 View Account Information

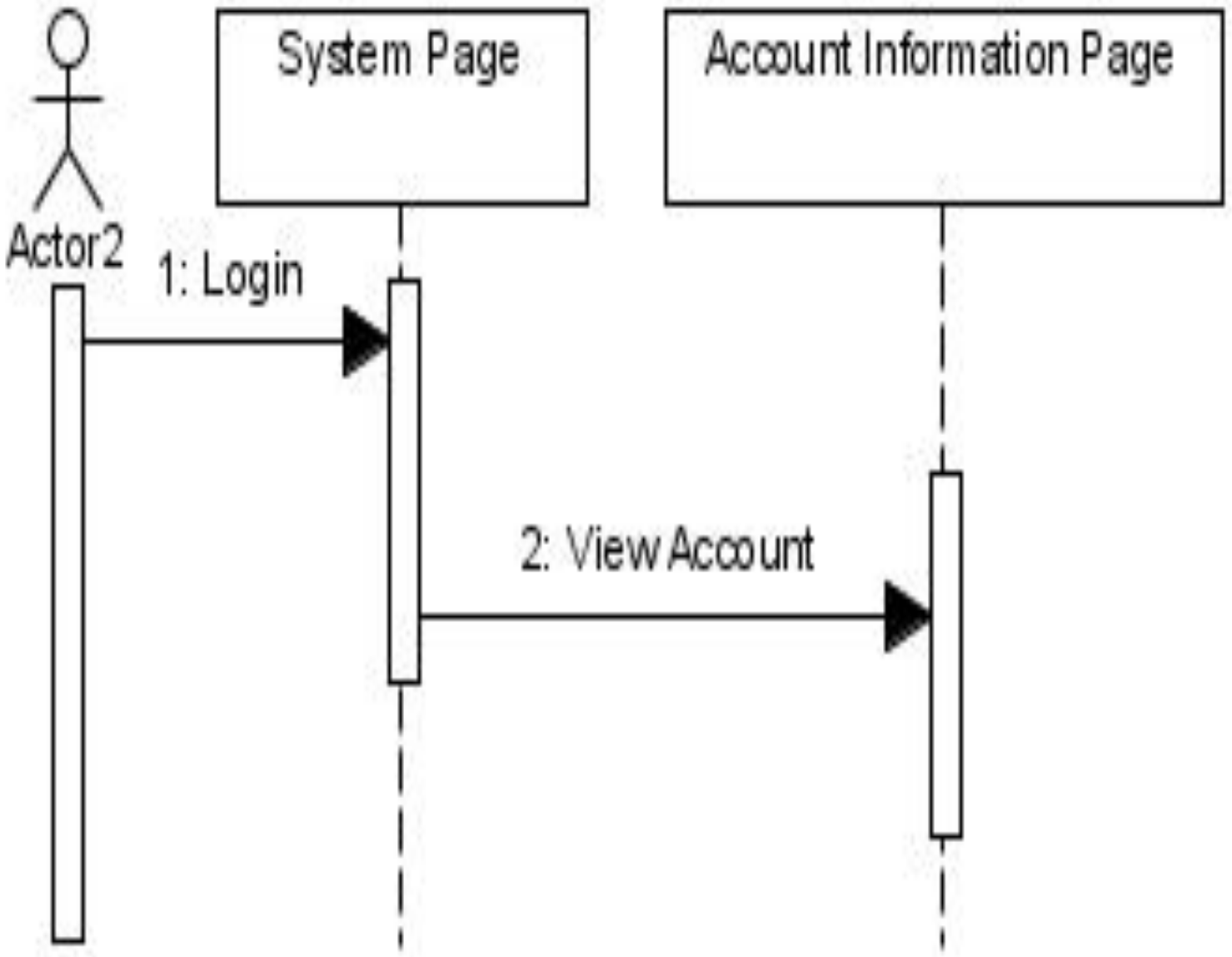

**Figure 16:** View Account Information – Sequence diagram

### 5.9.10 View update Software

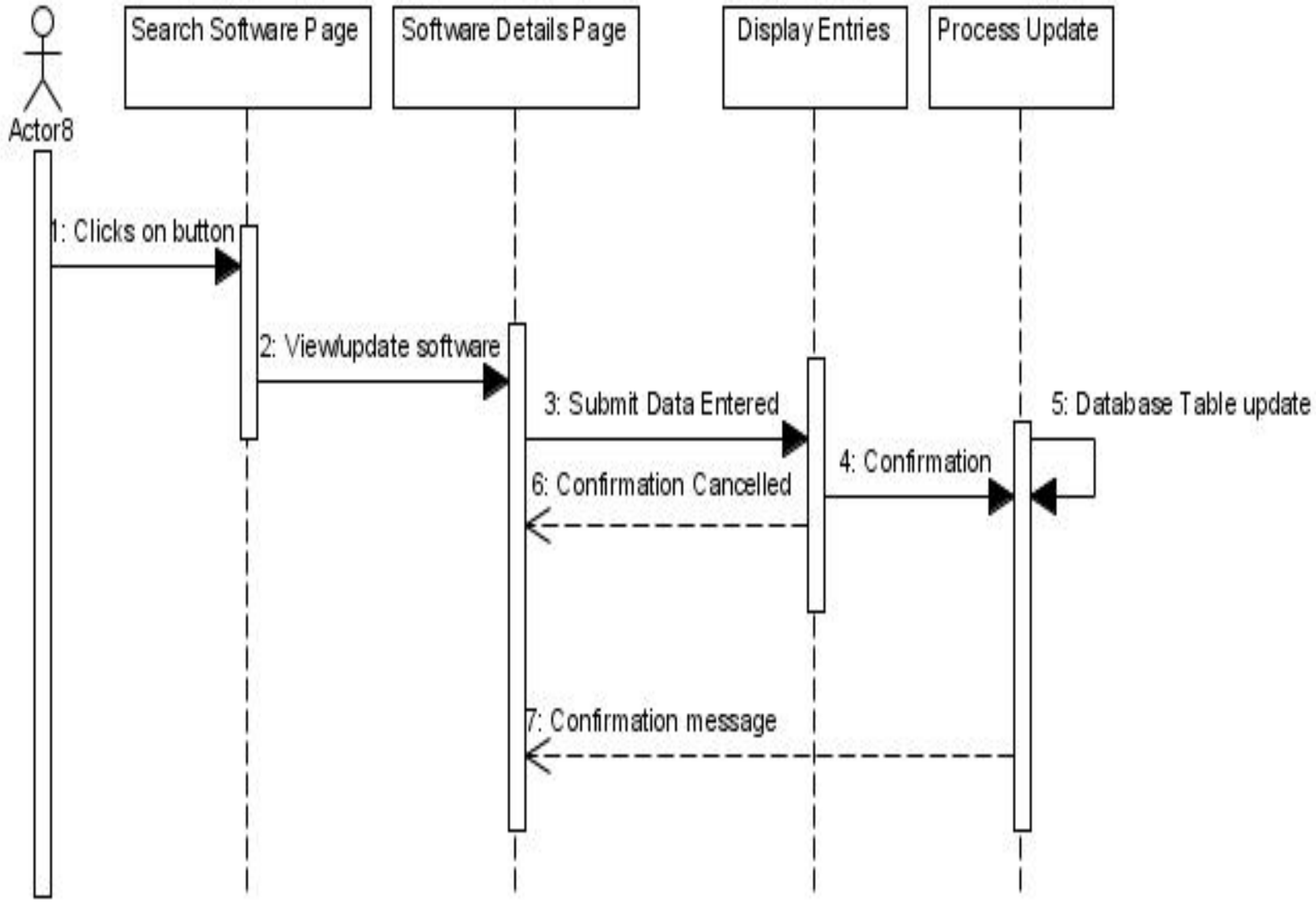

**Figure 17:** View update Software – Sequence diagram

## 5.10 State Diagrams

### 5.10.1 Log In

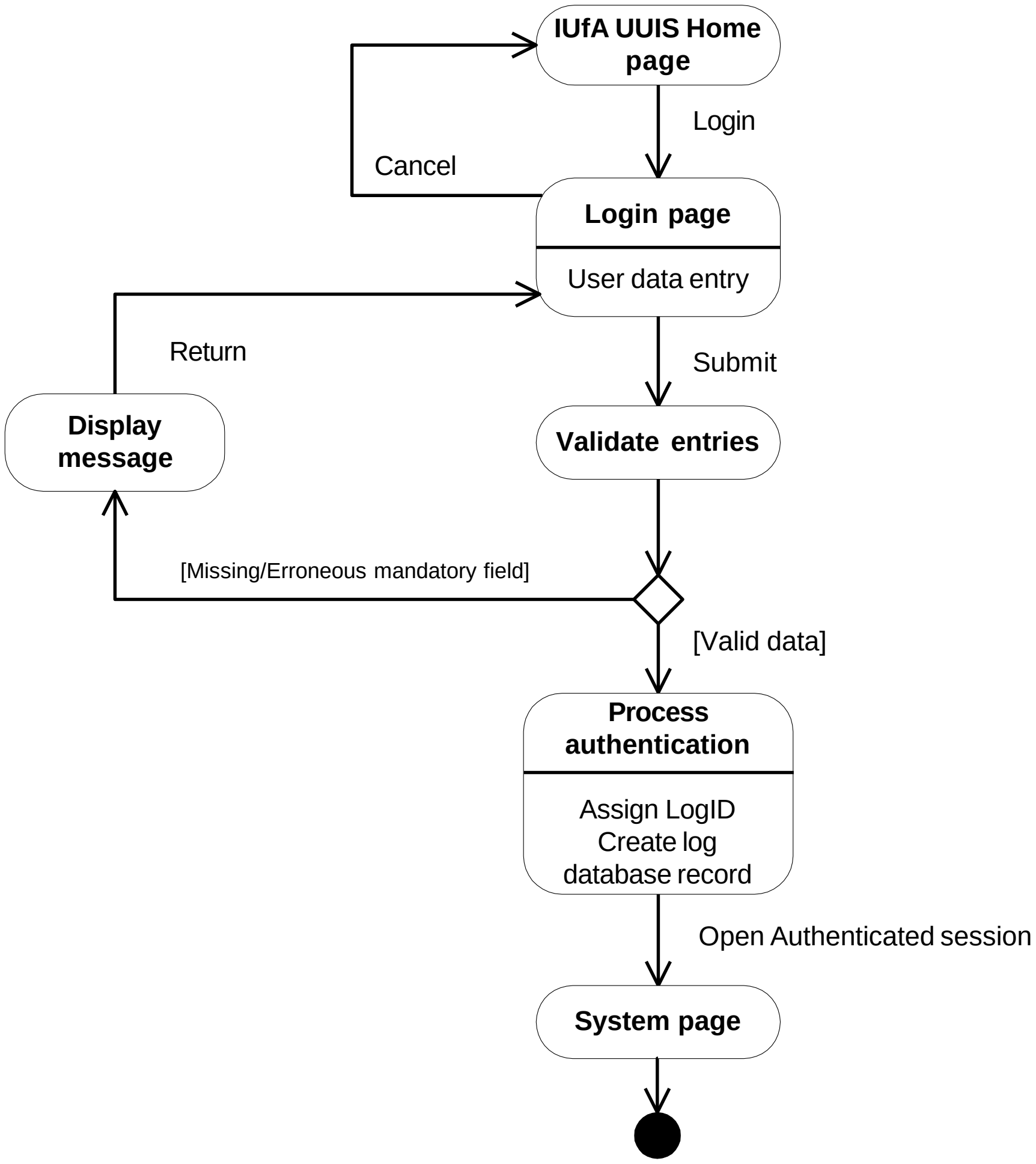

**Figure 18 -** Login – State diagram

### 5.10.2 View Account Information

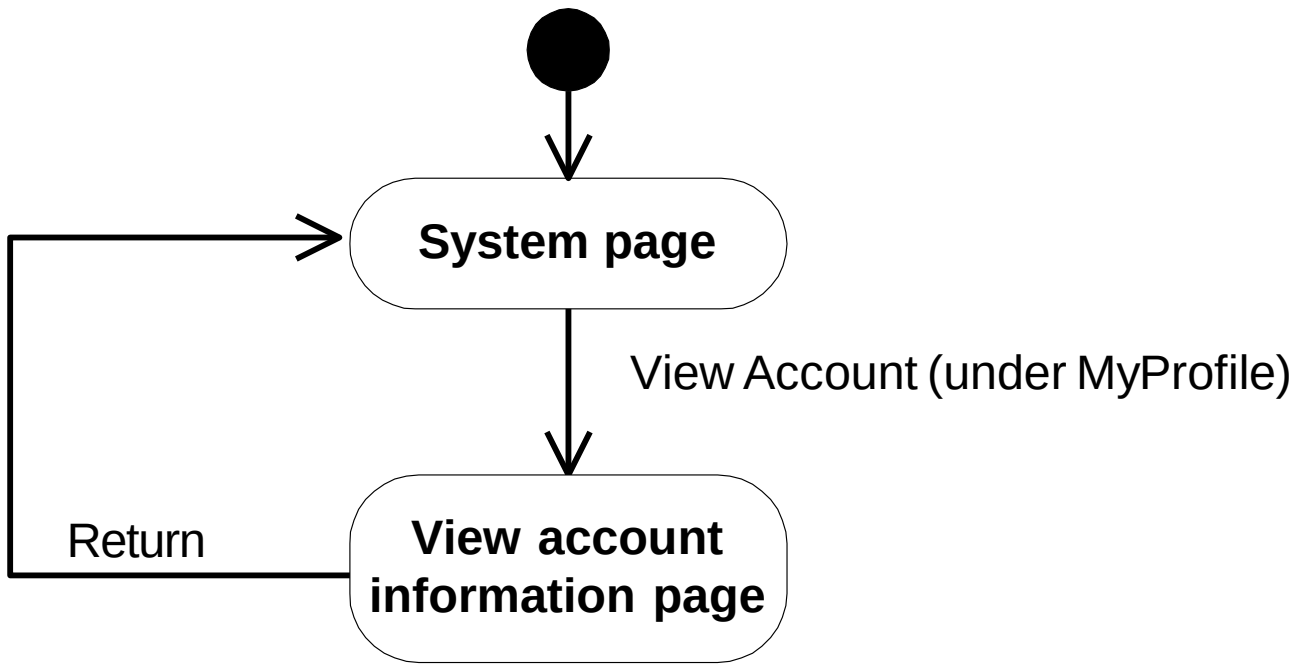

**Figure 19 -** View Account Information – State diagram

### 5.10.3 Update Account Inormation

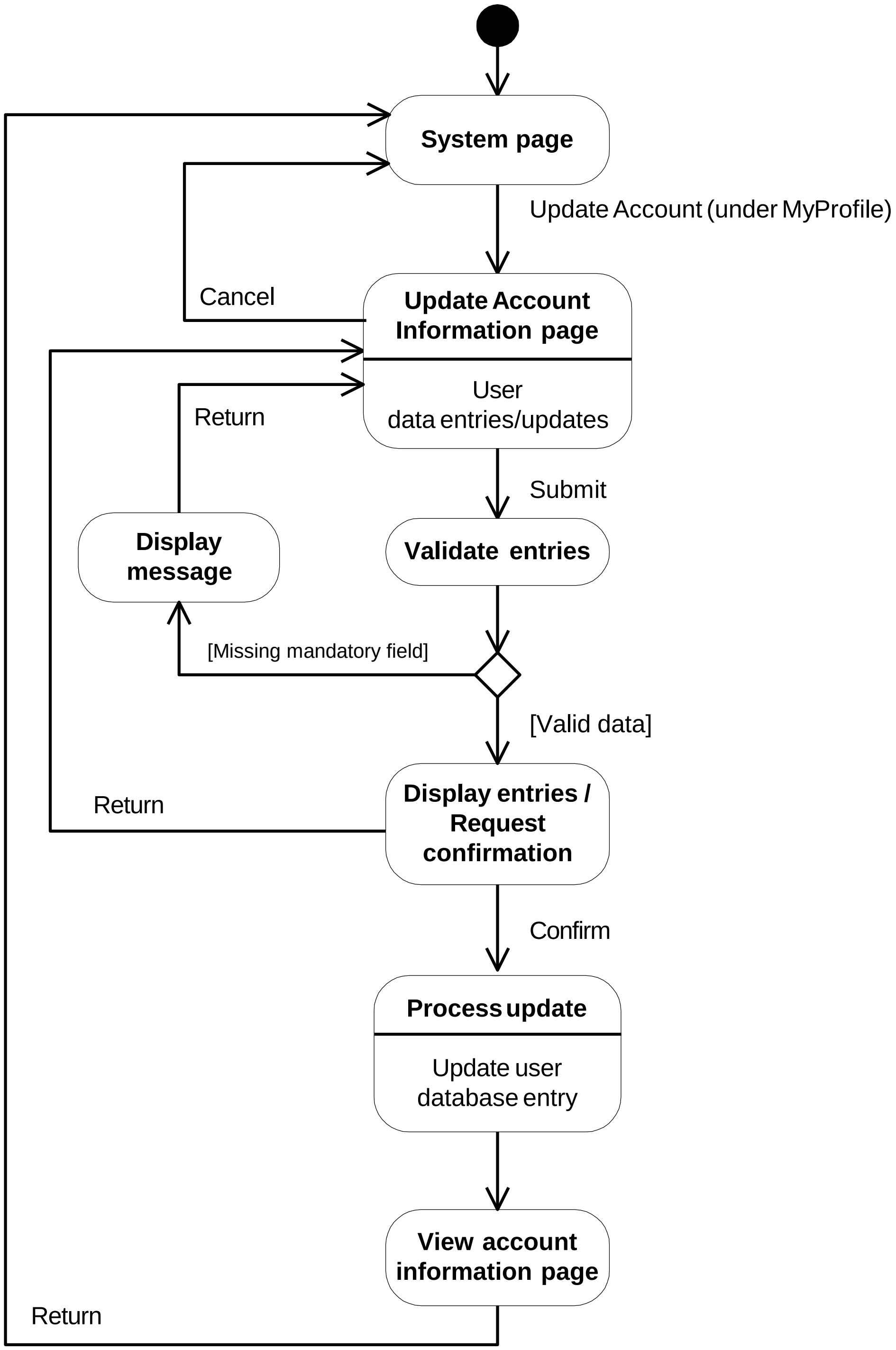

**Figure 20 -** Update Account Information – State diagram

### 5.10.4 Change Password

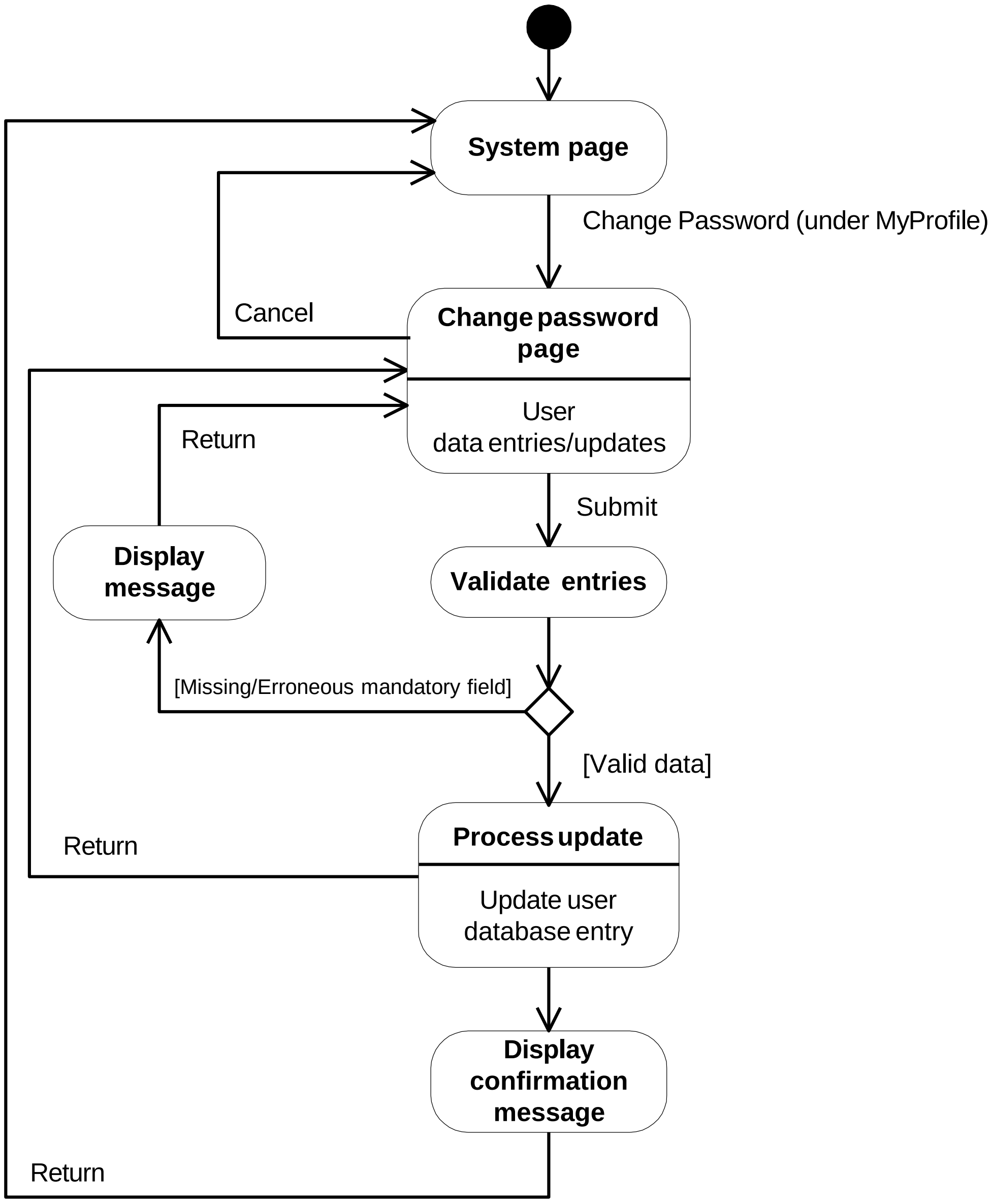

**Figure 21 -** Change Password – State diagram

### 5.10.5 Log Out

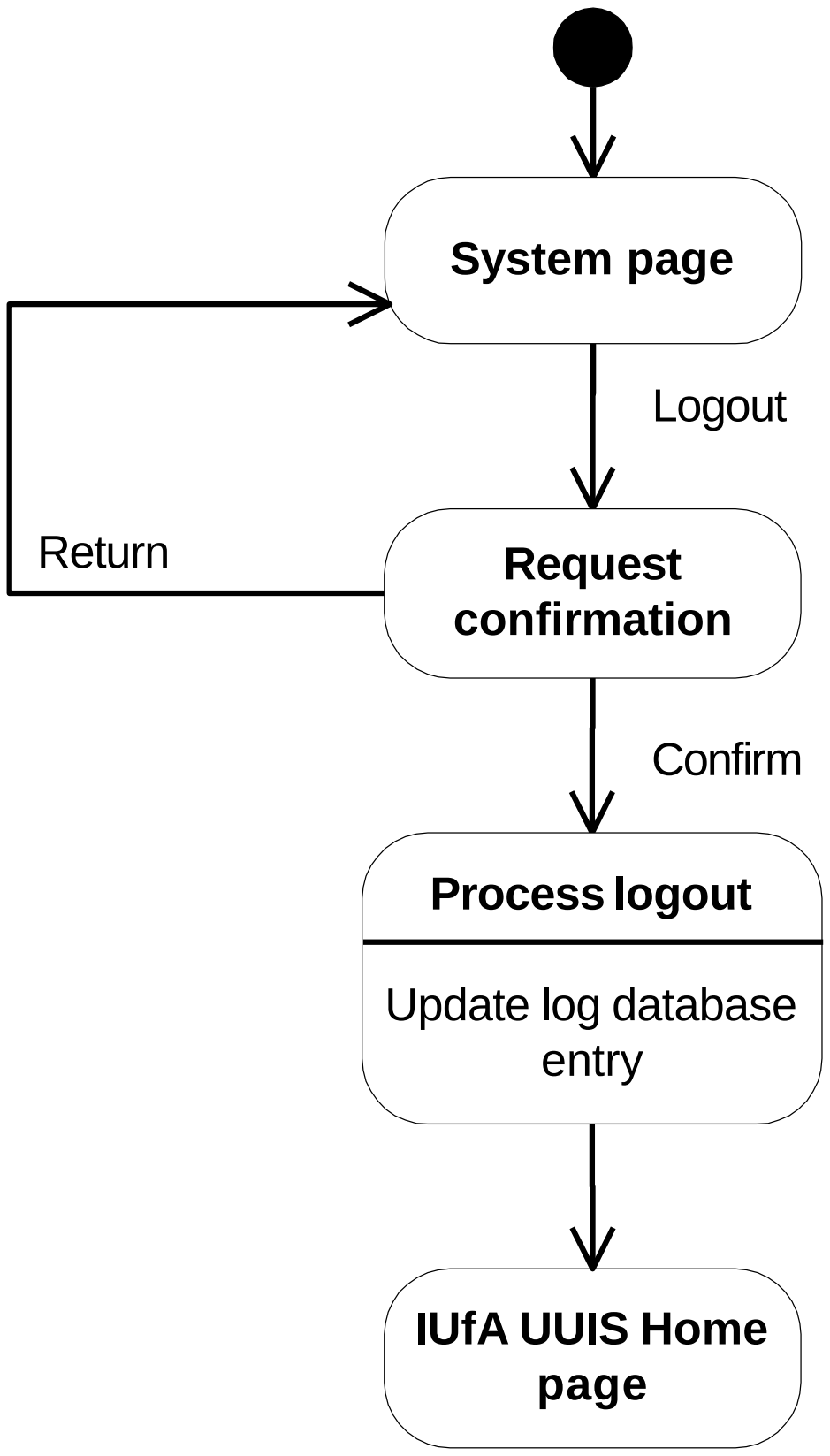

**Figure 22 -** Logout – State diagram

### 5.10.6 Grant/Revoke Permission

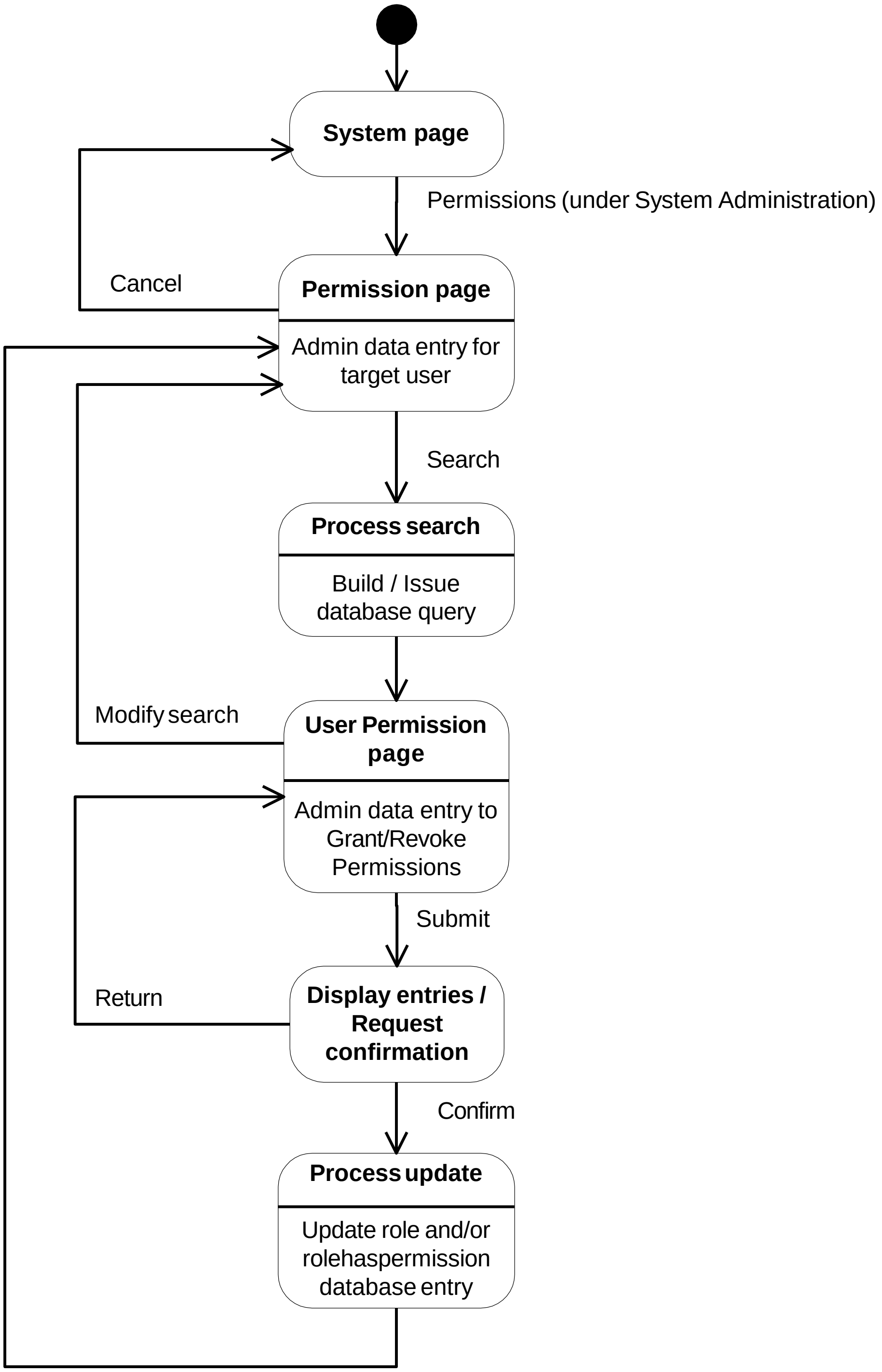

**Figure 23 -** Grant/Revoke Permission – State diagram

### 5.10.7 Edit Default Role Permissions

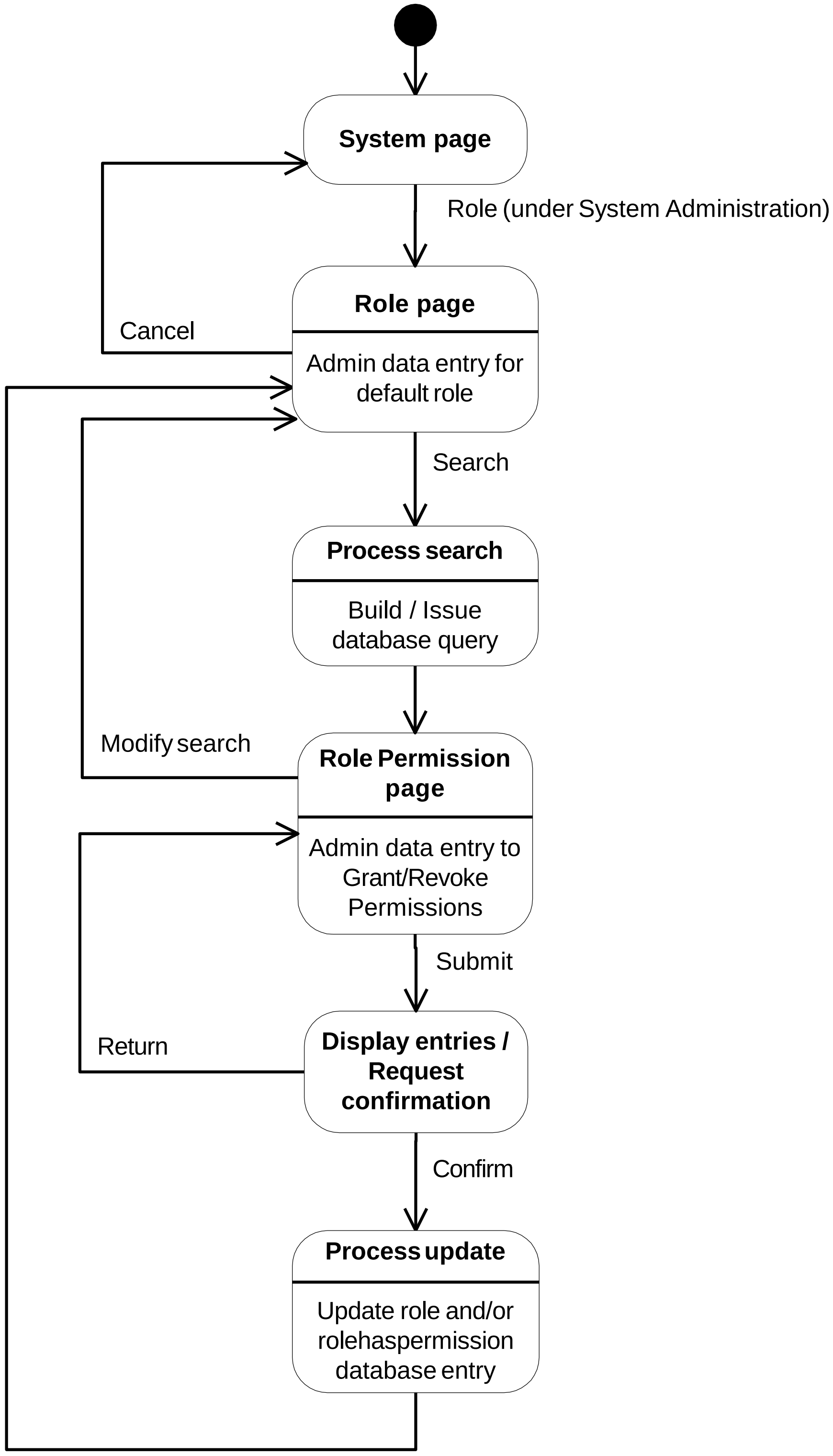

**Figure 24 -** Edit Default Role Permissions – State diagram

### 5.10.8 Add Physical Asset

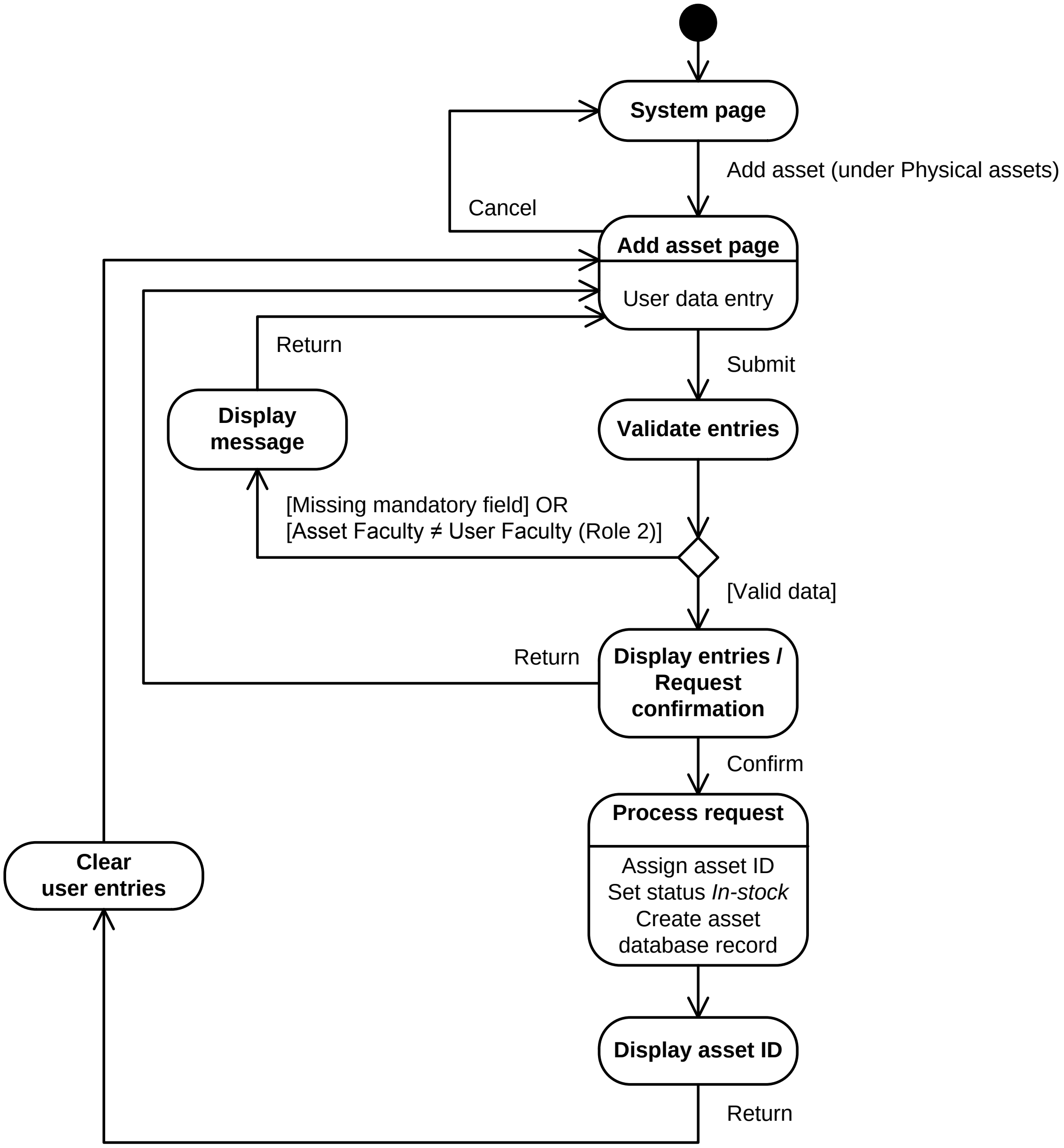

**Figure 25 -** Add physical asset – State diagram

### 5.10.9 Search Physical Asset

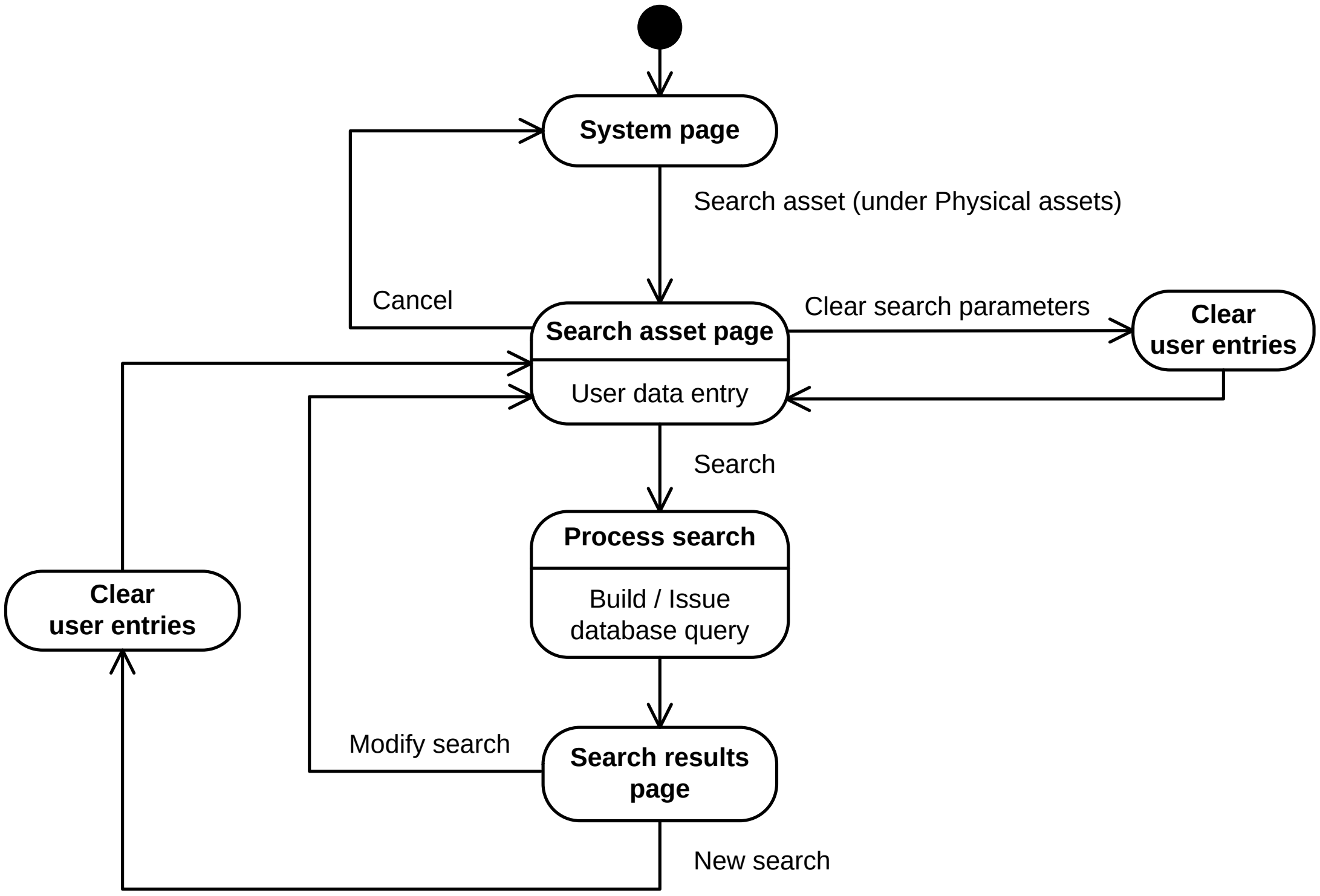

**Figure 26 -** Search physical asset – State diagram

### 5.10.10 View Physical Asset

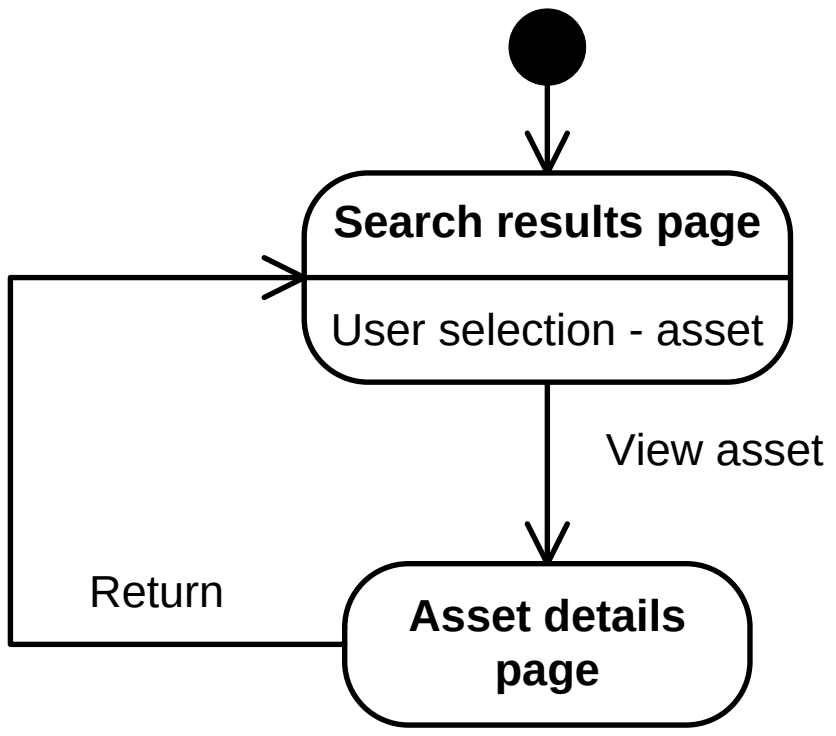

**Figure 27 -** View physical asset – State diagram

### 5.10.11 Update Physical Asset

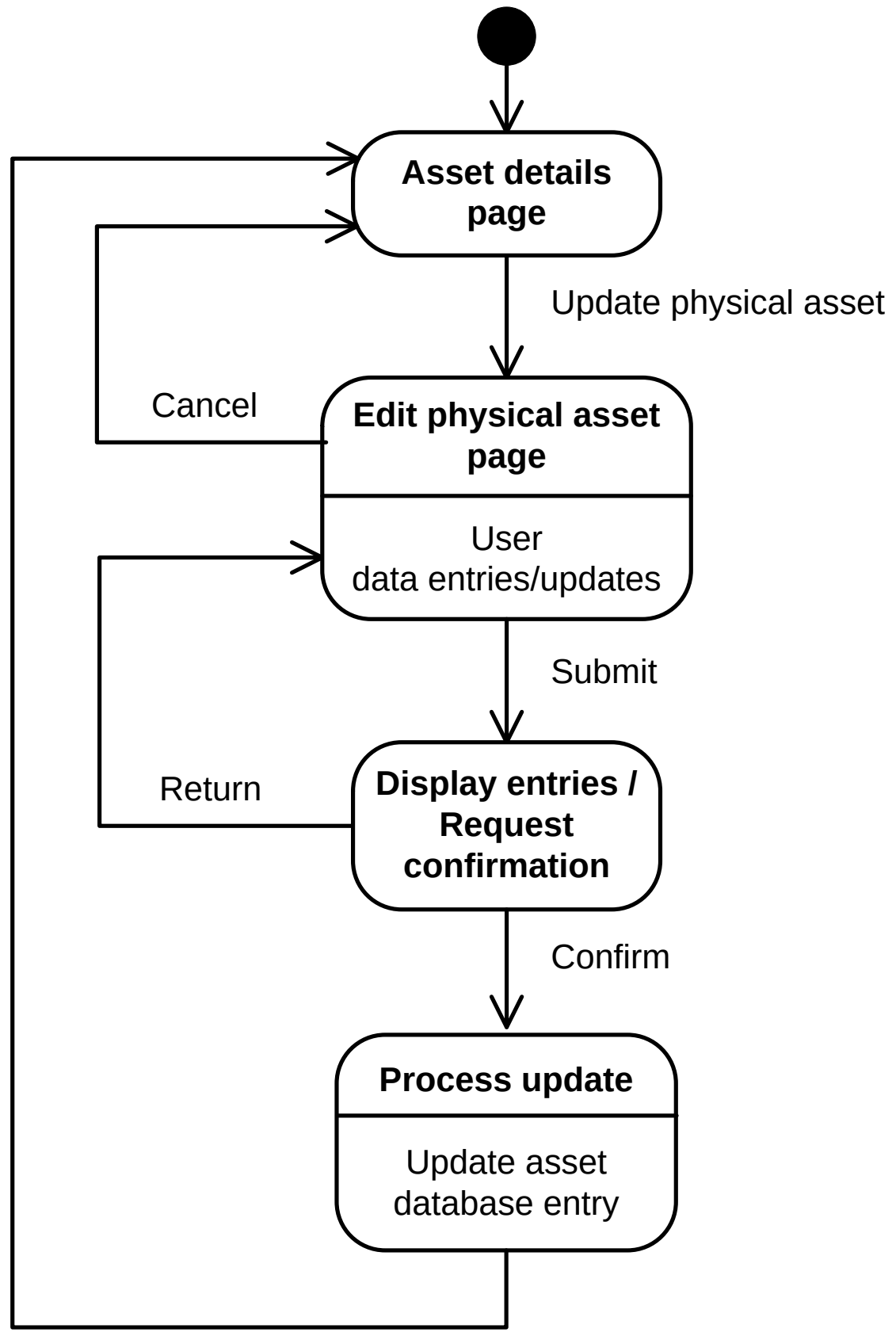

**Figure 28 -** Update physical asset – State diagram

### 5.10.12 Create Group

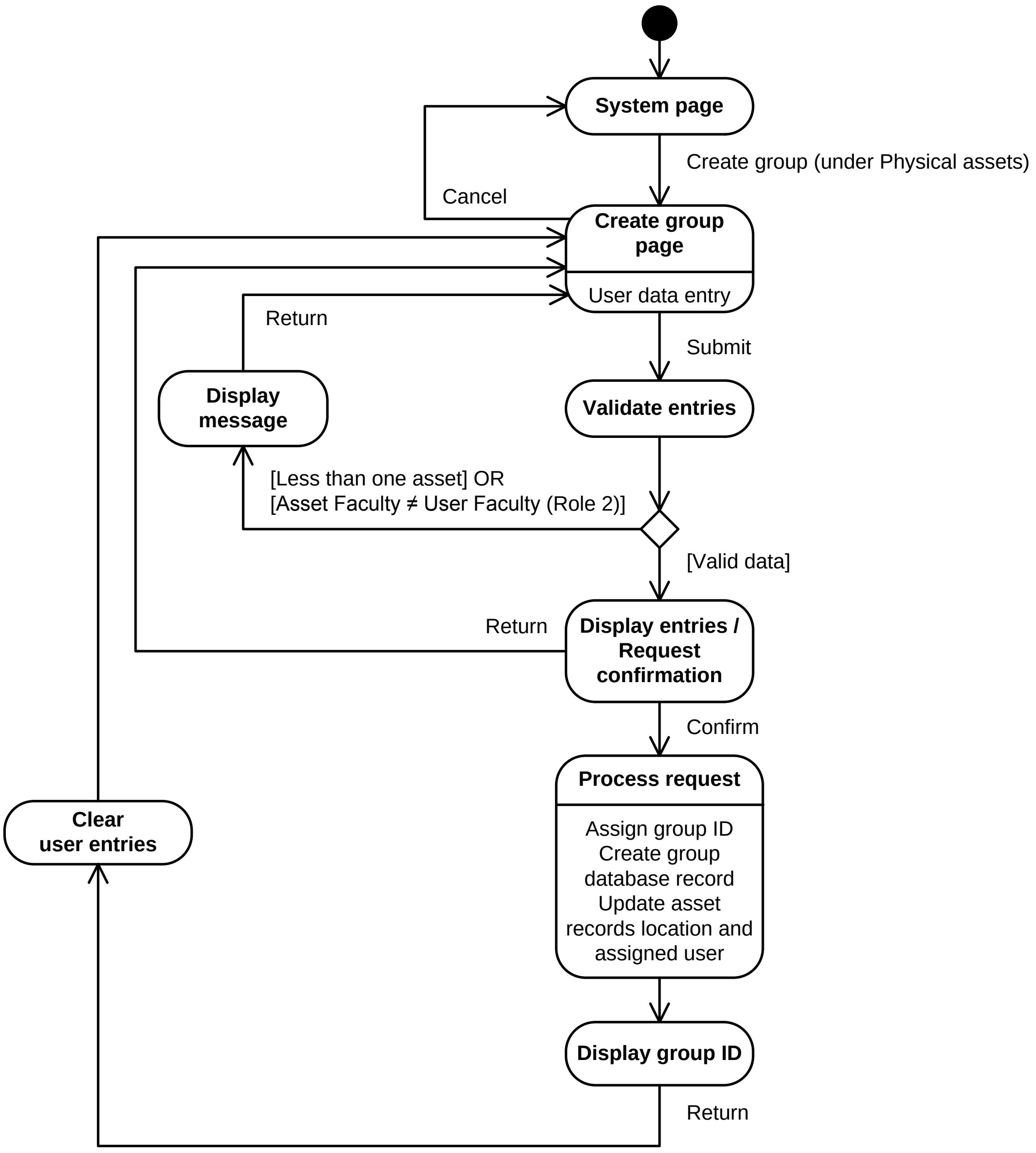

**Figure 29 -** Create group – State diagram

### 5.10.13 View/Update/Delete Group

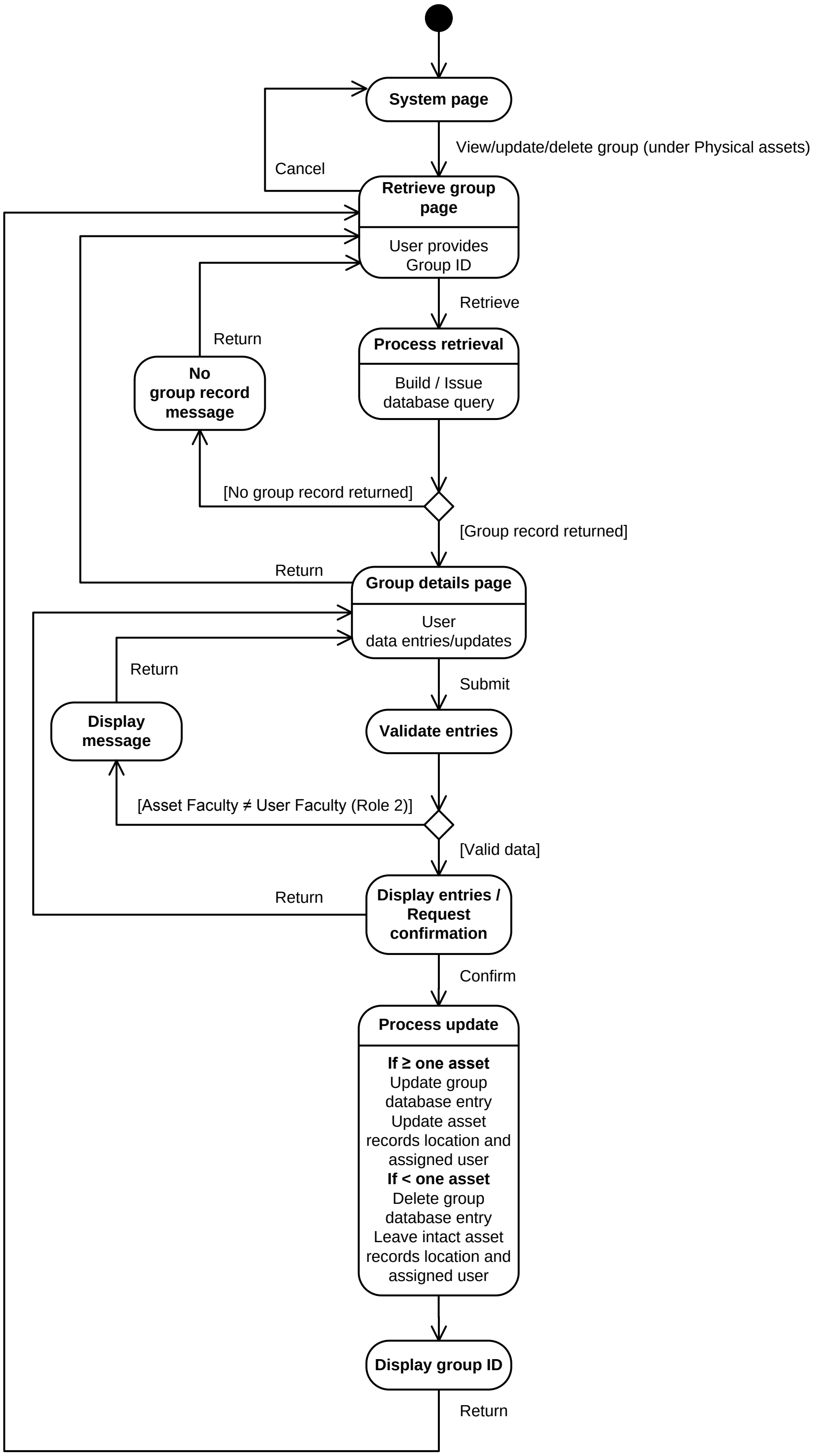

**Figure 30 -** View/update/delete group – State diagram

### 5.10.14 Add Software

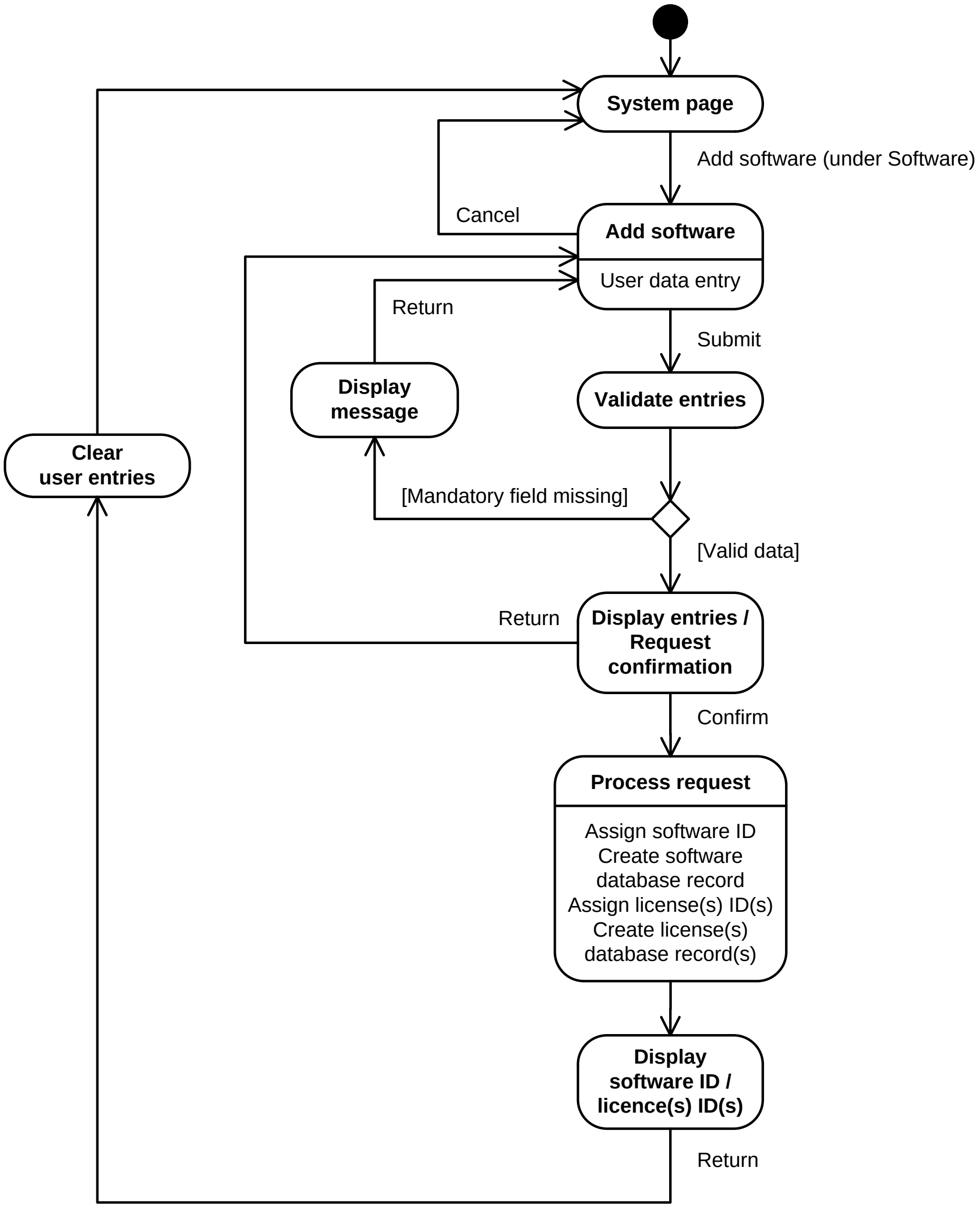

**Figure 31 -** Add software – State diagram

### 5.10.15 Search Sotfware

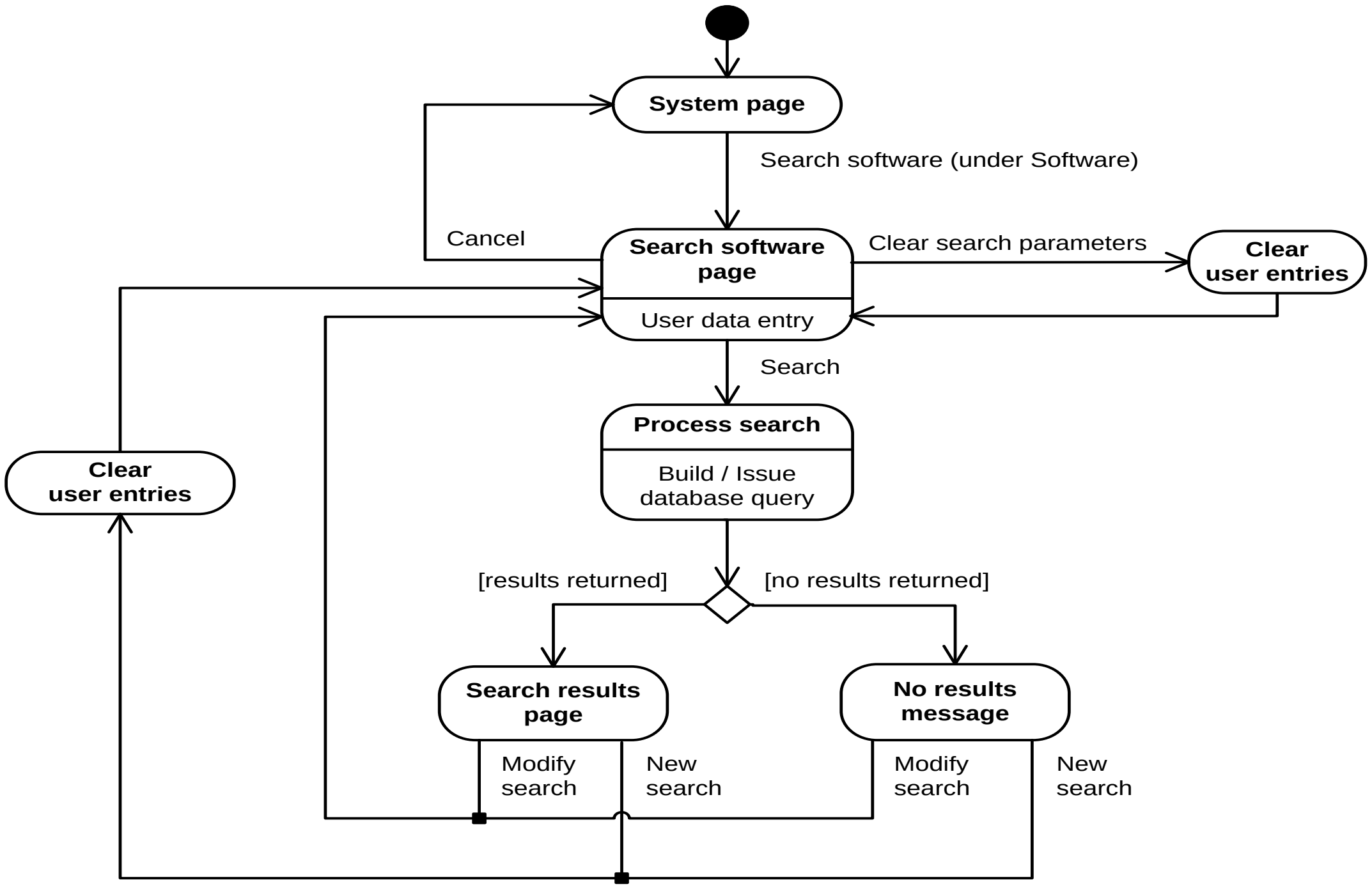

**Figure 32 -** Search software – State diagram

### 5.10.16 View/Update Software

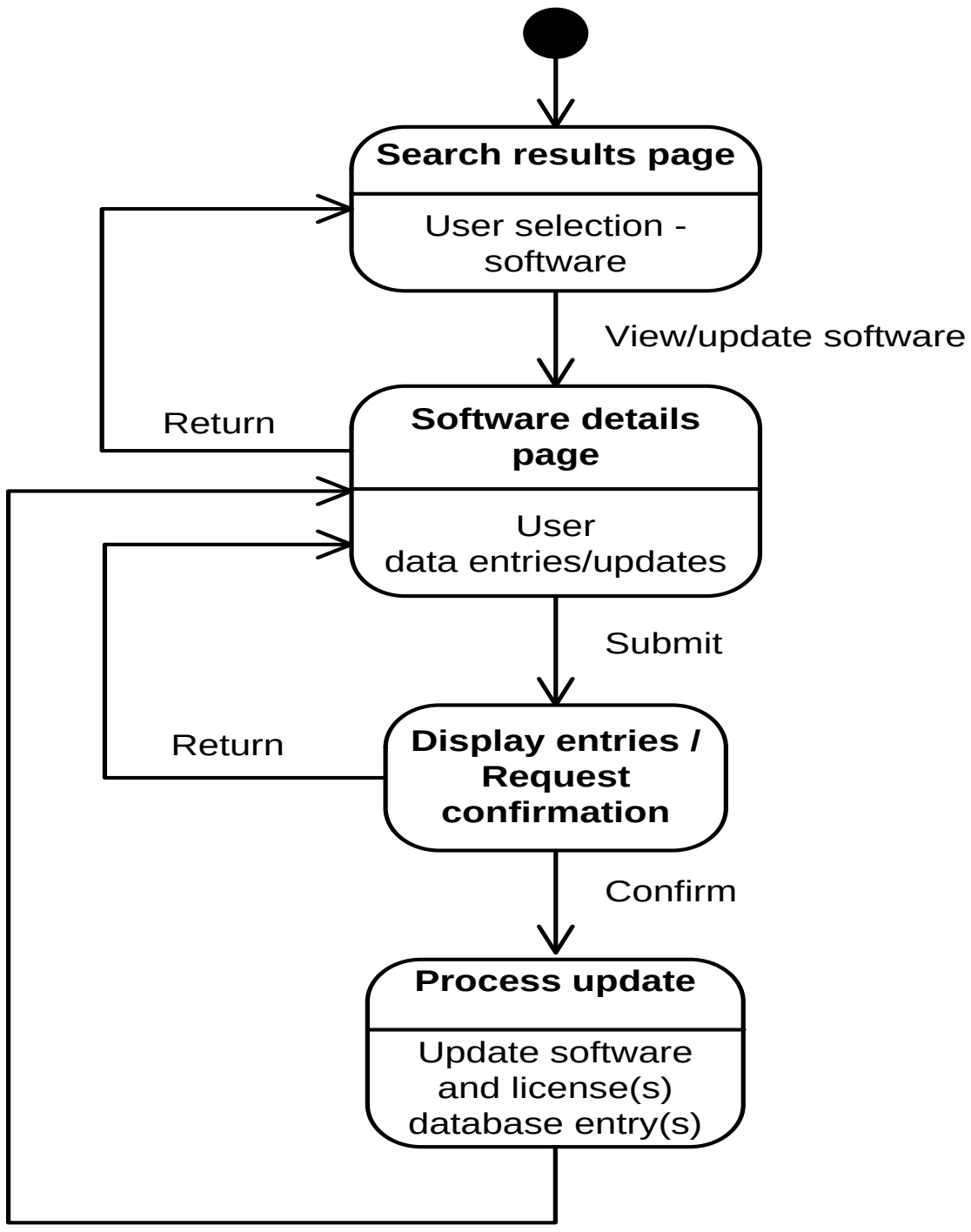

**Figure 33 -** View/update software – State diagram

### 5.10.17 Notification of License near Expiry

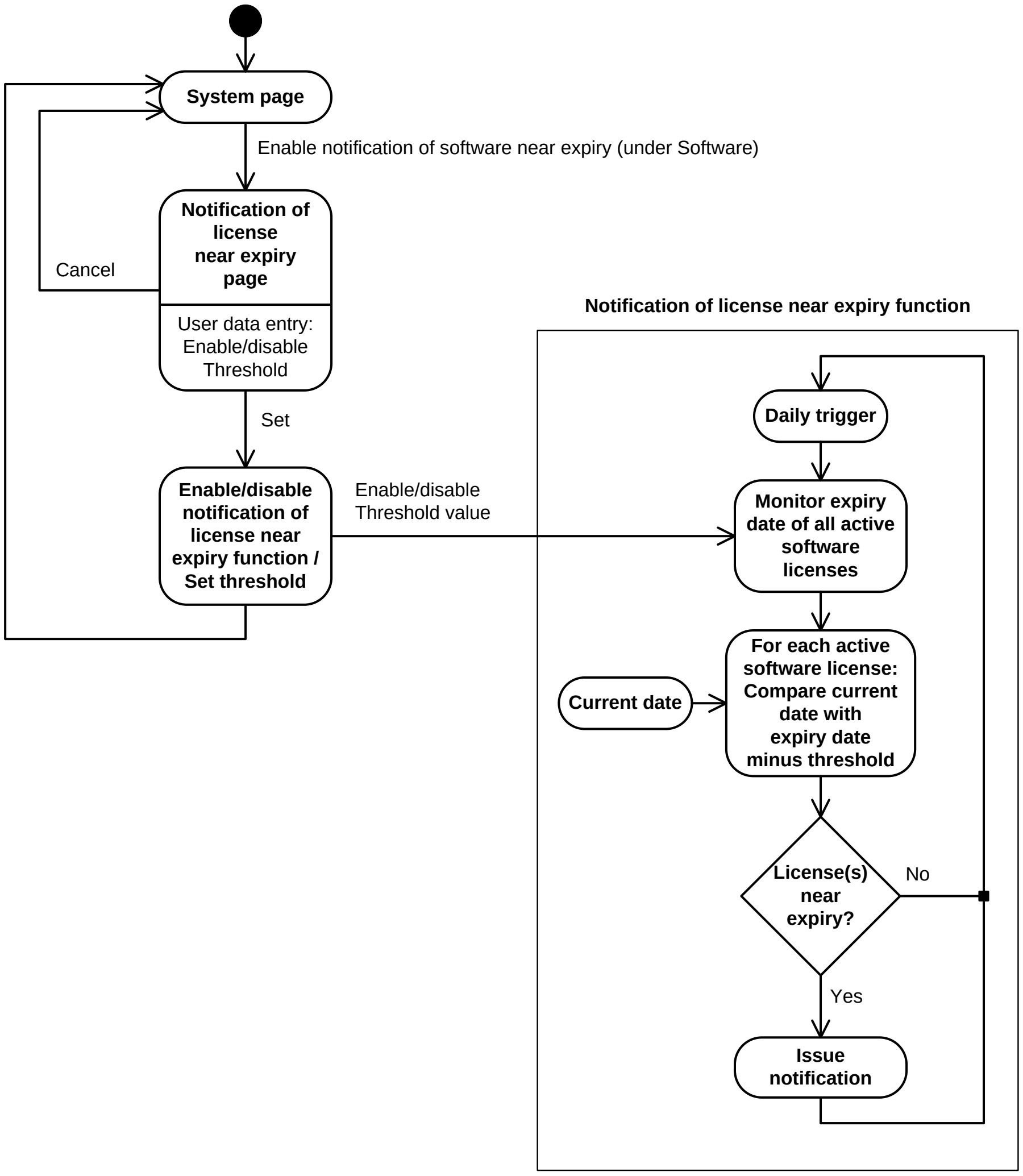

**Figure 34 -** Notification of license near expiry – State diagram

### 5.10.18 Search Location

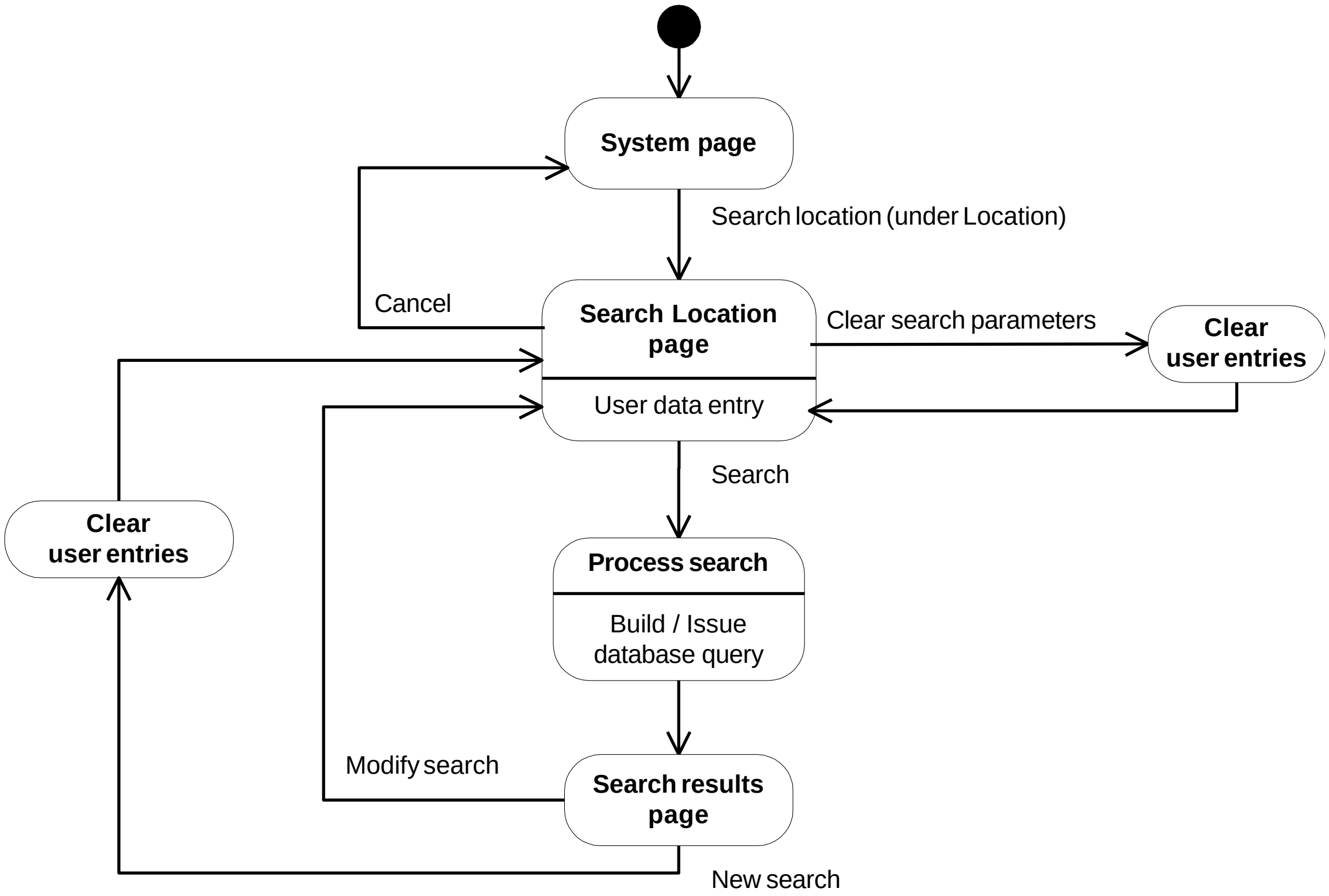

**Figure 35 –** Search Location – State diagram

### 5.10.19 View Location

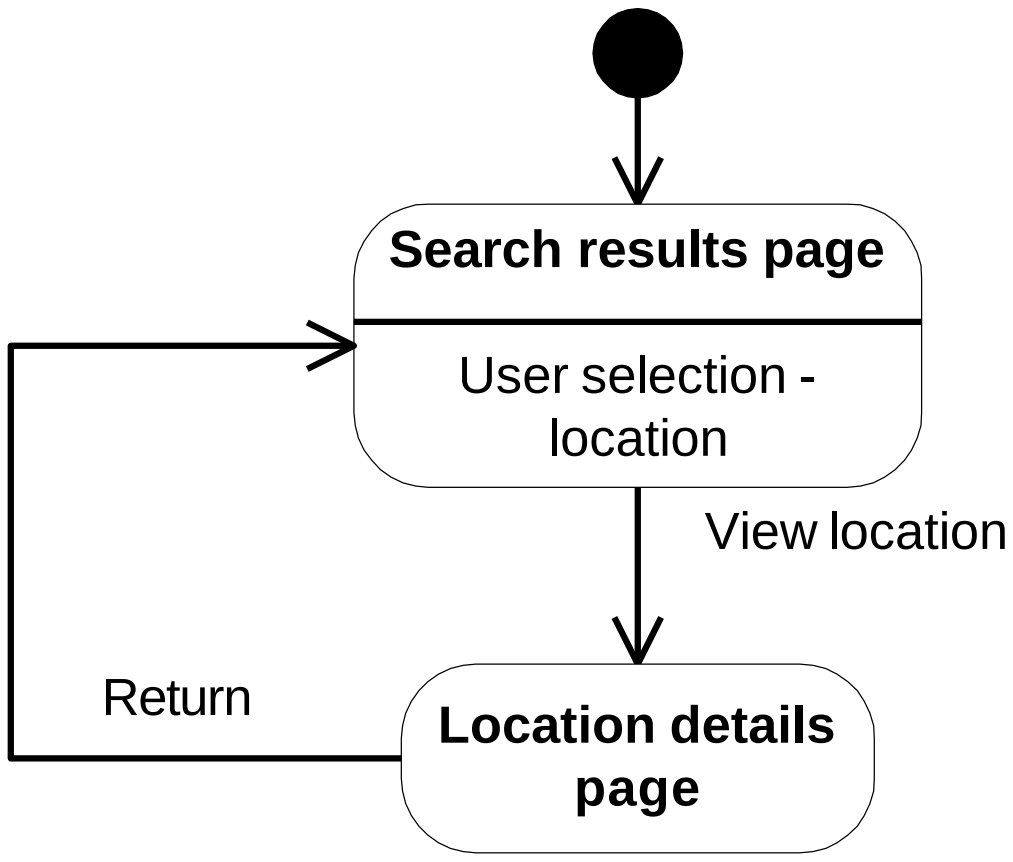

**Figure 36 –** View Location – State diagram

### 5.10.20 Edit Location

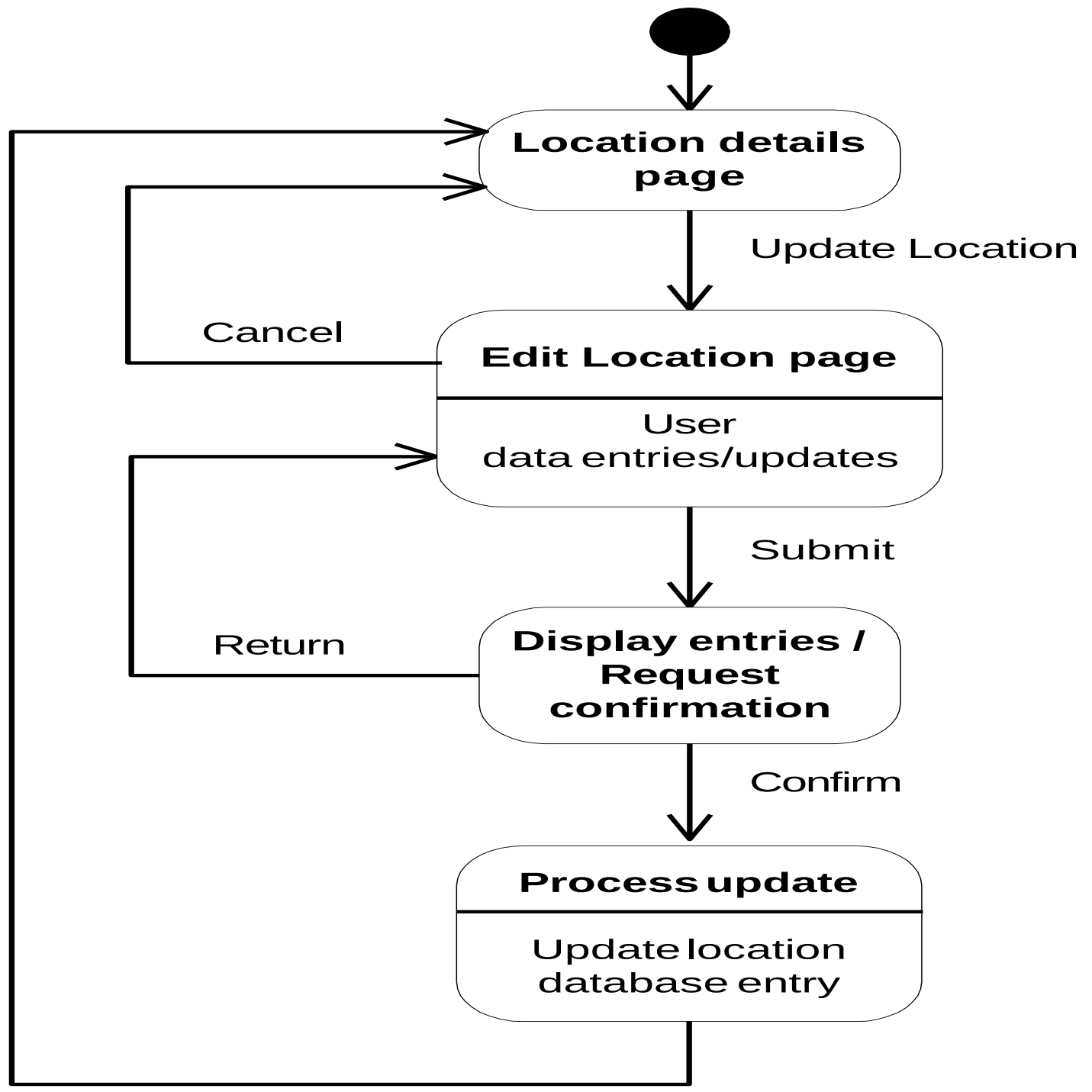

**Figure 37 -** Edit Location – State diagram

### 5.10.21 Assign Responsible to Lab

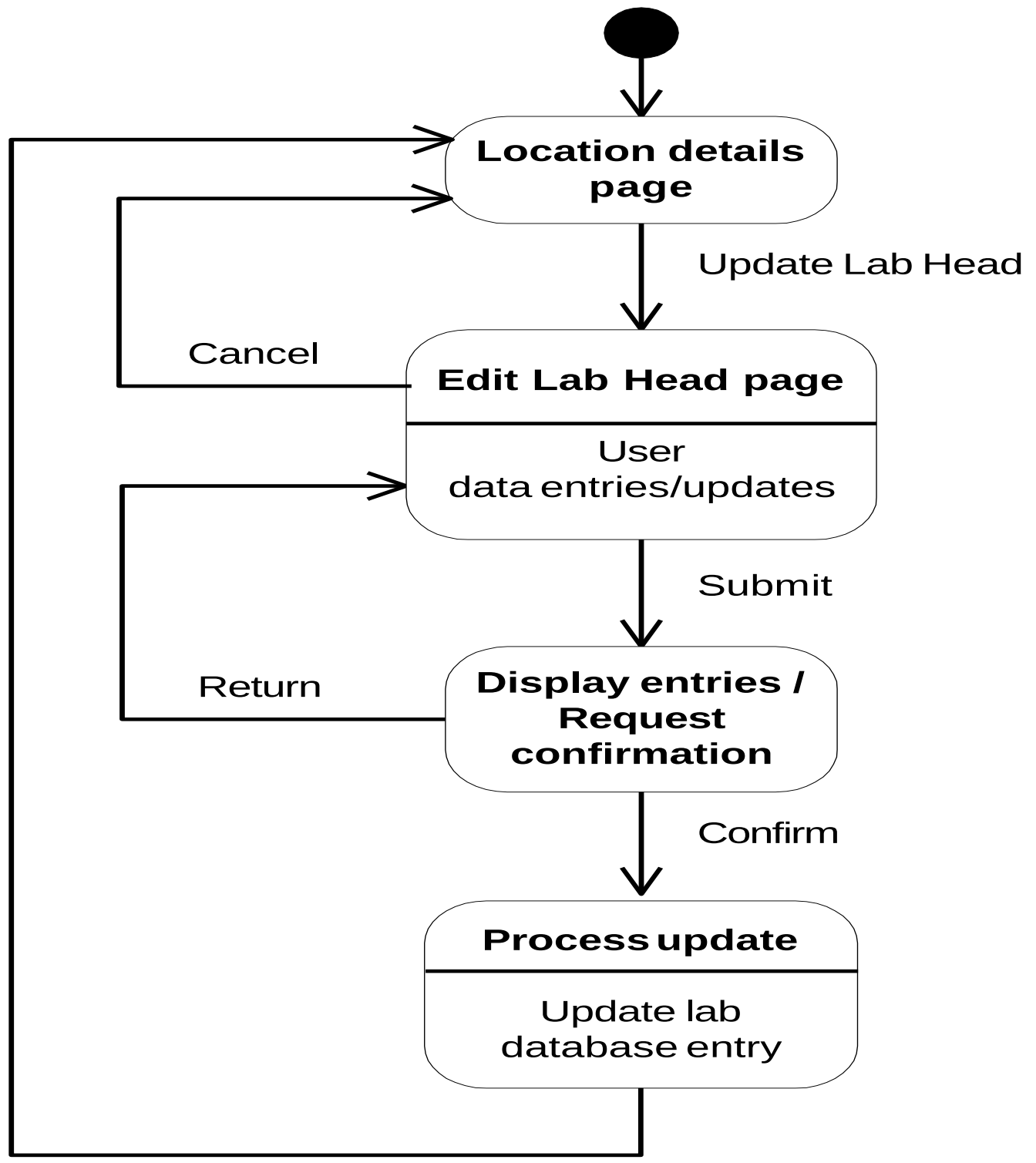

**Figure 38–** Assign Responsible to Lab – State diagram

### 5.10.22 Create a Building

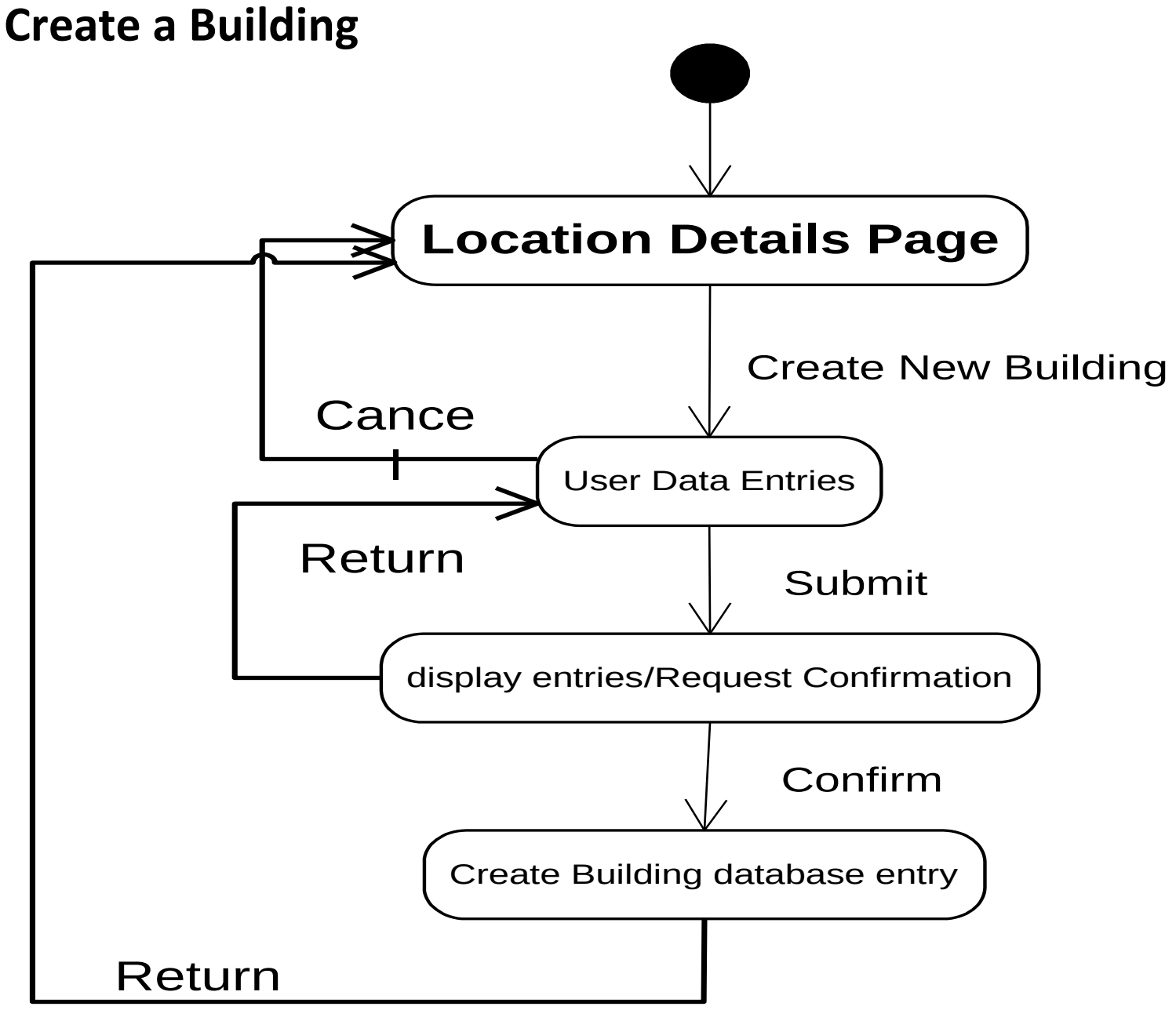

**Figure 39 –** Create Building – State diagram

### 5.10.23 Submit General Request

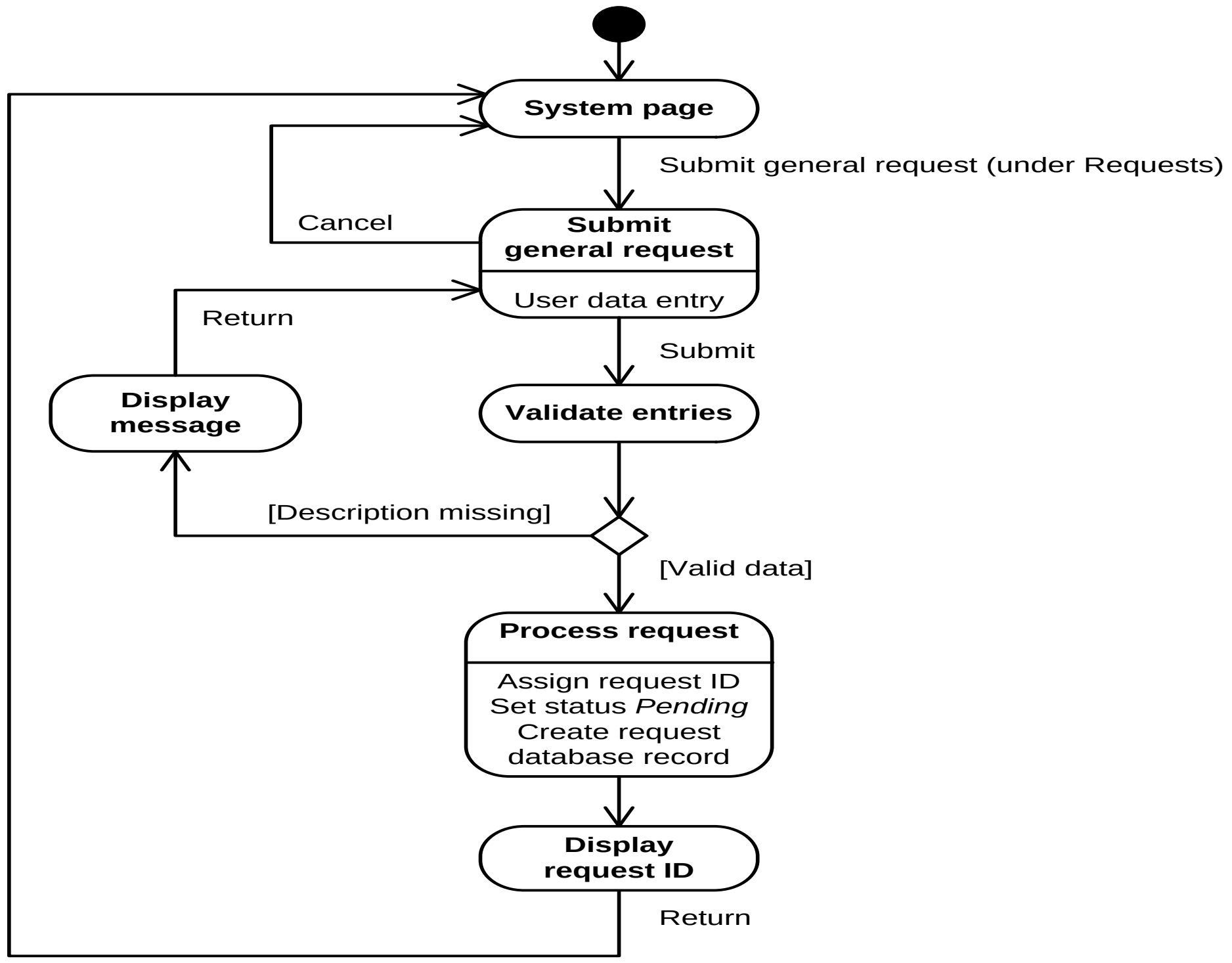

Figure 40 - **Submit general request – State diagram**

## 5.10.24 Submit Specific Request

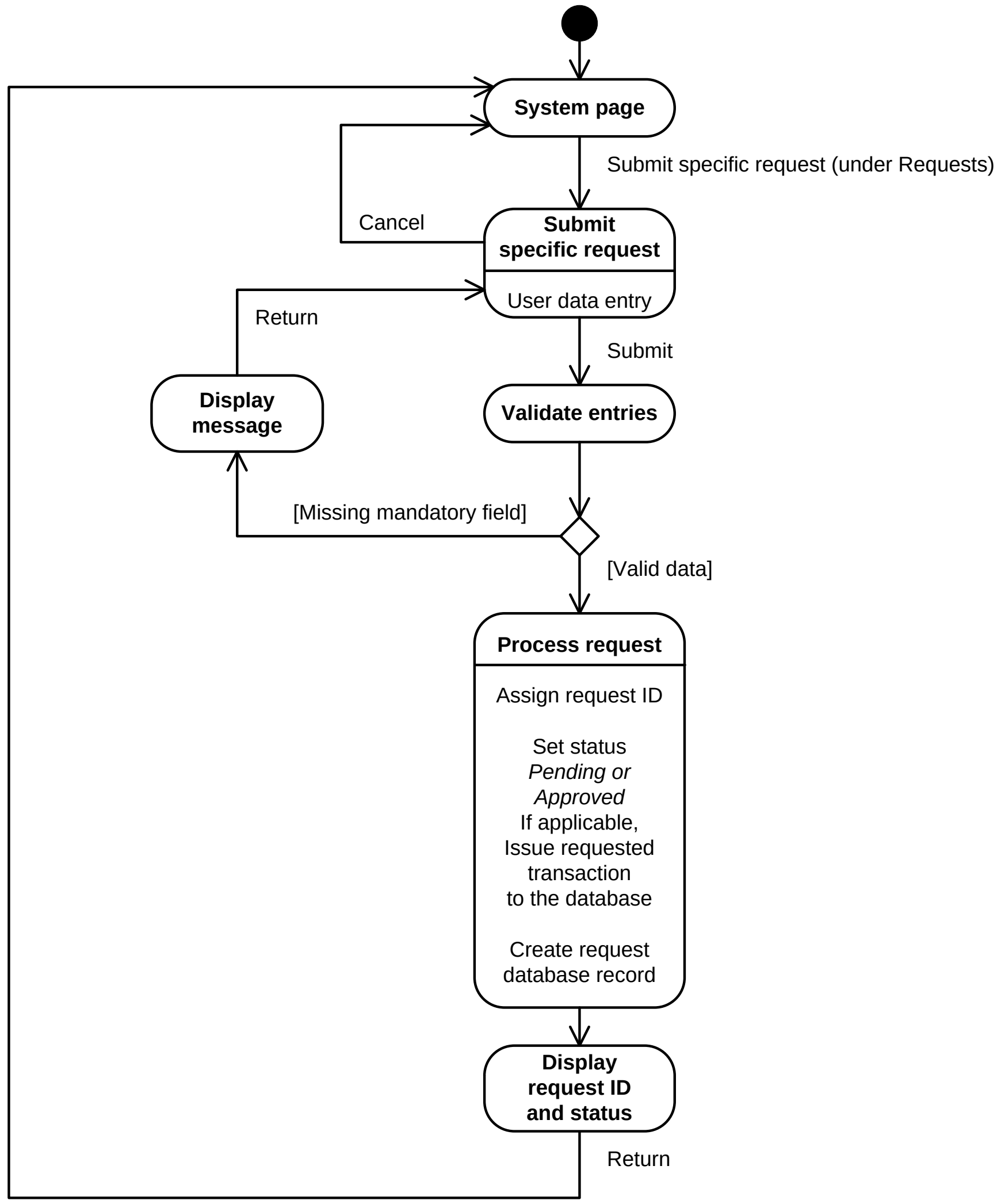

**Figure 41 -** Submit specific request – State diagram

### 5.10.25 Search Request

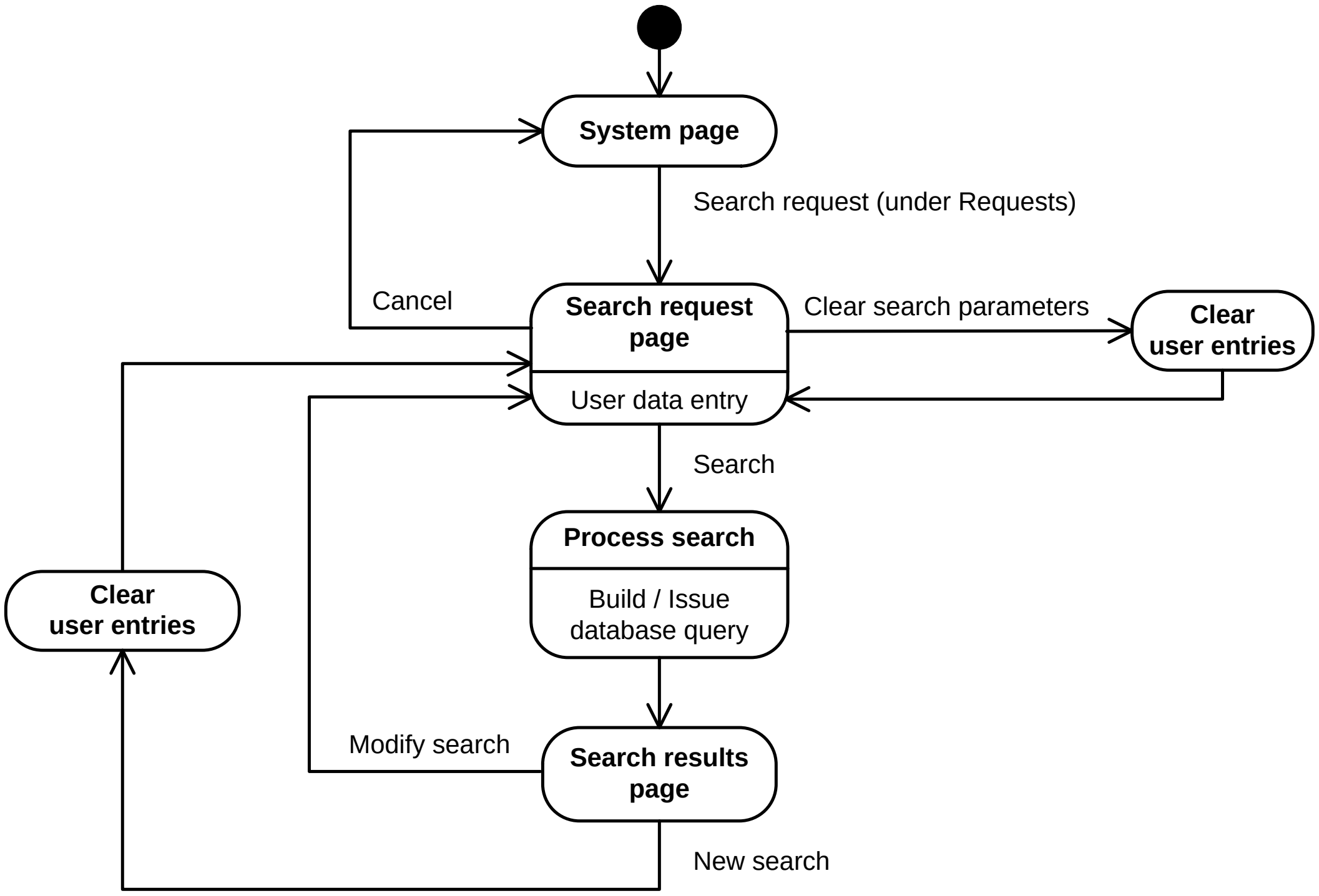

**Figure 42 -** Search request – State diagram

### 5.10.26 View General Request

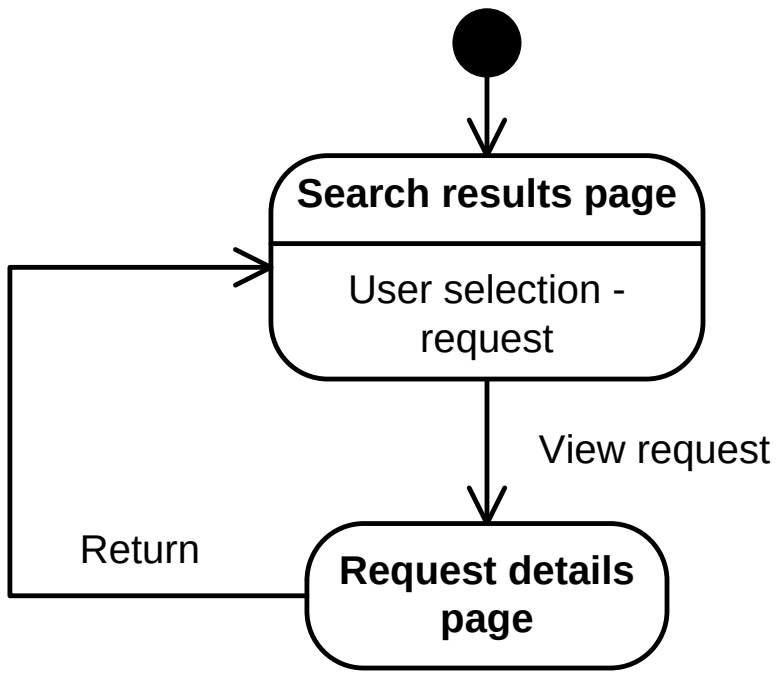

**Figure 43 -** View request – State diagram

### 5.10.27 Close General Request

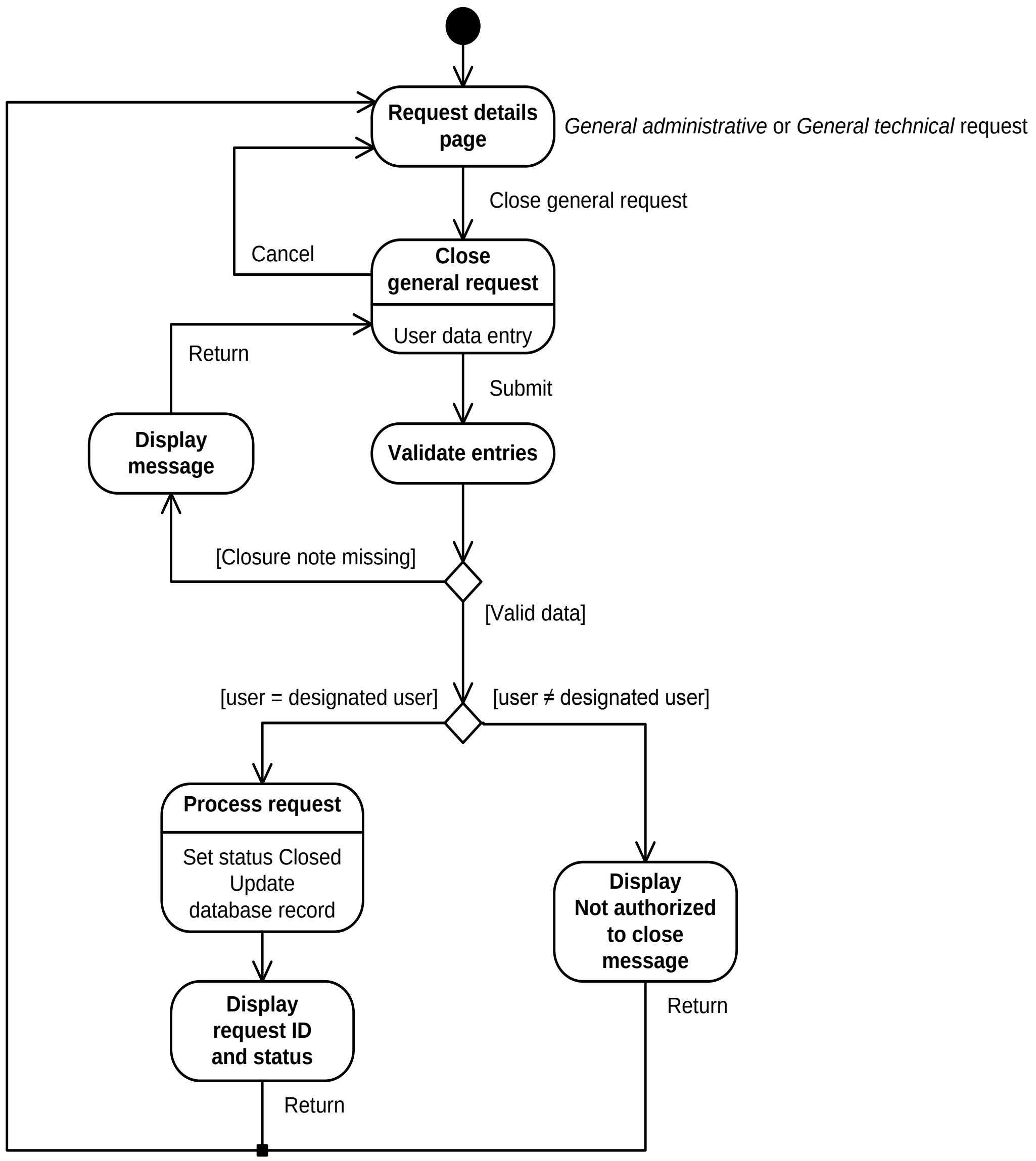

**Figure 44 -** Close general request – State diagram

### 5.10.28 Aprove Specific Request

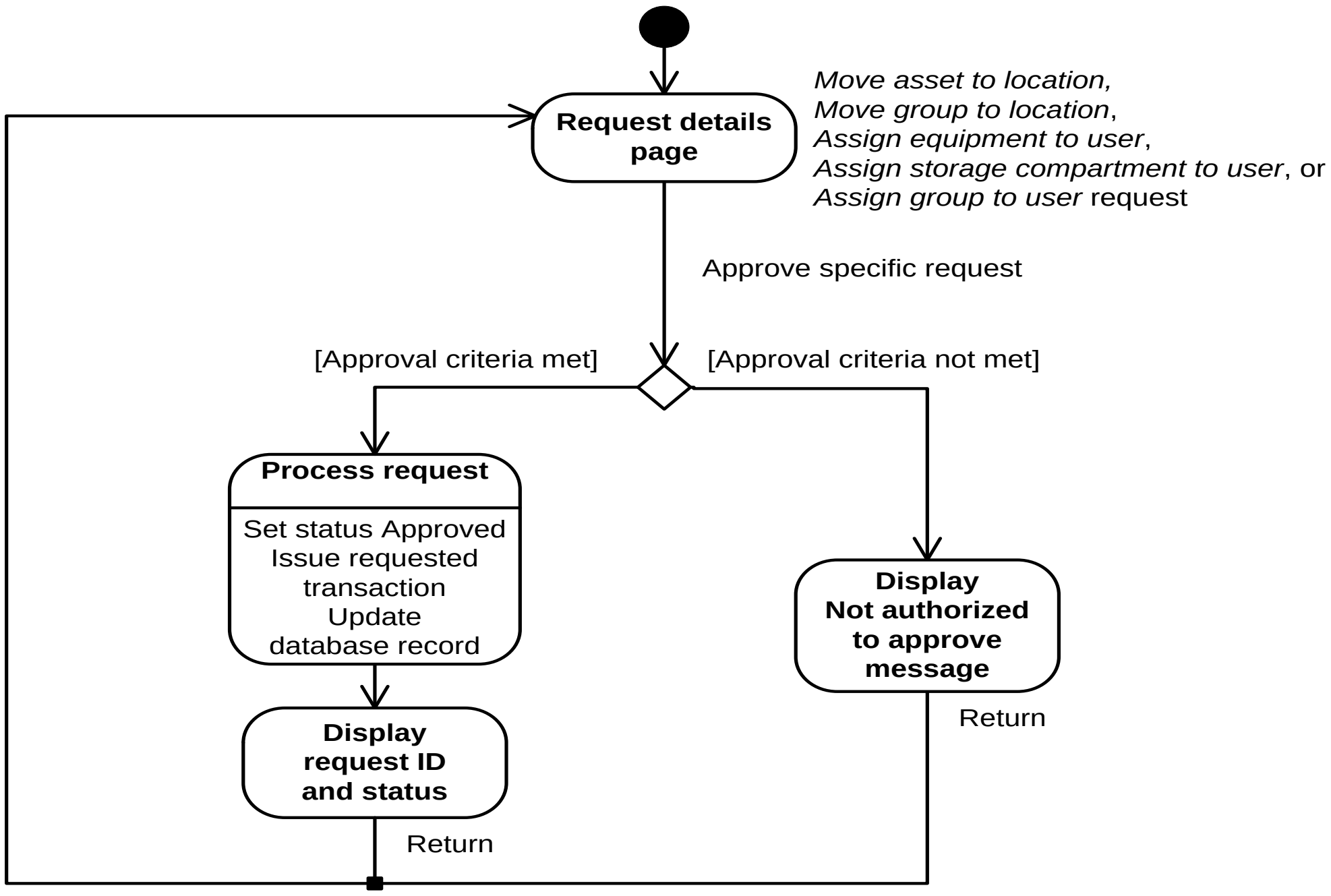

**Figure 45 -** Approve specific request – State diagram

## 5.11 Data Detailed Design

This section provides, in the form of data dictionary, a description of data elements internal to each entity of the UUIS system. The entities and their relationships are shown in Figure B. 1, Figure B. 2, Figure B. 3 and Figure B. 4 of Appendix B. The database model, in first normal form, was issued based on the ER diagram and the project requirements, and is shown in the following figure:

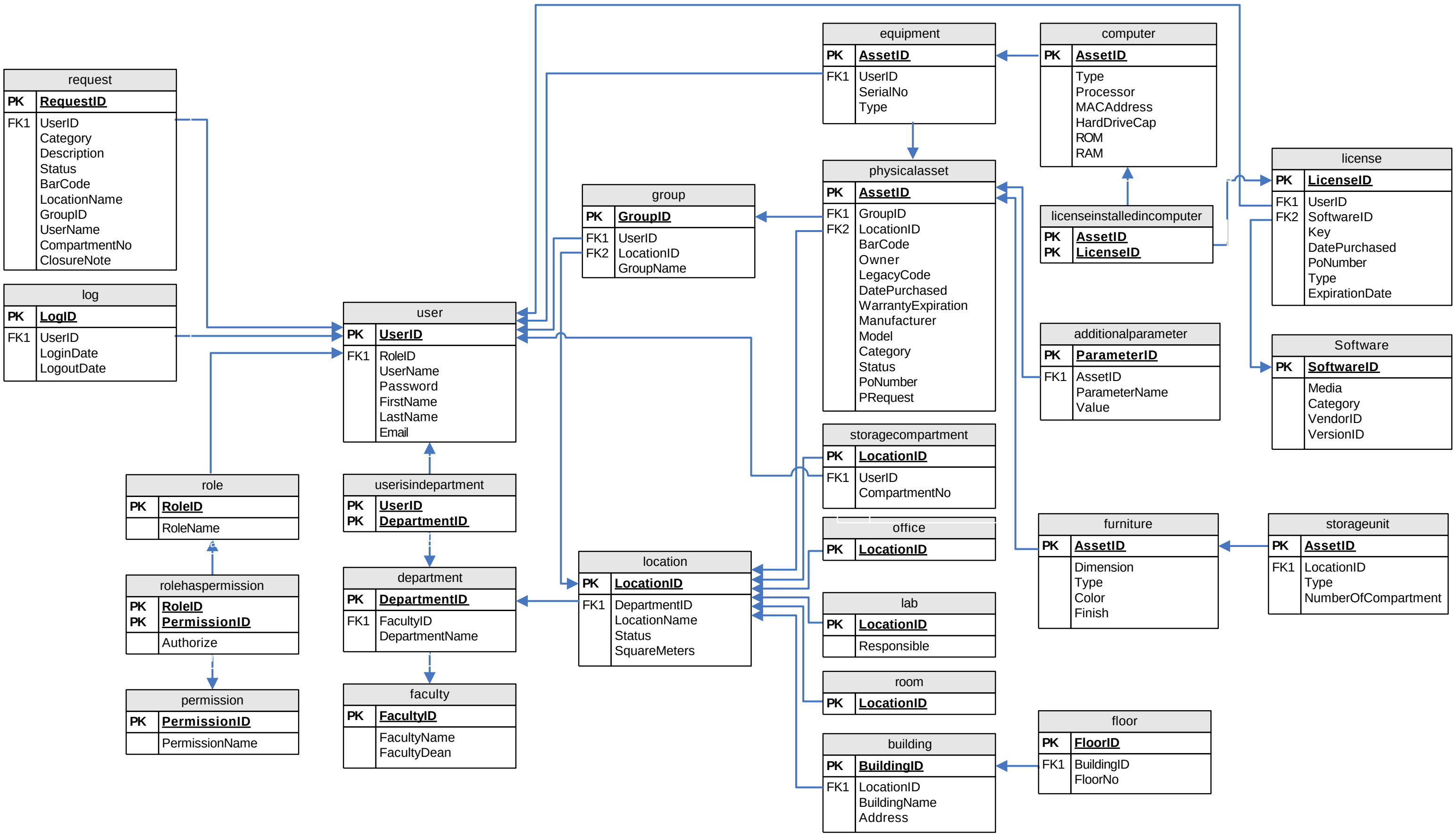

**Figure 46** – IUfA UUIS Database model

### 5.11.1 User Entity Description

The following table provides the data dictionary for the *User* entity.

**Table 1** – User Table Data Dictionary

| Column Name | Data Type | NULL | Default | Description |
|---|---|---|---|---|
| UserID (PK) | integer | No | | Unique value created at each new entry |
| RoleID (FK) | integer | No | | Foreign Key to Role Table |
| UserName | varchar(32) | No | | |
| Password | varchar(32) | No | | |
| FirstName | varchar(64) | Yes | Null | |
| LastName | varchar(64) | Yes | Null | |
| Email | varchar(64) | Yes | Null | |

### 5.11.2 Role Entity Description

The following table provides the data dictionary for the *Role* entity.

**Table 2** – Role Entity Data Dictionary

| Column Name | Data Type | NULL | Default | Description |
|---|---|---|---|---|
| RoleID (PK) | integer | No | | Unique value created at each new entry |
| RoleName | varchar(32) | Yes | | |

### 5.11.3 Permission Entity Description

The following table provides the data dictionary for the *Permission* entity.

**Table 3** – Permission Entity Data Dictionary

| Column Name | Data Type | NULL | Default | Description |
|---|---|---|---|---|
| PermissionID (PK) | integer | No | | Unique value created at each new entry |
| PermissionName | varchar(128) | No | | |

### 5.11.4 Request Entity Description

The following table provides the data dictionary for the *Request* entity.

**Table 4 –** Request Entity Data Dictionary

| Column Name | Data Type | NULL | Default | Description |
|---|---|---|---|---|
| RequestID (PK) | integer | No | | Unique value created at each new entry |
| UserID (FK) | integer | No | | Foreign Key to User Table |
| Category | varchar(64) | No | | |
| Description | varChar(1024) | Yes | Null | |
| Status | varchar(32) | No | “Pending” | |
| BarCode | varchar(64) | Yes | Null | |
| LocationName | varchar(128) | Yes | Null | |
| GroupID | integer | Yes | Null | |
| UserName | varchar(64) | Yes | Null | |
| CompartmentNo | integer | Yes | Null | |
| ClosureNote | varchar(256) | Yes | Null | 1..n |

### 5.11.5 Log Entity Description

The following table provides the data dictionary for the *Log* entity.

**Table 5 –** Log Entity Data Dictionary

| Column Name | Data Type | NULL | Default | Description |
|---|---|---|---|---|
| LogID (PK) | integer | No | | Unique value created at each new entry |
| UserID (FK) | integer | No | | Foreign Key to User Table |
| LoginDate | timestamp | No | | |
| LogoutDate | timestamp | Yes | Null | |

### 5.11.6 Department Entity Description

The following table provides the data dictionary for the *Department* entity.

**Table 6** – Department Entity Data Dictionary

| Column Name | Data Type | NULL | Default | Description |
|---|---|---|---|---|
| DepartmentID (PK) | integer | No | | Unique value created at each new entry |
| FacultyID | integer | No | | Foreign Key to Faculty Table |
| DepartmentName | varchar(128) | No | | |

### 5.11.7 Faculty Entity Description

The following table provides the data dictionary for the *Faculty* entity.

**Table 7** – Faculty Entity Data Dictionary

| Column Name | Data Type | NULL | Default | Description |
|---|---|---|---|---|
| FacultyID (PK) | integer | No | | Unique value created at each new entry |
| FacultyName | varchar(128) | No | | |
| FacultyDean | varchar(128) | Yes | Null | |

### 5.11.8 Location Entity Description

The following table provides the data dictionary for the *Location* entity.

**Table 8** – Location Entity Data Dictionary

| Column Name | Data Type | NULL | Default | Description |
|---|---|---|---|---|
| LocationID (PK) | integer | No | | Unique value created at each new entry |
| DepartmentID (FK) | integer | No | | Foreign Key to Department Table |
| LocationName | varchar(128) | No | | |
| Status | varchar(32) | Yes | Null | |
| SquareMeters | integer | Yes | Null | |

### 5.11.9 Lab Entity Description

The following table provides the data dictionary for the *Lab* entity.

**Table 9** – Lab Entity Data Dictionary

| Column Name | Data Type | NULL | Default | Description |
|---|---|---|---|---|
| LocationID (PK) | integer | No | | Primary Key from Location Table |
| Reponsible | varchar(128) | Yes | | |

### 5.11.10 Room Entity Description

The following table provides the data dictionary for the *Room* entity.

**Table 10** – Room Entity Data Dictionary

| Column Name | Data Type | NULL | Default | Description |
|---|---|---|---|---|
| LocationID (PK) | integer | No | | Primary Key from Location Table |

### 5.11.11 Office Entity Description

The following table provides the data dictionary for the *Office* entity.

**Table 11** – Office Entity Data Dictionary

| Column Name | Data Type | NULL | Default | Description |
|---|---|---|---|---|
| LocationID (PK) | integer | No | | Primary Key from Location Table |

### 5.11.12 StorageCompartment Entity Description

The following table provides the data dictionary for the *StorageCompartment* entity.

**Table 12** – StorageCompartment Entity Data Dictionary

| Column Name | Data Type | NULL | Default | Description |
|---|---|---|---|---|
| LocationID (PK) | integer | No | | Primary Key from Location Table |
| UserID (FK) | integer | No | | Foreign Key to User Table |
| CompartmentNo | integer | No | | |

### 5.11.13 Building Entity Description

The following table provides the data dictionary for the *Building* entity.

**Table 13** – Building Entity Data Dictionary

| Column Name | Data Type | NULL | Default | Description |
|---|---|---|---|---|
| BuildingID (PK) | integer | No | | Unique value created at each new entry |
| LocationID (FK) | integer | No | | Foreign Key to Location Table |
| BuildingName | varchar(128) | No | | |
| Address | varchar(128) | Yes | Null | |

### 5.11.14 Floor Entity Description

The following table provides the data dictionary for the *Floor* entity.

**Table 14** – Floor Entity Data Dictionary

| Column Name | Data Type | NULL | Default | Description |
|---|---|---|---|---|
| FloorID (PK) | integer | No | | Unique value created at each new entry |
| BuildingID (FK) | integer | No | | Foreign Key to Building Table |
| FloorNo | integer | No | | |

### 5.11.15 Group Entity Description

The following table provides the data dictionary for the *Group* entity.

**Table 15** – Group Entity Data Dictionary

| Column Name | Data Type | NULL | Default | Description |
|---|---|---|---|---|
| GroupID (PK) | integer | No | | Unique value created at each new entry |
| UserID (FK) | integer | No | | Foreign Key to User Table |
| LocationID (FK) | integer | No | | Foreign Key to Location Table |
| GroupName | varchar(128) | Yes | Null | |

### 5.11.16 AdditionalParameter Entity Description

The following table provides the data dictionary for the *AdditionalParameter* entity.

**Table 16** – AdditionalParameter Entity Data Dictionary

| Column Name | Data Type | NULL | Default | Description |
|---|---|---|---|---|
| ParameterID (PK) | int(255) | No | | Primary Key from PhysicalAsset Table |
| AssetID (FK) | varchar(64) | No | | Foreign Key to PhysicalAsset Table |
| ParameterName | varchar(128) | No | | |
| Value | varchar(64) | Yes | Null | |

### 5.11.17 PhysicalAsset Entity Description

The following table provides the data dictionary for the *PhysicalAsset* entity.

**Table 17** – PhysicalAsset Entity Data Dictionary

| Column Name | Data Type | NULL | Default | Description |
|---|---|---|---|---|
| AssetID (PK) | integer | No | | Unique value created at each new entry |
| LocationID (FK) | integer | No | | Foreign Key to Location Table |
| GroupID (FK) | integer | No | | Foreign Key to Group Table |
| BarCode | varchar(64) | No | | |
| Owner | varchar(128) | No | | |
| LegacyCode | varchar(64) | Yes | Null | |
| DatePurchased | timestamp | Yes | Null | |
| WarrantyExpiration | timestamp | Yes | Null | |
| Manufacturer | varchar(128) | Yes | Null | |
| Model | varchar(128) | Yes | Null | |
| Category | varchar(64) | Yes | Null | |
| Status | varchar(32) | No | “In-stock” | |
| PoNumber | varchar(64) | Yes | Null | |
| PRequest | varchar(64) | Yes | Null | |

### 5.11.18 Furniture Entity Description

The following table provides the data dictionary for the *Furniture* entity.

**Table 18** – Furniture Entity Data Dictionary

| Column Name | Data Type | NULL | Default | Description |
|---|---|---|---|---|
| AssetID (PK) | integer | No | | Primary Key from PhysicalAsset Table |
| Dimension | varchar(64) | Yes | Null | |
| Type | varchar(64) | Yes | Null | |
| Color | varchar(64) | Yes | Null | |
| Finish | varchar(64) | Yes | Null | |

### 5.11.19 StorageUnit Entity Description

The following table provides the data dictionary for the *StorageUnit* entity.

**Table 19** – StorageUnit Entity Data Dictionary

| Column Name | Data Type | NULL | Default | Description |
|---|---|---|---|---|
| AssetID (PK) | integer | No | | Primary Key from PhysicalAsset Table |
| LocationID (PK) | integer | No | | Foreign Key to Location Table |
| Type | varchar(64) | Yes | Null | |
| NumberOfCompartment | integer | No | 1 | |

### 5.11.20 Equipment Entity Description

The following table provides the data dictionary for the *Equipment* entity.

**Table 20** – Equipment Entity Data Dictionary

| Column Name | Data Type | NULL | Default | Description |
|---|---|---|---|---|
| AssetID (PK) | integer | No | | Primary Key from PhysicalAsset Table |
| UserID (FK) | integer | No | | Foreign Key to User Table |
| SerialNo | varchar(64) | Yes | Null | |
| Type | varchar(64) | Yes | Null | |

### 5.11.21 Computer Entity Description

The following table provides the data dictionary for the *Computer* entity.

**Table 21** – Computer Entity Data Dictionary

| Column Name | Data Type | NULL | Default | Description |
|---|---|---|---|---|
| AssetID (PK) | integer | No | | Primary Key from PhysicalAsset Table |
| Type | varchar(64) | Yes | Null | |
| Processor | varchar(64) | Yes | Null | |
| MACAddress | varchar(64) | Yes | Null | |
| HardDriveCap | varchar(64) | Yes | Null | |
| ROM | varchar(64) | Yes | Null | |
| RAM | varchar(64) | Yes | Null | |

### 5.11.22 License Entity Description

The following table provides the data dictionary for the *License* entity.

**Table 22** – License Entity Data Dictionary

| Column Name | Data Type | NULL | Default | Description |
|---|---|---|---|---|
| LicenseID (PK) | integer | No | | Unique value created at each new entry |
| UserID (FK) | integer | No | | Foreign Key to User Table |
| SoftwareID (FK) | integer | No | | Foreign Key to SoftwareTable |
| Key | varchar(128) | No | | |
| DatePurchased | timestamp | No | | |
| PoNumber | varchar(64) | Yes | Null | |
| Type | varchar(64) | No | | |
| ExpirationDate | timestamp | No | | |

### 5.11.23 Software Entity Description

The following table provides the data dictionary for the *Software* entity.

**Table 23** – Software Entity Data Dictionary

| Column Name | Data Type | NULL | Default | Description |
|---|---|---|---|---|
| SoftwareID (PK) | integer | No | | Unique value created at each new entry |
| Media | varchar(128) | Yes | Null | |
| Category | varchar(64) | Yes | Null | |
| VendorID | varchar(64) | No | | |
| VersionID | varchar(64) | No | | |

### 5.11.24 Role Has Permission Relationship Description

The following table provides the data dictionary for the *Role Has Permission* relationship.

**Table 24** – RoleHasPermission Relationship Data Dictionary

| Column Name | Data Type | NULL | Default | Description |
|---|---|---|---|---|
| RoleID (PK) | integer | No | | Foreign Key to RoleTable |
| PermissionID (PK) | integer | No | | Foreign Key to Permission Table |
| Authorize | boolean | No | | |

### 5.11.25 User Is In Department Relationship Description

The following table provides the data dictionary for the *User Is In Department* relationship.

**Table 25 –** UserIsInDepartment Relationship Data Dictionary

| Column Name | Data Type | NULL | Default | Description |
|---|---|---|---|---|
| UserID (PK) | integer | No | | Foreign Key to User Table |
| DepartmentID (PK) | integer | No | | Foreign Key to Department Table |

### 5.11.26 License Installed In Computer Relationship Description

The following table provides the data dictionary for the *License Installed In computer* relationship.

**Table 26 –** LicenseInstalledInComputer Relationship Data Dictionary

| Column Name | Data Type | NULL | Default | Description |
|---|---|---|---|---|
| LicenseID (PK) | integer | No | | Foreign Key to License Table |
| AssetID (PK) | integer | No | | Foreign Key to Computer Table |

# Appendix A – User Interface Design – Prototype 11

The Prototype designs below show the position and contents of the menu and footer.

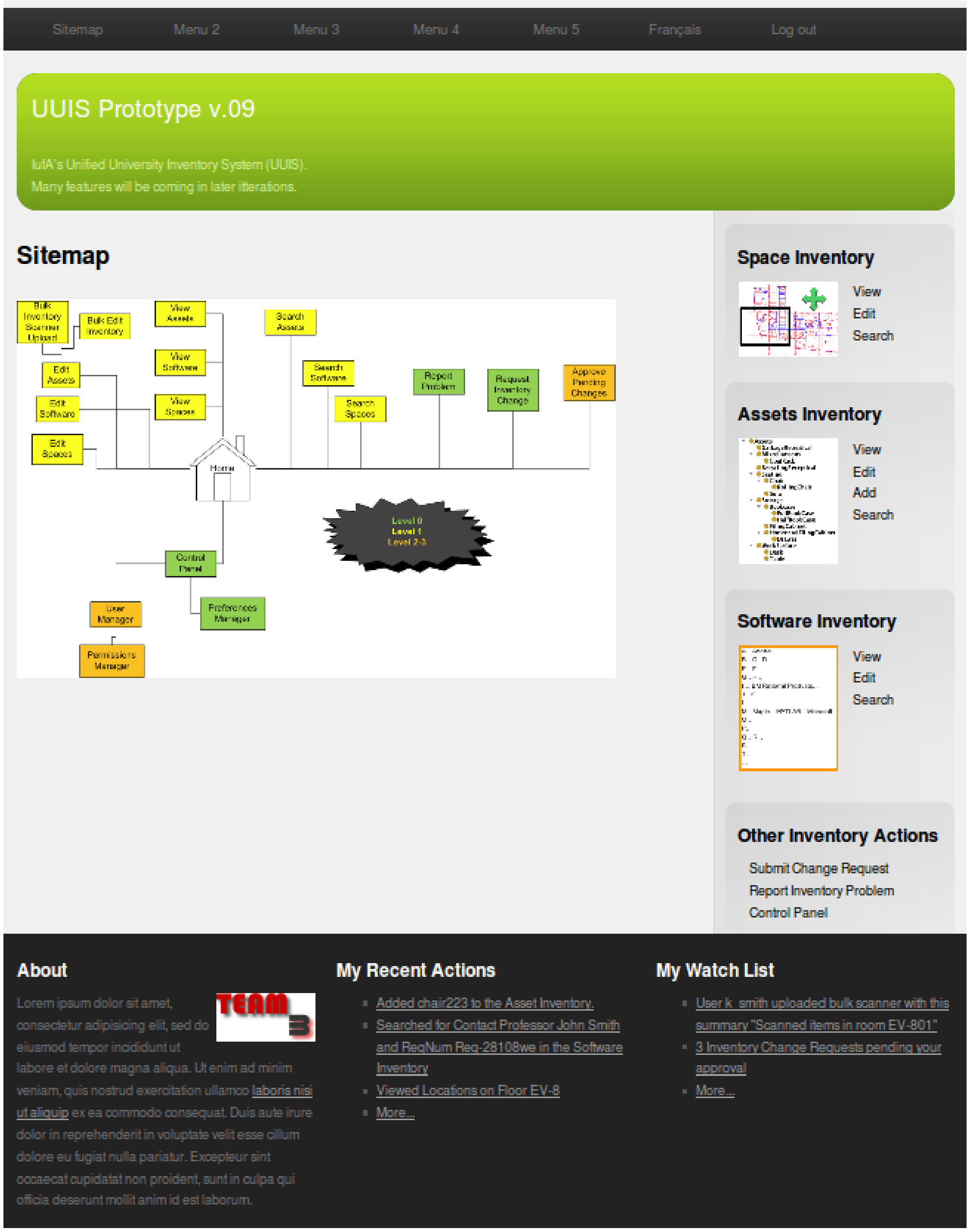

**Figure A. 1**- User Interface (1 of 12)

The following figures focus only on the changing content, namely the main display area.

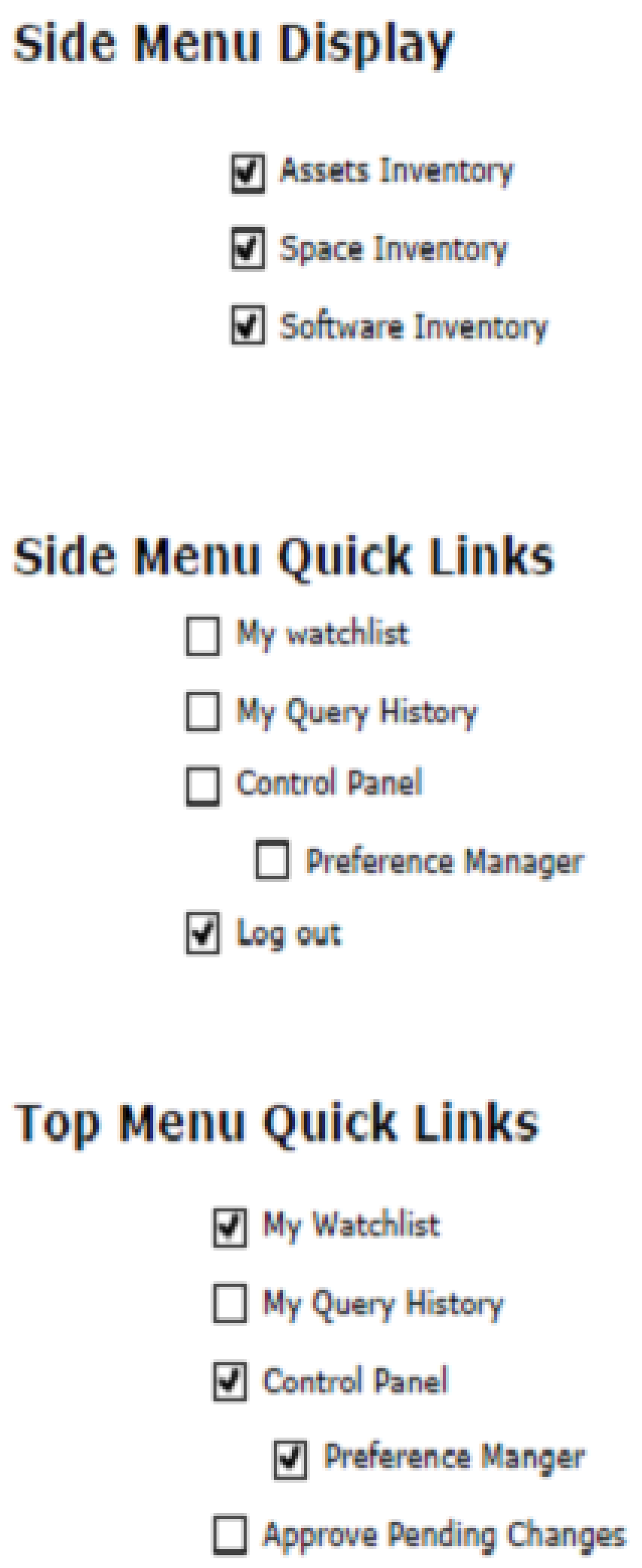

**Figure A. 2**- User Interface Customization of Menus (2 of 12)

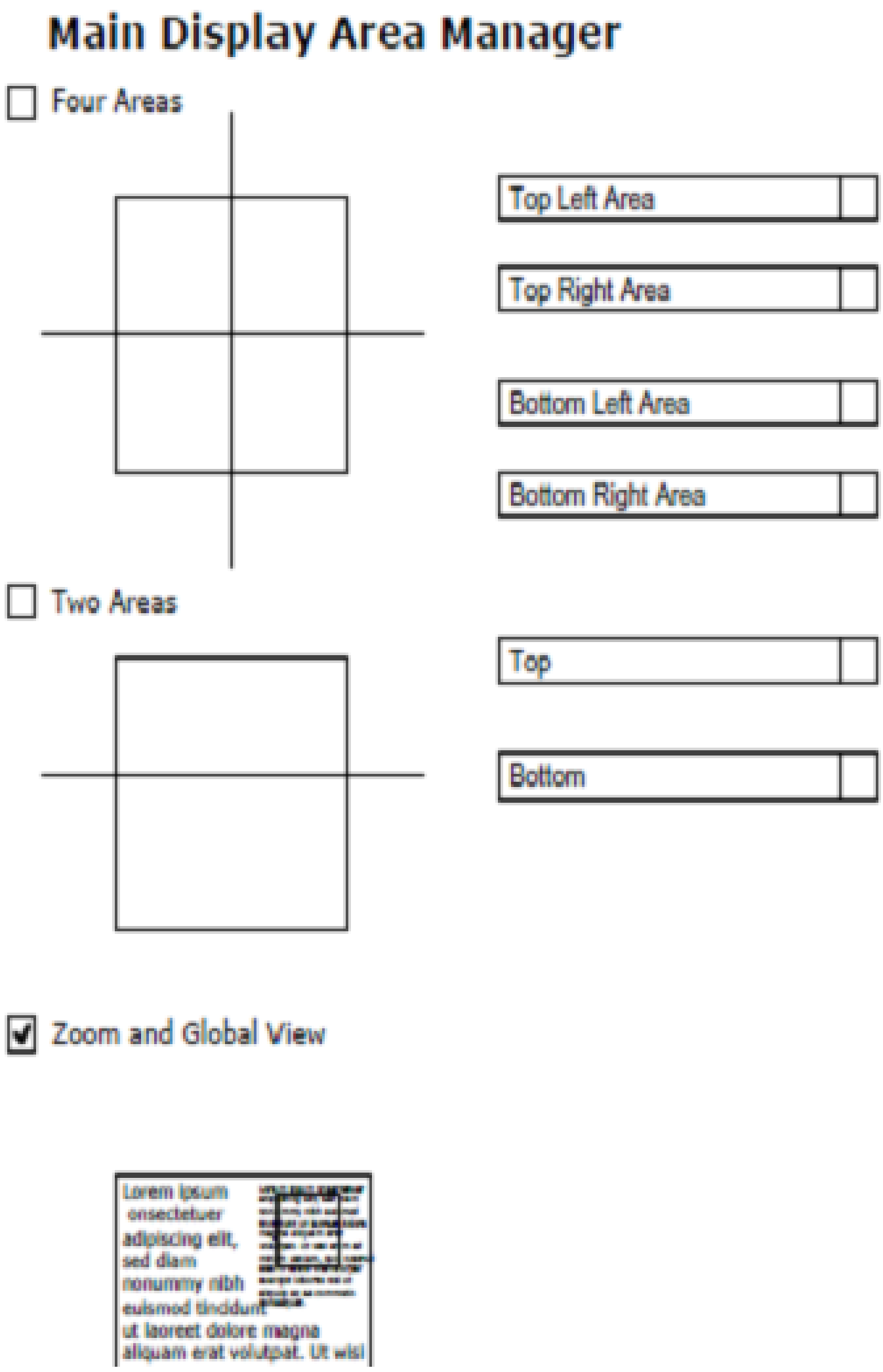

**Figure A. 3**- User Interface Customization of Main Display Area (3 of 12)

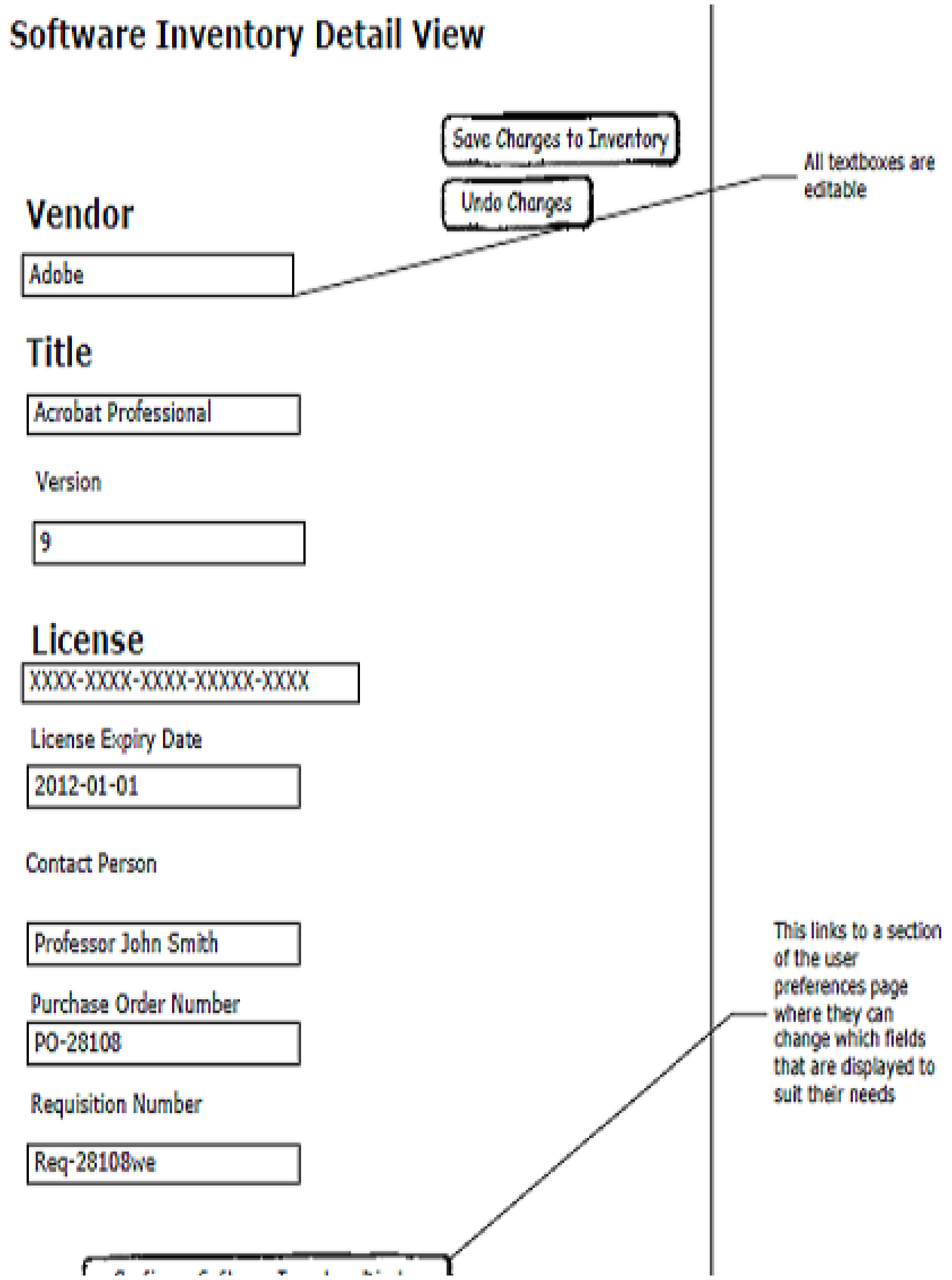

**Figure A. 4**- User Interface – Detail View (4 of 12)

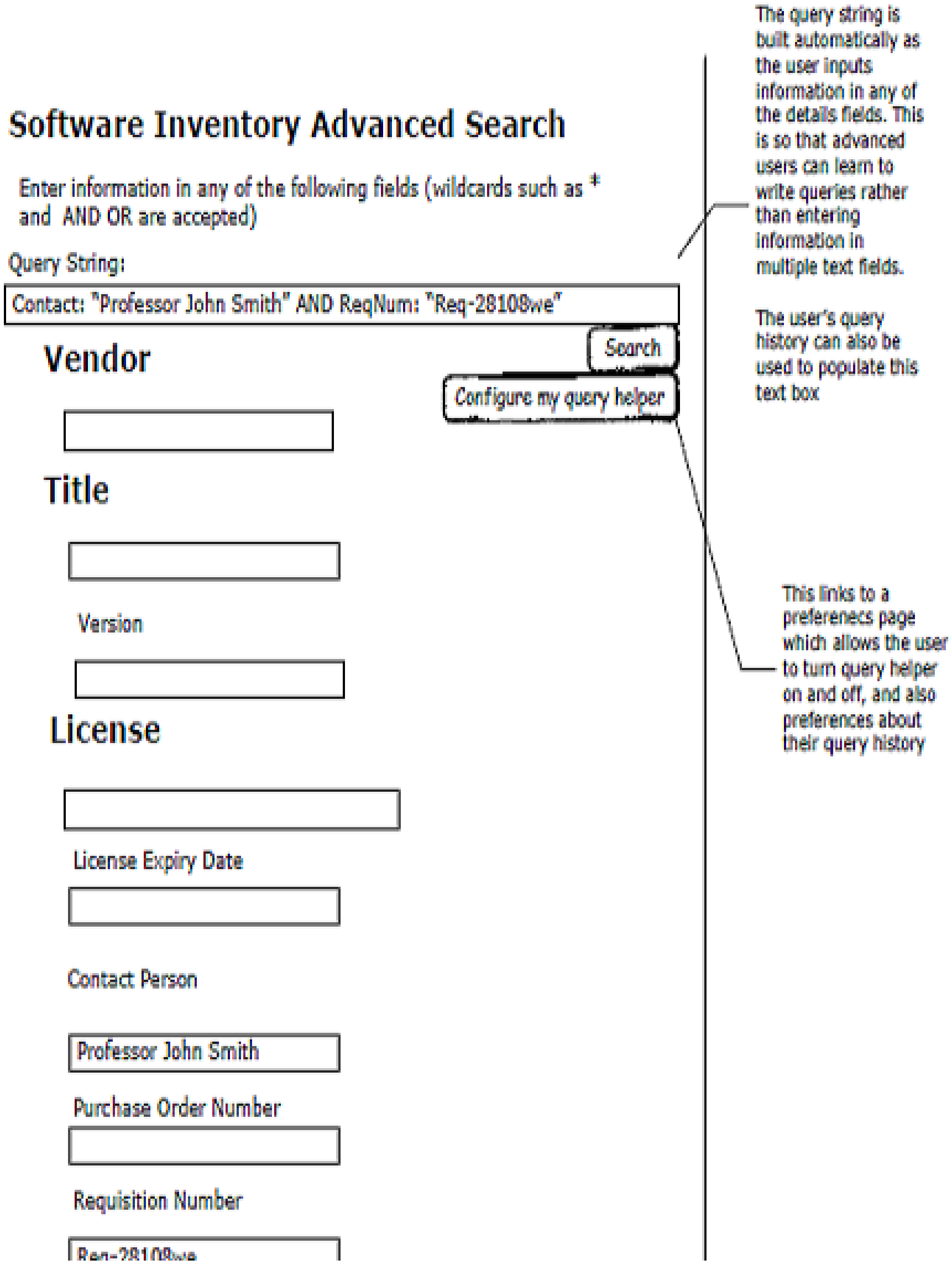

**Figure A. 5**- User Interface – Advanced Search (5 of 12)

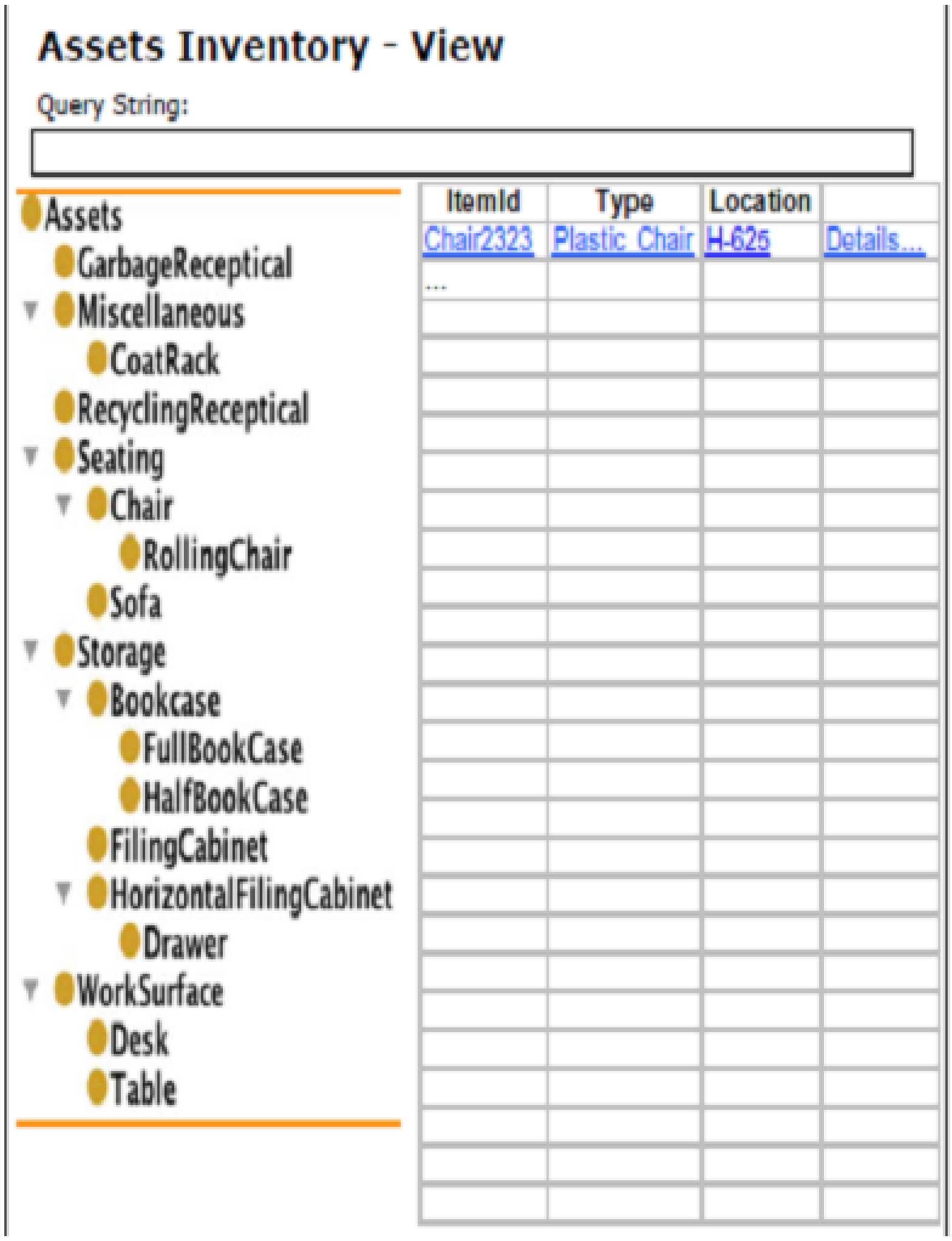

**Figure A. 6**- User Interface – Asset Ontology and Item List View (6 of 12)

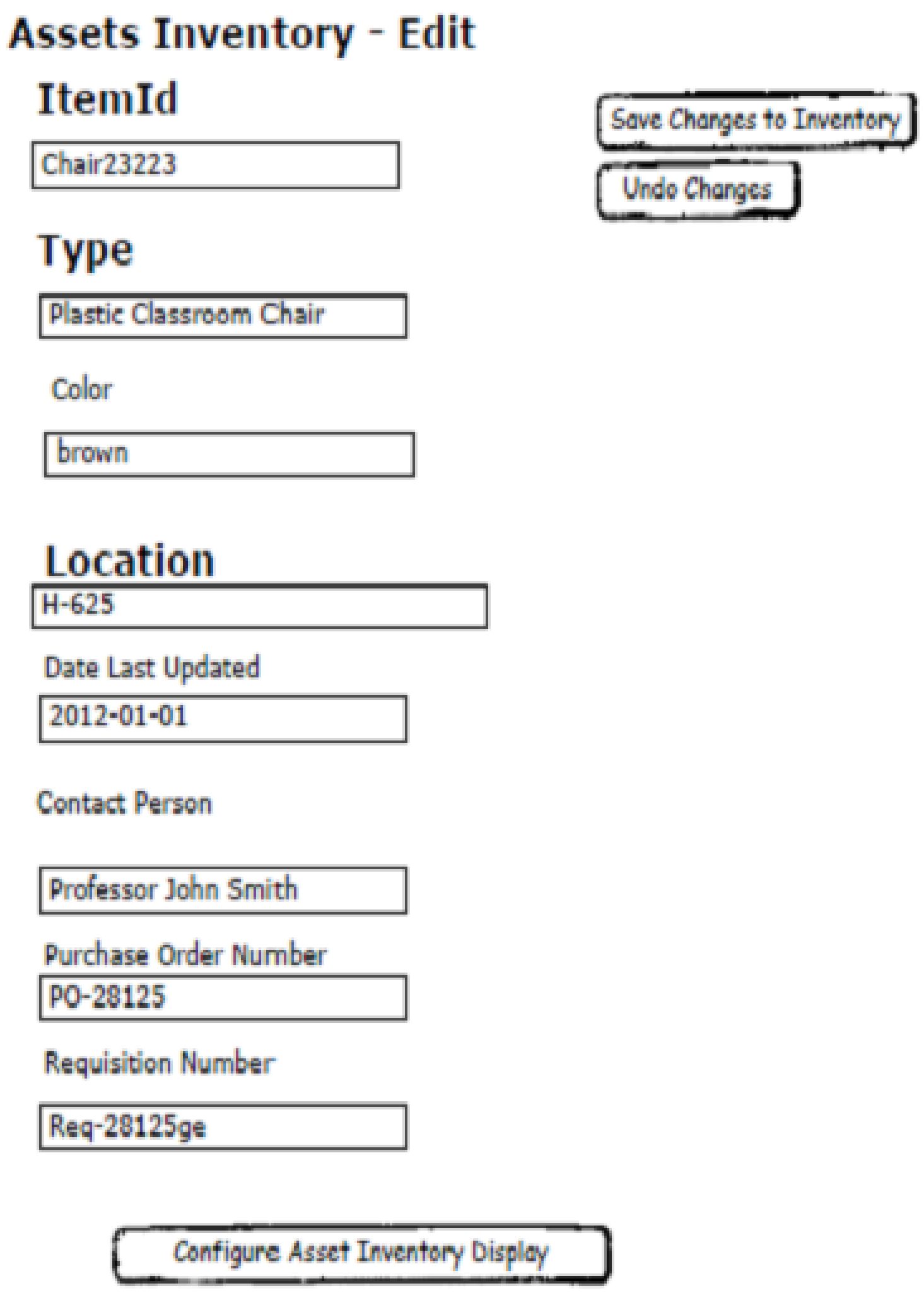

**Figure A. 7**- User Interface – Edit Asset Details (7 of 12)

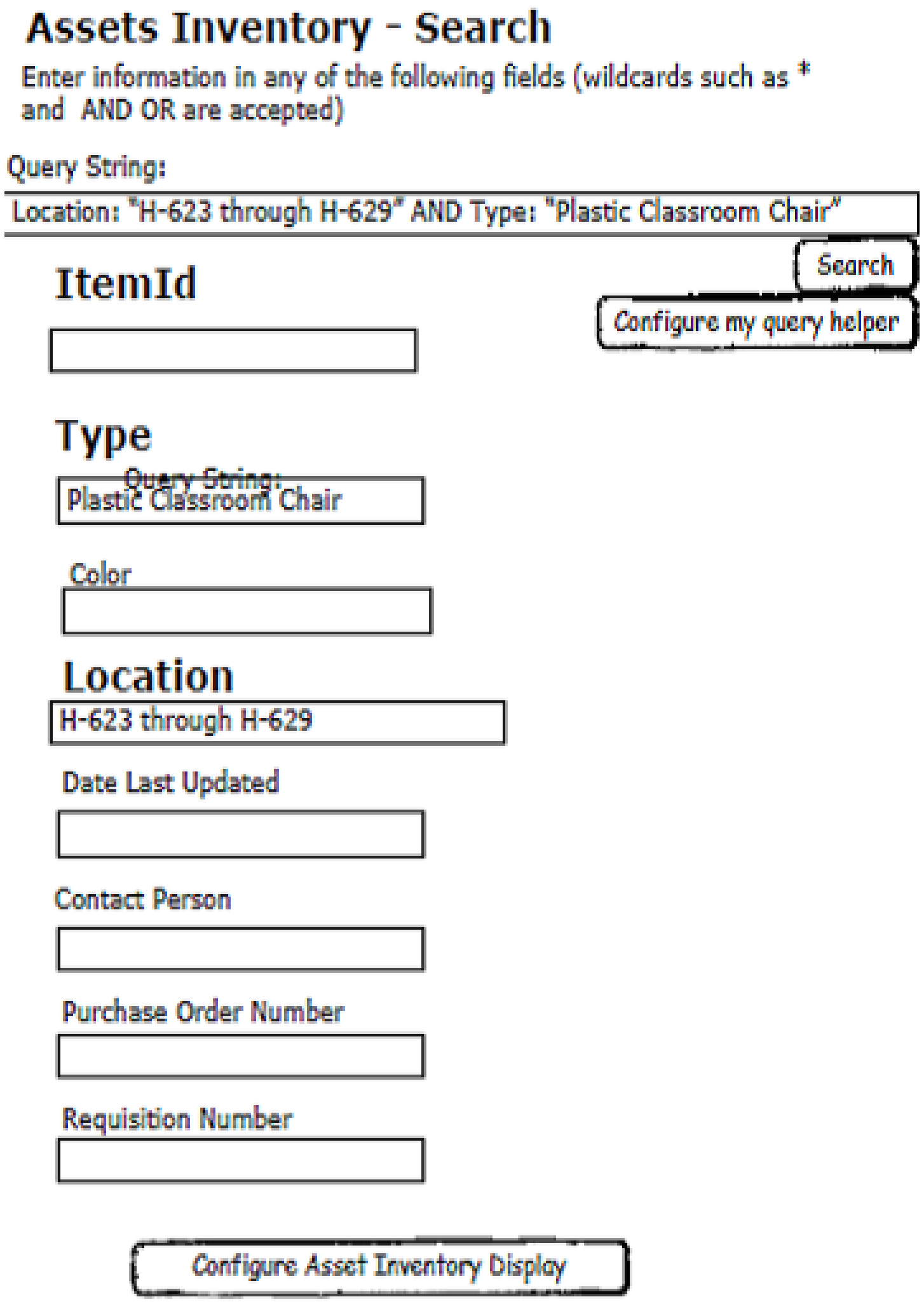

**Figure A. 8**- User Interface – Advanced Assets Search (8 of 12)

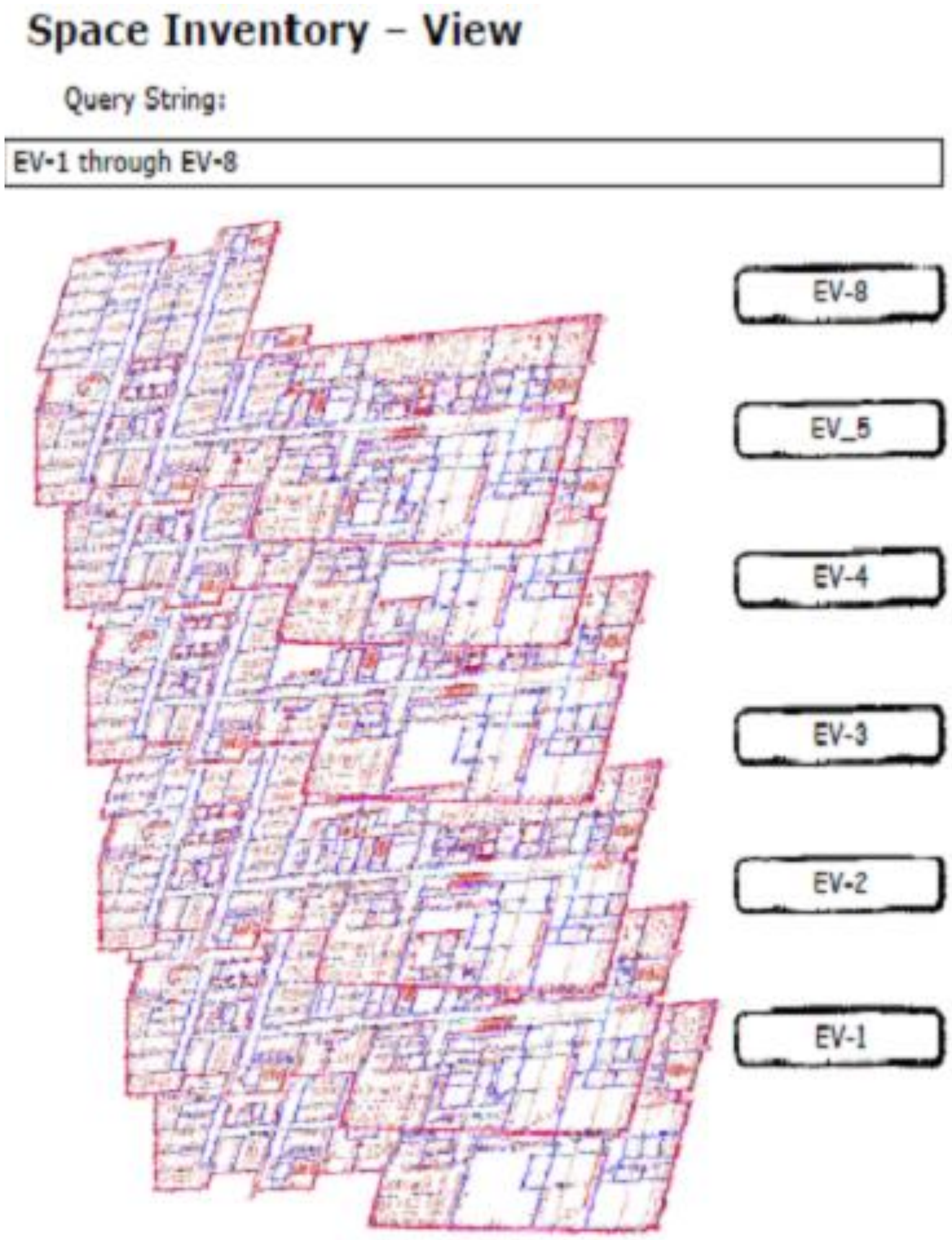

**Figure A. 9**- User Interface – Browse by floor visualization (9 of 12)

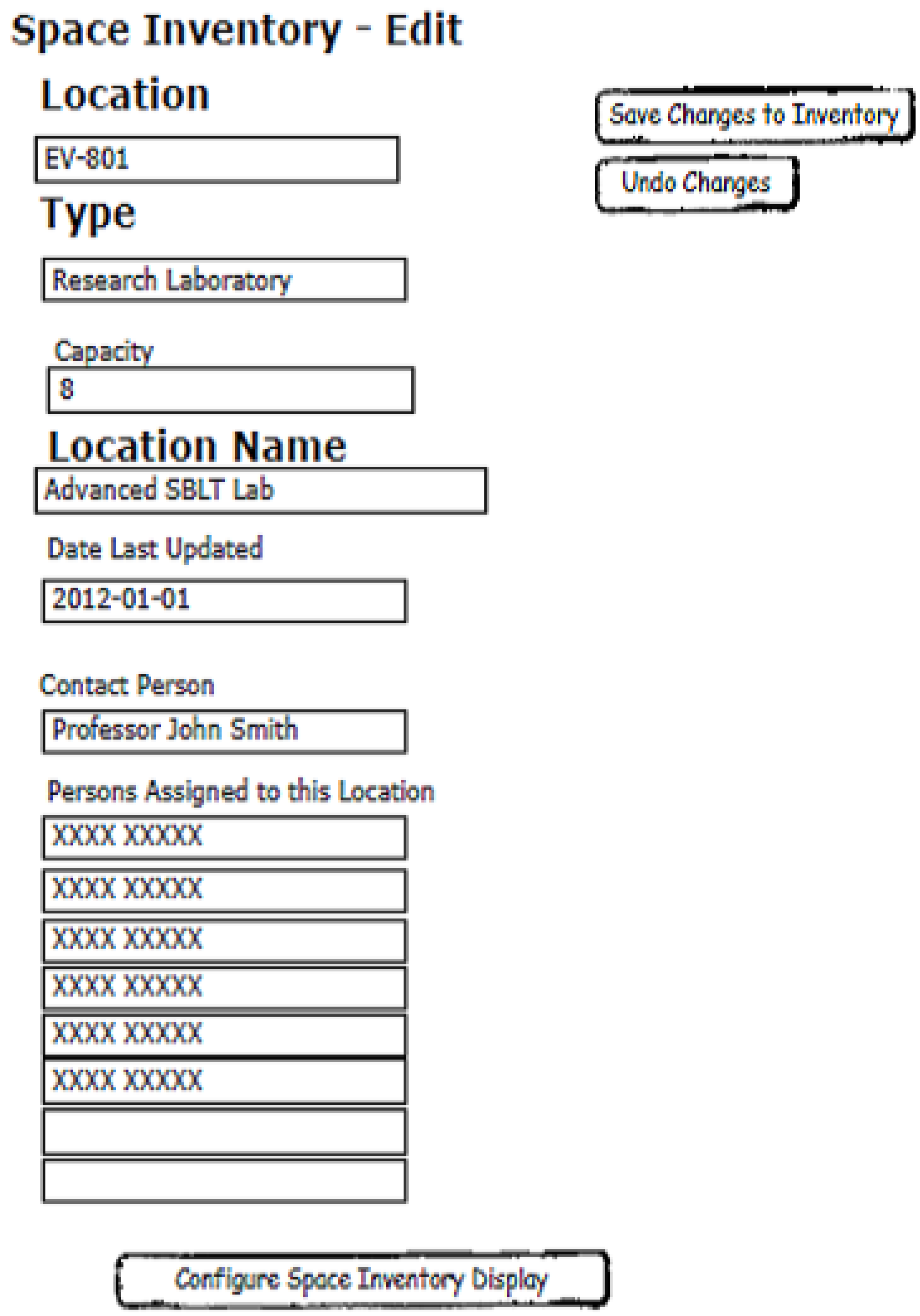

**Figure A. 10**- User Interface – Edit Locations Details (10 of 12)

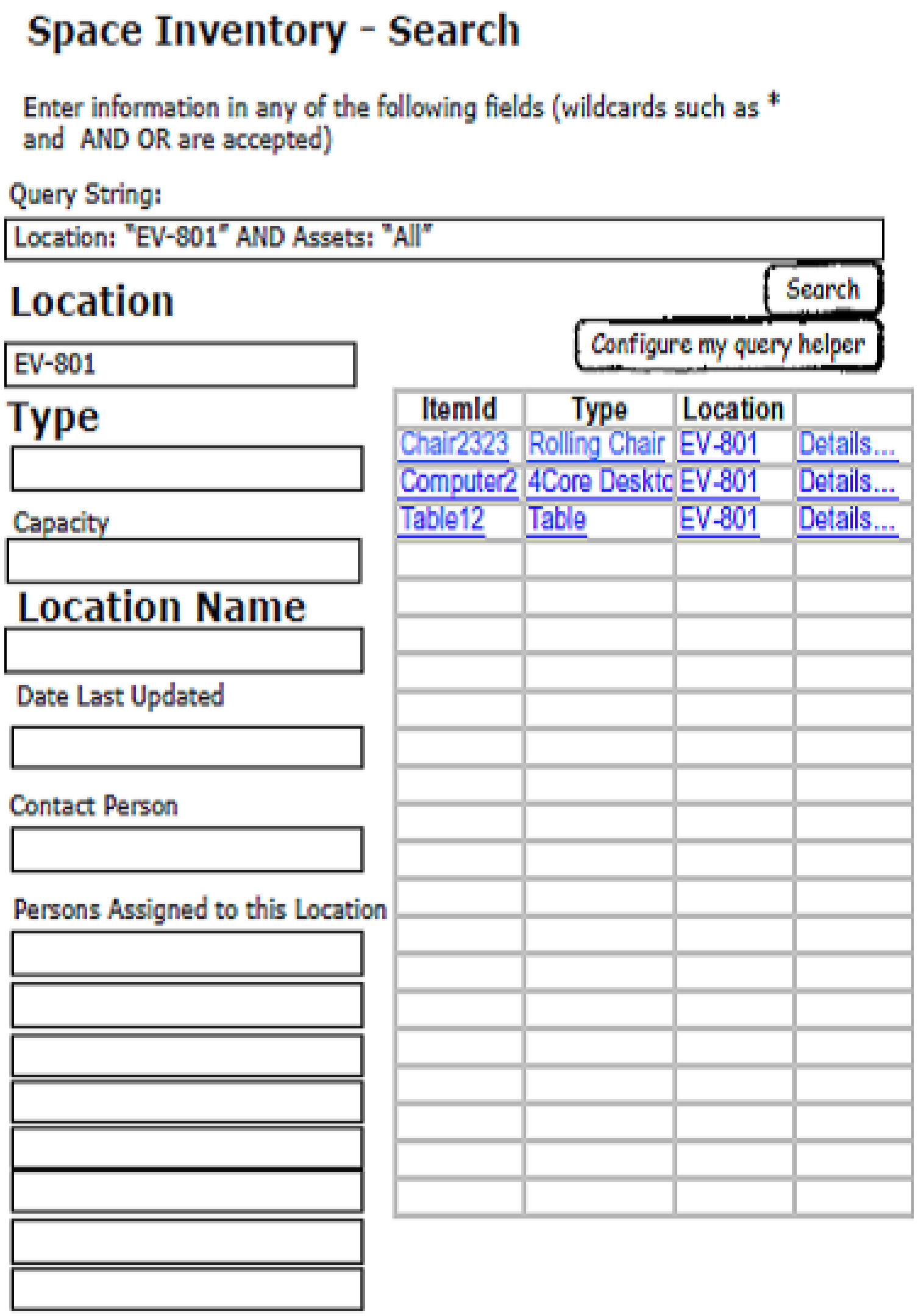

**Figure A. 11** - User Interface – Detail Location Search and Asset List View (11 of 12)

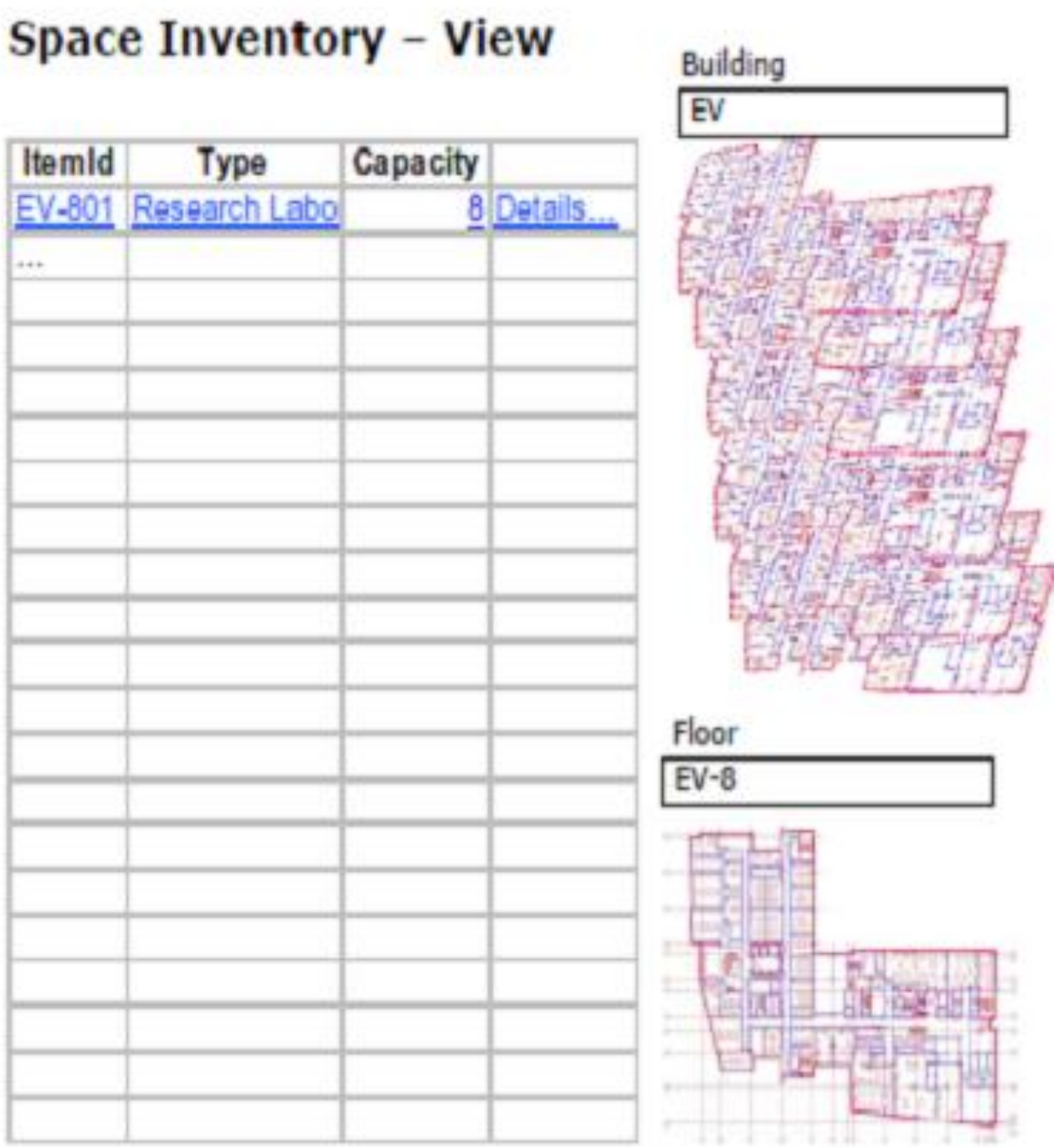

**Figure A. 12**- User Interface – Location list view and location selector visualization (12 of 12)

# Appendix B – E/R Diagram

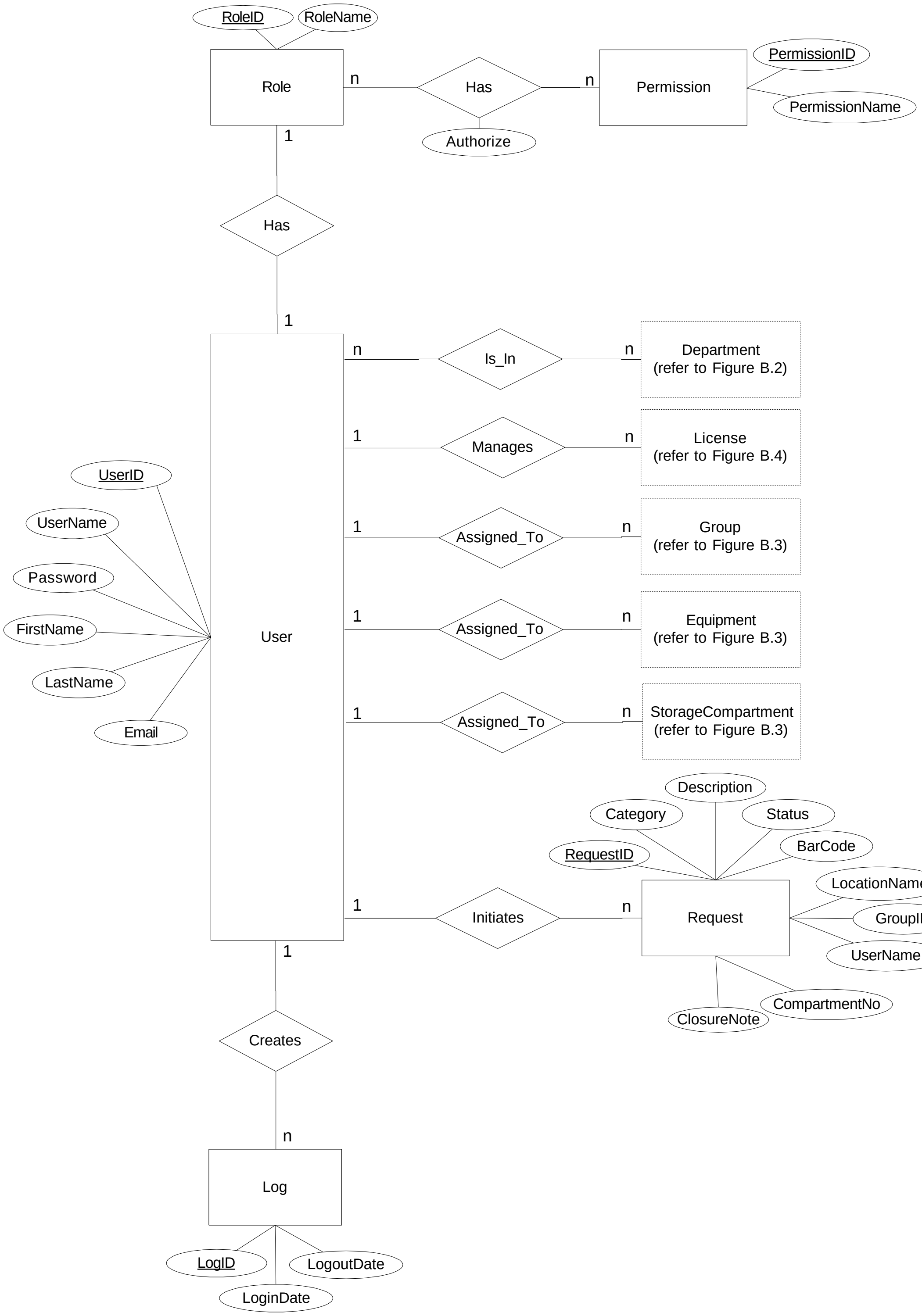

**Figure B. 1**– E/R Diagram (1 of 4)

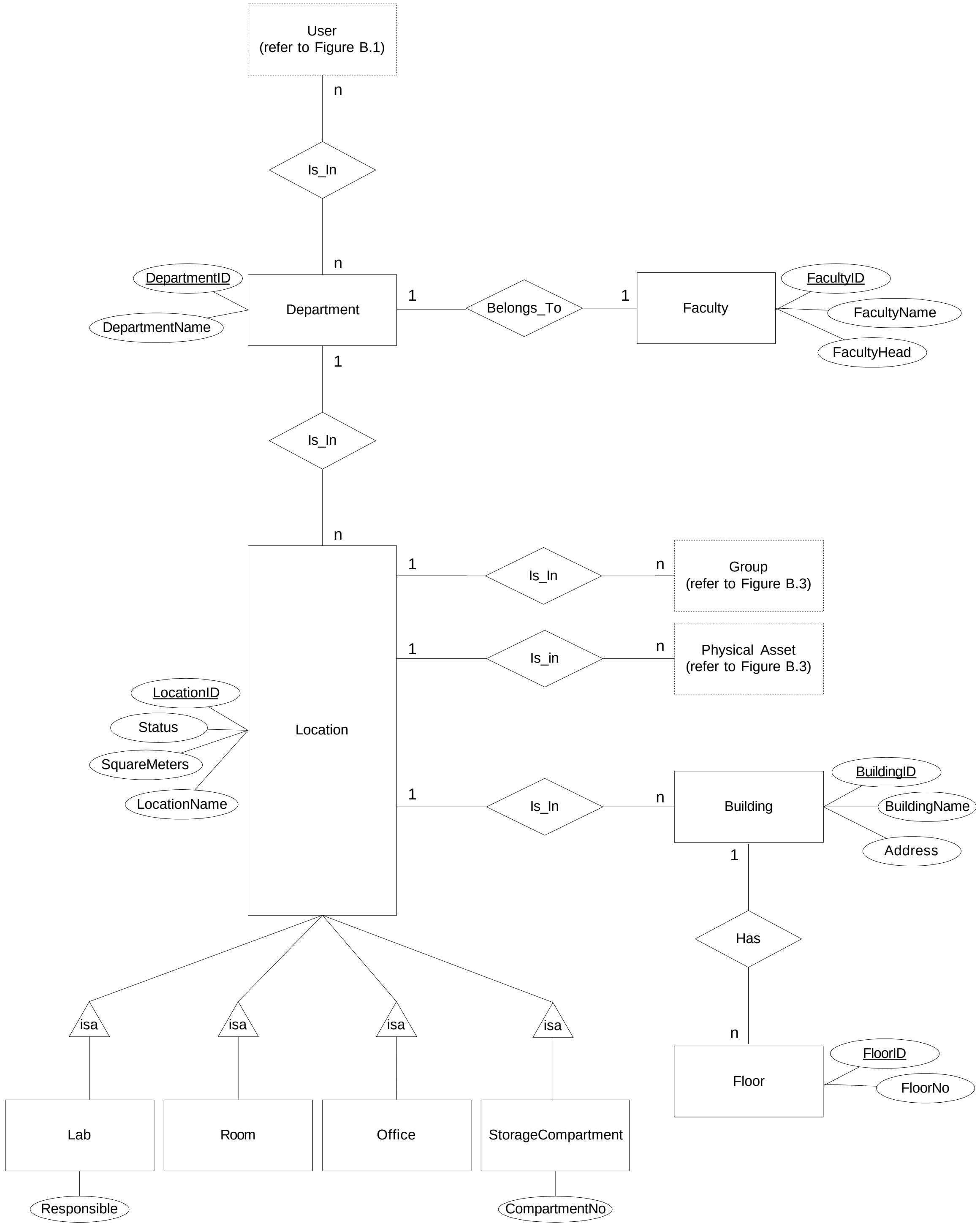

**Figure B. 2**– E/R Diagram (2 of 4)

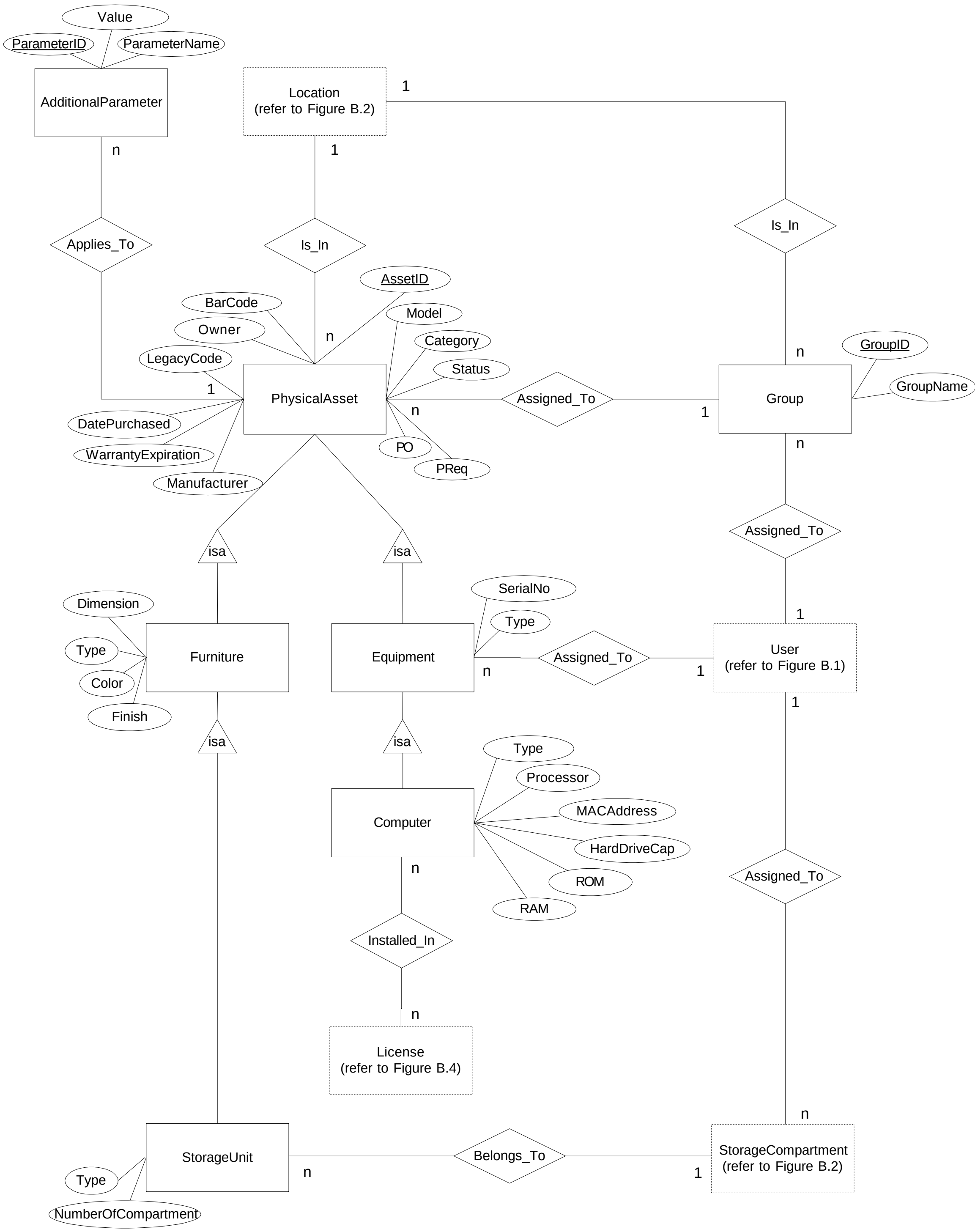

**Figure B. 3**– E/R Diagram (3 of 4)

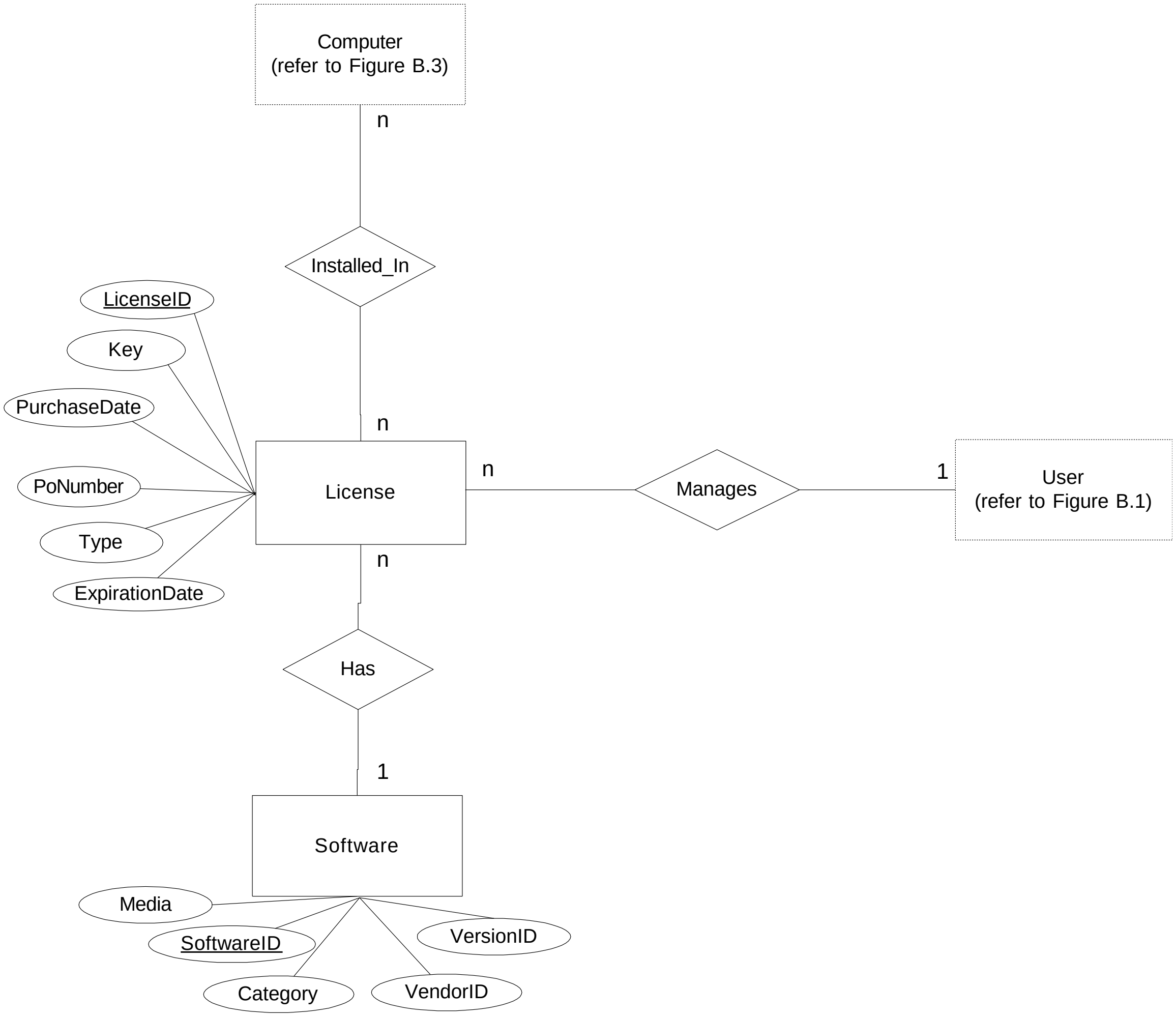

**Figure B. 4**– E/R Diagram (4 of 4)

# Appendix C – Test Cases

## C.1 Login Test Cases - all user roles

| Step No | Steps | Data | Expected Results | Pass/ Fail |
|---|---|---|---|---|
| 1 | Enter user name; no password; press Submit Button | User Name= test1 | Error Message | |
| 2 | Enter password; no user name; press Submit Button | Password= test1pass | Error Message | |
| 3 | Enter user name; wrong password; press Submit Button | User Name= test1 password=wrong | Error Message | |
| 4 | Enter wrong user name; password; press Submit Button | User name= wrong; password =test1pass | Error Message | |
| 5 | Press Cancel button | | Return to IUfA Home Page | |
| 6 | Enter user name; password; press Submit Button | User name= test1; password =test1pass | Go to UUIS main page with permission appropriate menus | |

## C.2. Logout Test Cases - all user roles

| Step No | Steps | Data | Expected Results | Pass/ Fail |
|---|---|---|---|---|
| 1 | Look on every page. Is the Logout option available? | | | |
| 2 | Press Logout Button | | Message to confirm logout | |
| 3 | Press Return Button | | Original screen appears; user still logged in | |
| 4 | Press Confirm Button | | Return to IUfA Home Page; user logged out | |

## C.3 Request Test Cases - enter general request - all user roles

| Step No | Steps | Data | Expected Results | Pass/ Fail |
|---|---|---|---|---|
| 1 | User selects Requests tab | | Submit General Request Button is | |
| | | | available | |
| 2 | User selects Submit General Request button | | Submit General Request Page is onscreen; | |
| 3 | User enters data, selects Submit | Technical button selected; no string in description field | Error message; allowed to enter data again | |
| 4 | User enters data, selects Submit | Technical button selected; 50 char string in description field | Request Entered, original page returns | |
| 5 | User enters data, selects Submit | Technical button selected; 260 char string in description field | Description field stops accepting new data after 256 chars. First 256 are used, request entered, original page returns. | |
| 6 | User enters data, selects Submit | Administrative button selected; 50 char string in description field | Request Entered, original page returns | |
| 7 | User enters data, selects Cancel | Administrative button selected; 50 char string in description field | No Request Entered, original page returns | |
| 8 | User enters no data, selects Cancel | | No Request Entered, original page returns | |

### C.4 Request Test Cases - search request - all user roles

| **Step No** | **Steps** | **Data** | **Expected Results** | **Pass/ Fail** |
|---|---|---|---|---|
| 1 | User selects Requests tab | | Search Request Button is available | |
| 2 | User selects Search Request button | | Search Request Page is onscreen; | |
| 3 | Role 3 User selects Search button | No data entered | Search request results page displays no items | |
| 4 | Role 0 User enters data; selects Search button | Pending is checked; all category boxes checked | Search request results page displays all requests where the user is the originator AND status is pending | |
| 5 | Role 1 User enters data; selects Search button | Closed is checked; all category boxes checked | Search request results page displays all requests submitted by any role 0 or role 1 user where the originator of the request is in the same department as the current user AND status is checked | |
| 6 | Role 2 User enters data; selects Search button | Approved is checked; all category boxes checked | Search request results page displays all requests submitted by any role 0, role 1, or role 2 user where the originator of the request is in the same faculty as the current user AND status is approved | |
| 7 | Role 3 User enters data; selects Search button | Pending, closed AND Approved is checked; all category boxes checked | Search request results page displays all requests in the database | |
| 8 | Role 0 User enters data; selects Search button | Originator username = "wrong"(not the test users username) AND all three status boxes checked; all category boxes checked | Search request results page displays empty list | |
| 9 | Role 1 User enters data; selects Search button | Originator department = "wrong"(not the test users department) AND all three status boxes checked; all category boxes checked | Search request results page displays empty list | |
| 10 | Role 2 User enters data; selects Search button | Originator faculty = "wrong"(not the test users faculty) AND all three status boxes checked; all category boxes checked | Search request results page displays empty list | |

## C.5 Request Test Cases - close general request - designated users - only accessible via search request results page

| <u>**Step No**</u> | <u>**Steps**</u> | <u>**Data**</u> | <u>**Expected Results**</u> | <u>**Pass/ Fail**</u> |
|---|---|---|---|---|
| 1 | From Request Details Page: designated user selects Close Request Button | | Close Request Page is displayed with an editable note field | |
| 2 | Designated user selects Return Button | | Previous View Request Details Page is displayed | |
| 2 | Designated user selects Submit button | No data entered | Error message: note field NOT NULL | |
| 3 | Designated user enters data; selects Submit Button | Note ="test" | Request Details Page is displayed with the status changed to closed AND "test" written in the note field. | |
| 4 | From Request Details Page: non-designated user selects Close Request Button | | Close Request Page is displayed with an editable note field | |
| 5 | Designated user enters data; selects Submit Button | Note ="test" | Error message: user does not have permission to close requests. | |

## C.6 Physical Assets Test Cases - add physical asset - designated users Role 2 and Role 3-

| Step No | Steps | Data | Expected Results | Pass/ Fail |
|---|---|---|---|---|
| 1 | Non Designated User selects Physical Assets Tab | | Add Asset Button is NOT available | |
| 2 | Designated User selects Physical Assets tab | | Add Asset Button is available | |
| 3 | Designated User selects Add Asset button | | Add Asset Page is displayed | |
| 4 | Designated User enters data; selects Submit Button | Barcode = NULL Owner="ENCS" Category="Equipment" | Error message: must fill out mandatory fields | |
| 5 | Designated User enters data; selects Submit Button | Barcode="test1234" Owner="NULL" Category="Equipment" | Error message: must fill out mandatory fields | |
| 6 | Designated User enters data; selects Submit Button | Barcode="test1234" Owner="ENCS" Category=NULL | Error message: must fill out mandatory fields | |
| 7 | Designated User enters data; selects Submit Button | Barcode="test1234" Owner="ENCS" Category="Furniture" Furniture Type=NULL | Error message: must fill out mandatory fields | |
| 8 | Designated Role 2 User enters data; selects Submit Button | Barcode="test1234" Owner="wrong"(not user's faculty) Category="Equipment" | Error Message: User can add asset in user's faculty only | |
| 9 | Designated User enters data; selects Submit Button | Barcode="test1234" Owner="ENCS" Category="Equipment" | System prompts for confirmation | |
| 10 | Designated User selects Return Button | | Add Asset Page is displayed with current information | |
| 11 | Designated User selects Confirm Button | | Confirmation Message: Asset ID # was created; add asset page is displayed with empty fields | |

## C.7 Physical Assets Test Cases - search physical asset - designated users Role 2 and Role 3

| Step No | Steps | Data | Expected Results | Pass/ Fail |
|---|---|---|---|---|
| 1 | Non Designated User selects Physical Assets Tab | | Search Asset Button is NOT available | |
| 2 | Designated User selects Physical Assets tab | | Search Asset Button is available | |
| 3 | Designated User selects Search Asset Button | | Search Asset Page is displayed | |
| 4 | Designated Role 2 User enters data; selects Search Button | Owner="wrong" (enters faculty other than user's faculty) | Search result is NULL | |
| 5 | Designated Role 2 User enters data; selects Search Button | Various | Search Results Page Displayed; Search result contains ONLY those assets in the user's own faculty, regardless of other constraints | |
| 6 | Designated Role 3 User enters data; selects Search Button | Various | Search results page displayed; No additional constraints | |

## C.8 Physical Asset Test Cases - view physical asset - designated users Role 2 and Role 3- only accessible via search Asset results page

| Step No | Steps | Data | Expected Results | Pass/ Fail |
|---|---|---|---|---|
| 1 | User selects radio button next to the asset list entry; selects View Asset Button | | View Asset Details Page is displayed; | |
| 2 | User selects Return Button | | Search asset results page displays the results from the last query | |

## C.9 Physical Asset Test Cases - update physical asset - designated users Role 2 and Role 3- only accessible via search View Asset Details page

| Step No | Steps | Data | Expected Results | Pass/ Fail |
|---|---|---|---|---|
| 1 | User selects Update Physical Asset Button | | Edit Physical Asset Page is displayed; | |
| 2 | Examine the following fields: i. Asset ID;<br>ii. Barcode;<br>iii. Purchase requisition number;<br>iv. Purchase order number;<br>v. Manufacturer;<br>vi. Model;<br>vii. Category;<br>viii. Furniture type if applicable;<br>ix. Storage unit type if applicable;<br>x. Equipment type if applicable;<br>xi. Equipment serial number if applicable;<br>xii. Computer type if applicable; | | These fields are NOT editable. All other fields are editable. | |
| 3 | Role 2 User enters data; Selects Submit Button | Owner ="edit" | Error message: Faculty User doesn't have permission for this function | |
| 4 | User enters data; Selects Submit Button | Various | System prompts for confirmation | |
| 5 | Designated User selects Return Button | | Edit Physical Asset Page is displayed with current information | |
| 6 | Designated User selects Confirm Button | | Confirmation Message: Asset ID # was updated Asset Details Peg is displayed with updated information | |

## C.10 Physical Assets Test Cases - create group - designated users Role 2 and Role 3

| Step No | Steps | Data | Expected Results | Pass/ Fail |
|---|---|---|---|---|
| 1 | Non Designated User selects Physical Assets Tab | | Create Group Button is NOT available | |
| 2 | Designated User selects Physical Assets Tab | | Create Group Button is available | |
| 3 | Designated User selects Create Group Button | | Create Group page is displayed | |
| 4 | Designated User selects Submit | No data entered | Error Message: Group must contain at least one asset | |
| 5 | Designated Role 2 User enters data; selects Submit | One or more assets entered belong to a different faculty than the user | Error Message: Group must contain only assets from your faculty | |
| 6 | Designated User enters data; selects Submit | Asset#1="invalid asset ID" | Error Message: Asset #1 has invalid Asset ID | |
| 7 | Designated User enters data; selects Submit | Location="invalid location ID" | Error Message: Invalid Location ID | |
| 8 | Designated User enters data; selects Submit | Assigned User= "invalid user ID" | Error Message: Invalid User ID | |
| 9 | Designated User enters data; selects Submit | Various | Systems prompts for confirmation | |
| 10 | Designated User selects Confirm Button | | Confirmation Message: Group ID # was updated Create Group Page is displayed with empty fields | |

## C.11 Physical Assets Test Cases - view edit update group - designated users Role 2 and Role 3-

| Step No | Steps | Data | Expected Results | Pass/ Fail |
|---|---|---|---|---|
| 1 | Non Designated User selects Physical Assets Tab | | View/Update/Delete Group Button is NOT available | |
| 2 | Designated User selects Physical Assets Tab | | View/Update/Delete Group Button is available | |
| 3 | Designated User selects View/Update/Delete Group Button | | Retrieve Group page is displayed | |
| 4 | Designated User enters data; selects Retrieve Button | Group id = "invalid id" | Error Message: Invalid ID | |
| 5 | Designated User enters data; selects Retrieve Button | Group id = "valid id" | Group Details page is displayed | |
| 6 | Designated User enters data; selects Submit | Asset#1="invalid asset ID" | Error Message: Asset #1 has invalid Asset ID | |
| 7 | Designated User enters data; selects Submit | Location="invalid location ID" | Error Message: Invalid Location ID | |
| 8 | Designated User enters data; selects Submit | Assigned User= "invalid user ID" | Error Message: Invalid User ID | |
| 9 | Designated Role 2 User enters data; selects Submit | One or more assets entered belong to a different faculty than the user | Error Message: Group must contain only assets from your faculty | |
| 11 | Designated User enters data; Selects Submit Button | Various | System prompts for confirmation | |
| 12 | Designated User selects Return Button | | Group Details Page is displayed with current information | |
| 13 | Designated User selects Confirm Button | | Confirmation Message: Group ID # was updated; Group Details Page is displayed with current information | |
| | | | | |

# TEST CSS

The validation and testing of our CSS code is done using the W3C validation tool at the following address:
http://jigsaw.w3.org/css-validator/
The CSS validator will help us determine if we have respected the syntax as well as the rules of CSS code.

First run on the CSS validation tool highlights an error on line 160.

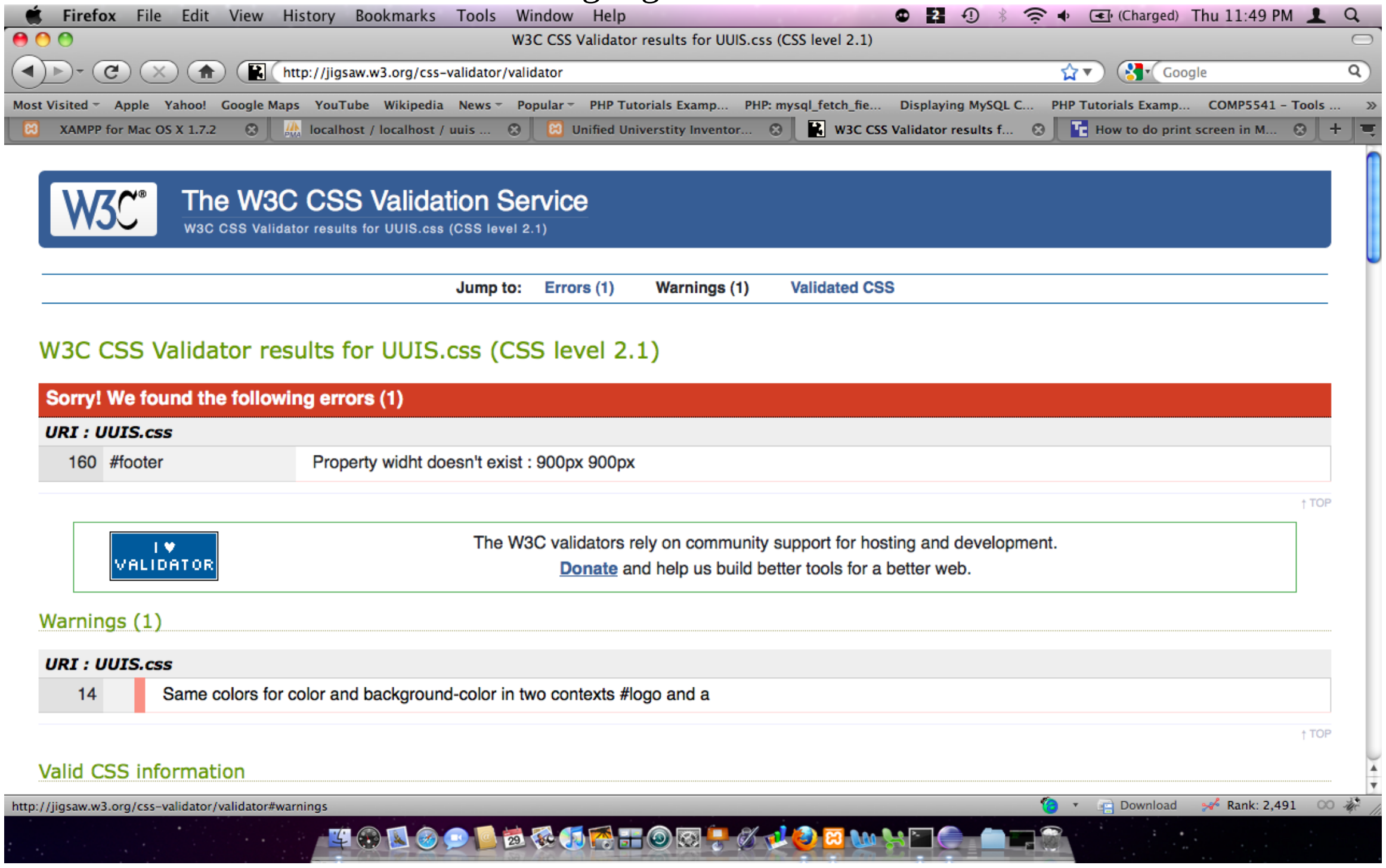

The error highlighted was due to a typo on the width.

Our second instance of validation returned the following screen:

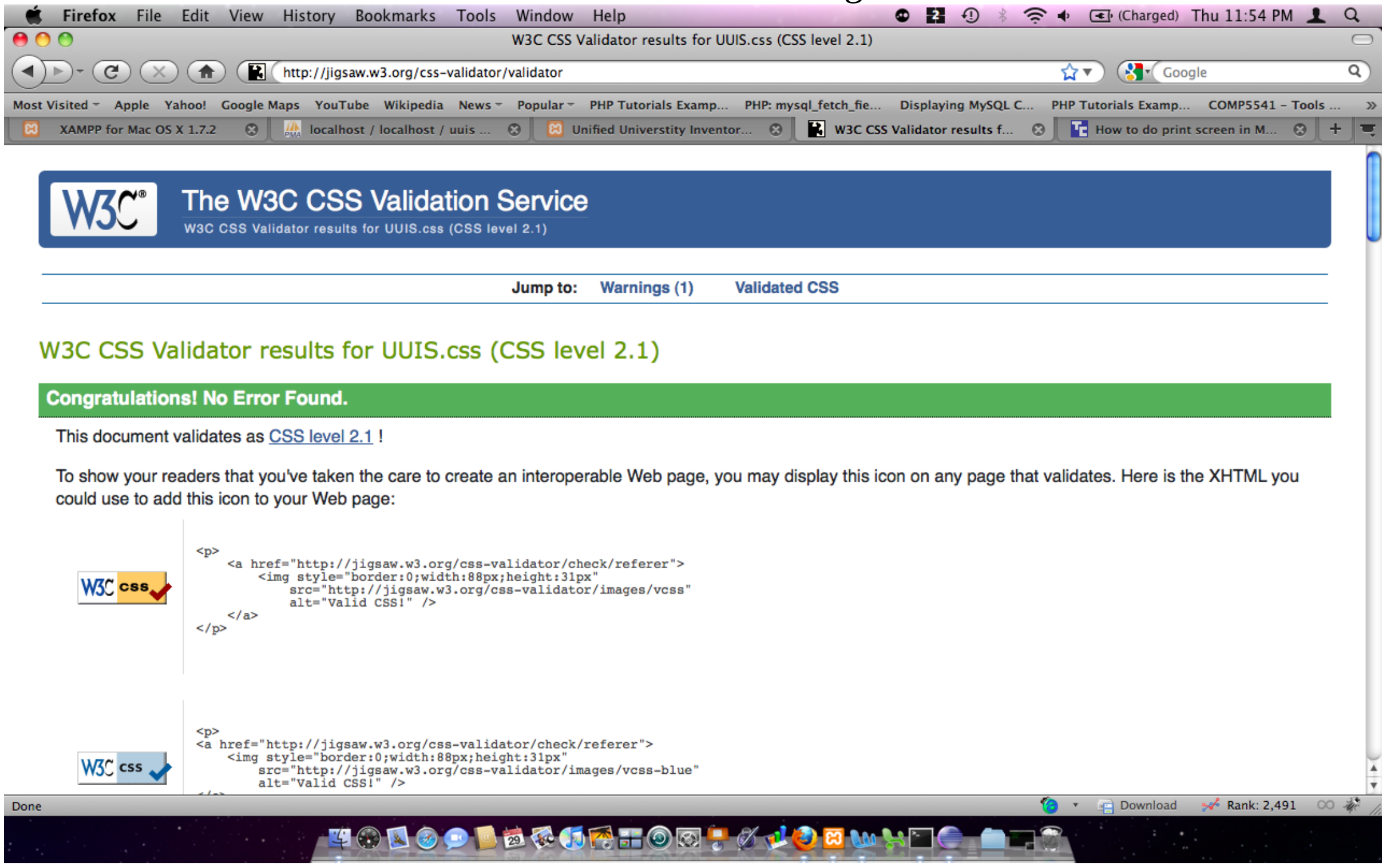

As we can see the second test returns no errors. Our CSS follows the W3C standard of CSS coding.

# Appendix D – COCOMO Estimate

## Estimate1 - Schedule Report

Costar 7.0 Demo 03/08/2010 15:53:21 Page: 1

Estimate Name: Estimate1
Model Name: COCOMO II 2000
Process Model: COCOMO II Model
Estimate ID:
Model ID: 2000
Phases: Waterfall

| Month | Effort this Month (Person-Months) RQ | PD | DD | CT | IT | Total | Cumulative Effort | Cost (K$) This Month | Cumulative Cost (K$) |
|---|---|---|---|---|---|---|---|---|---|
| 1 | 0.5 | 0.0 | 0.0 | 0.0 | 0.0 | 0.5 | 0.5 | 2.6 | 2.6 |
| 2 | 0.1 | 0.6 | 0.0 | 0.0 | 0.0 | 0.7 | 1.2 | 4.0 | 6.6 |
| 3 | 0.0 | 0.6 | 0.2 | 0.0 | 0.0 | 0.8 | 2.0 | 4.7 | 11.3 |
| 4 | 0.0 | 0.0 | 1.2 | 0.0 | 0.0 | 1.2 | 3.2 | 6.8 | 18.1 |
| 5 | 0.0 | 0.0 | 0.6 | 0.6 | 0.0 | 1.2 | 4.4 | 6.9 | 25.0 |
| 6 | 0.0 | 0.0 | 0.0 | 1.2 | 0.0 | 1.2 | 5.6 | 7.0 | 32.0 |
| 7 | 0.0 | 0.0 | 0.0 | 0.8 | 0.3 | 1.2 | 6.8 | 6.6 | 38.6 |
| 8 | 0.0 | 0.0 | 0.0 | 0.0 | 1.0 | 1.0 | 7.8 | 5.7 | 44.3 |
| 9 | 0.0 | 0.0 | 0.0 | 0.0 | 0.1 | 0.1 | 7.9 | 0.7 | 45.0 |

**Figure C. 1**– UUIS Schedule – COCOMO Report (1 of 3)

## Estimate1 - Activity Report

Costar 7.0 Demo 03/08/2010 15:55:06 Page: 1

Estimate Name: Estimate1
Model Name: COCOMO II 2000
Process Model: COCOMO II Model
Estimate ID:
Model ID: 2000
Phases: Waterfall

| Activity | Effort in Person-Months RQ | PD | DD | CT | IT | Total RQ to IT | MN |
|---|---|---|---|---|---|---|---|
| Requirements | 0.2 | 0.2 | 0.1 | 0.1 | 0.0 | 0.6 | 0.0 |
| Product Design | 0.1 | 0.5 | 0.2 | 0.2 | 0.1 | 1.0 | 0.0 |
| Programming | 0.0 | 0.2 | 1.1 | 1.5 | 0.5 | 3.3 | 0.0 |
| Test Plans | 0.0 | 0.1 | 0.1 | 0.1 | 0.0 | 0.3 | 0.0 |
| V & V | 0.0 | 0.1 | 0.1 | 0.2 | 0.5 | 0.9 | 0.0 |
| Project Office | 0.1 | 0.2 | 0.1 | 0.2 | 0.1 | 0.7 | 0.0 |
| CM/QA | 0.0 | 0.0 | 0.1 | 0.2 | 0.1 | 0.5 | 0.0 |
| Manuals | 0.0 | 0.1 | 0.1 | 0.2 | 0.1 | 0.5 | 0.0 |
| Totals | 0.5 | 1.3 | 2.0 | 2.7 | 1.4 | 7.9 | 0.0 |

**Figure C. 2** – UUIS Activity – COCOMO Report (2 of 3)

## Estimate1 - Detail Report

Costar 7.0 Demo 03/08/2010 15:52:24 Page: 1

| | | | |
|---|---|---|---|
| Estimate Name: | Estimate1 | Estimate ID: | |
| Model Name: | COCOMO II 2000 | Model ID: | 2000 |
| Process Model: | COCOMO II Model | Phases: | Waterfall |
| Component Name: | Component1 | Component ID: | |
| Increment: | 1 | Level: | 1 |
| Developed Size: | 3,000 | EAF: | 0.7355 |

| Phase | Effort (Person-Months) | Cost (K$) | Duration (Months) | Staffing |
|---|---|---|---|---|
| RQ -- Requirements | 0.5 | 2.9 | 1.1 | 0.5 |
| PD -- Product Design | 1.3 | 7.2 | 1.7 | 0.7 |
| DD -- Detailed Design | 2.0 | 11.3 | 1.7 | 1.2 |
| CT -- Code & Unit Test | 2.7 | 15.4 | 2.2 | 1.2 |
| IT -- Integration & Test | 1.4 | 8.2 | 1.4 | 1.0 |
| Development (PD+DD+CT+IT) | 7.4 | 42.1 | 7.0 | |
| Totals (RQ+PD+DD+CT+IT) | 7.9 | 45.0 | 8.1 | |
| MN -- Maintenance (per year) | 0.0 | 0.0 | | 0.0 |

**Figure C. 3 – UUIS Detail – COCOMO Report (3 of 3)**